\documentclass[a4paper,11pt]{article}
\pdfoutput=1 % if your are submitting a pdflatex (i.e. if you have
\usepackage{jheppub} % for details on the use of the package, please
\usepackage[T1]{fontenc} % if needed
\usepackage{subfigure}
\usepackage{float}
\usepackage{amssymb}
\usepackage{subfigure}
\usepackage[utf8]{inputenc}
\usepackage{cleveref}
\usepackage{soul}

\newcommand{\wpg}{\omega_{\rm Q}}
\newcommand{\wdiff}{\omega_{\rm D}}
\newcommand{\wgl}{\omega_{\rm G}}
\newcommand{\wus}{\omega_{\rm U}}

\newcommand{\hatbeta}{\hat{\beta}}

\newcommand{\psg}{quasi-hydro}

\title{%\boldmath 
Quasinormal modes of a semi-holographic black brane and thermalization}

\author[a]{Sukrut Mondkar,}
\author[a]{Ayan Mukhopadhyay,}
\author[b]{Anton Rebhan}
\author[c]{and Alexander Soloviev}

\affiliation[a]{Center for Strings, Gravitation and Cosmology, Department of Physics, Indian Institute of Technology Madras, Chennai 600036, India}
\affiliation[b]{Institut f\"{u}r Theoretische Physik, Technische Universit\"{a}t Wien, 
Wiedner Hauptstr.~8-10, A-1040 Vienna, Austria}
\affiliation[c]{Center for Nuclear Theory, Department of Physics and Astronomy, Stony Brook University, Stony Brook, New York 11794, USA}

\emailAdd{sukrut@physics.iitm.ac.in}
\emailAdd{ayan@physics.iitm.ac.in}
\emailAdd{anton.rebhan@tuwien.ac.at}
\emailAdd{alexander.soloviev@stonybrook.edu}

\abstract{We study  the quasinormal modes and non-linear dynamics of a simplified model of semi-holography, which consistently integrates mutually interacting perturbative and strongly coupled holographic degrees of freedom such that the full system has a total conserved energy. We show that the thermalization of the full system can be parametrically slow when the mutual coupling is weak. For typical homogeneous initial states, we find that initially %reverse 
energy is transferred from the black brane to the perturbative sector, 
later giving way to complete transfer of energy to the black brane at a slow and constant rate, while the entropy grows monotonically for all time. Larger mutual coupling between the two sectors leads to larger extraction of energy from the black brane by the boundary perturbative system, but also quicker irreversible transfer of energy back to the black brane. The quasinormal modes replicate features of a dissipative system with a softly broken symmetry including the so-called $k$-gap. Furthermore, when the mutual coupling is below a critical value, there exists a hybrid zero mode with finite momentum which becomes unstable at higher values of momentum, indicating a Gregory-Laflamme type instability. This could imply turbulent equipartitioning of energy between the boundary and the holographic degrees of freedom in the presence of inhomogeneities. }
\begin{document} 

\maketitle
\flushbottom

\section{Introduction}
\label{sec:intro}
Semi-holography introduces a way to model the complex dynamics of quantum field theories with asymptotic freedom. In this approach, one combines the perturbative description of weakly self-interacting ultra-violet degrees of freedom with a holographic description for those at lower energies %\st{that are}
and strongly self-interacting. This approach allows for flexibility in phenomenologically modelling open quantum systems, particularly in the case of a weakly self-interacting system coupled to a strongly self-interacting quantum critical bath \cite{Faulkner:2010tq,Mukhopadhyay:2013dqa,
Doucot:2020fvy}. %,Kurkela:2018dku}.

The crucial ingredient in the semi-holographic construction is the ``democratic'' coupling between the two sectors which allows one to extract the low energy dynamics of the full system from the effective dynamics of the subsectors at any scale \cite{Banerjee:2017ozx,Kurkela:2018dku}. In this coupling scheme, the subsectors are subject to marginal/relevant deformations in their couplings and effective background metrics which are determined by the local operators of the other sector %in a way that leads to the explicit construction of a 
such that there is a local and conserved energy-momentum tensor of the full system in the physical background metric. The full dynamics can be obtained by solving the dynamics of both systems self-consistently in an iterative procedure \cite{Iancu:2014ava,Mukhopadhyay:2015smb,Ecker:2018ucc}. Note that although one considers both subsectors at any scale, it is expected that the perturbative sector should dominate the ultraviolet behavior while the infrared behavior will be governed by the dynamics of the dynamical black hole horizon of the holographic sector.\footnote{We explicitly find that the perturbative sector dominates the dynamics of the hybrid hydrodynamic attractor at early times when the energy densities are large as discussed below.}

Recently, the hydrodynamic attractor of such a hybrid system has been constructed by simplifying the description of both sectors to fluids \cite{Mitra:2020mei}. It was found that the ratio of the energy densities of the strongly self-interacting to the weakly self-interacting sectors universally diverges as one approaches early proper time in Bjorken flow, confirming the expectation borne out of studies of perturbative QCD \cite{Baier:2000sb} that suggests such a bottom-up thermalization scenario. In explicit numerical simulations involving a black hole, it was found that there is irreversible transfer of energy from both the perturbative sector and the mutual interaction energy to the black hole whose apparent horizon grows monotonically. The rate of this irreversible transfer is very slow when the coupling between the subsectors is weak \cite{Ecker:2018ucc}.

In this work, we present a simplified model of semi-holography which allows us to {investigate} the low energy dynamics and thermalization via the study of quasinormal modes (QNMs). In particular, we obtain robust understanding of why homogeneous thermalization of the full system can be parametrically slow, and how one can have inverse transfer of energy from the holographic to the weakly coupled sector at intermediate stages as observed in the simplified context of the hybrid hydrodynamic attractor.  Another crucial question to which we gain new insights is whether one can have a mechanism in which there can be equipartitioning of energy between the holographic and the perturbative systems. We find instabilities that can lead to such a possibility in the presence of inhomogeneities. 

The presence of a weakly broken symmetry plays a crucial role in our setup. We are also able to make connections with the broad literature of applicability of the quasi-hydrodynamic paradigm \cite{Kovtun:2012rj,Grozdanov:2017ajz} in such systems \cite{Grozdanov:2018fic} along with bounds on transport coefficients \cite{Hartnoll:2014lpa,Blake:2016wvh,Blake:2016sud,Hartman:2017hhp,Lucas:2017ibu,Grozdanov:2020koi}. The novel feature of our model is that in addition to a diffusive Goldstone mode and a \psg\ mode, a third mode that is purely imaginary for small mutual coupling plays a crucial role in low energy dynamics creating potential instability towards turbulent/glassy dynamics.

The plan of the paper is as follows. In Section \ref{Sec:Model}, we introduce our simplified semi-holographic model explicitly detailing our motivation. 
% In Section \ref{Sec:Method}, we provide the technical details of the methodology of computation of quasinormal modes and full non-linear dynamics of this model. 
In Section \ref{Sec:QNMresults}, we provide the technical details of the methodology of computation of QNMs. We then present our results for the homogeneous and inhomogeneous QNMs, and discuss their implications. In Section \ref{Sec:Nonlinear}, we present the homogeneous non-linear dynamics of our model and investigate the parametrically slow thermalization. In Section \ref{Sec:Conclude}, we discuss the questions raised by our results. In the appendices %\ref{Sec:Appendix} 
we provide supplementary information about the numerical implementation.

\section{The simplified semi-holographic model}\label{Sec:Model}

The construction of the semi-holographic framework is based on the following principles:
\begin{enumerate}
\item The non-perturbative part of the dynamics can be described by a strongly coupled large $N$ holographic theory.
\item The interactions between the perturbative and holographic sectors can be described by the marginal and relevant deformations of the respective theories. The marginal and relevant couplings, and the effective background metric of each sector are promoted to ultralocal algebraic functions of the operators of the other sector in such a way that the full system has a local and conserved energy-momentum tensor in the physical background metric \cite{Banerjee:2017ozx,Kurkela:2018dku}. This coupling scheme is called \textit{democratic coupling} \cite{Banerjee:2017ozx}.
\item The full dynamics should be solved self-consistently. This can be achieved via an iterative procedure in which the dynamics of each system is solved with couplings and the effective background metrics set to the values in the previous iteration until convergence is reached. The initial conditions should be held fixed throughout the iteration \cite{Iancu:2014ava}.
\end{enumerate}
Numerous non-trivial examples explicitly demonstrate that the iterative procedure converges \cite{Mukhopadhyay:2015smb,Ecker:2018ucc} with the present article serving as another such demonstration. 

\subsection{{Review of the semiholographic glasma model}}

As an illustration, we consider a marginal scalar coupling between classical Yang-Mills theory and a holographic strongly coupled large $N$ conformal gauge theory dual to classical Einstein's gravity with a negative cosmological constant. The dynamics of this system have been studied in \cite{Ecker:2018ucc} to gain insights into the possible non-perturbative dynamics of the color glass condensate, i.e. for understanding how the strongly interacting soft sector affects the initially overoccupied gluons at saturation scale which can be described by the classical Yang-Mills field equations. 

The action for the full system  in $d$ spacetime dimensions is
\begin{equation}\label{Eq:Action1}
S= -\frac{1}{4g_{YM}^2}\int{\rm d}^d x\, \left(1 +\chi(x)\right) {\rm tr}(F^2) + W_{\rm CFT}[h(x)] - \frac{1}{\beta}\int{\rm d}^d x \chi(x) h(x),
\end{equation}
where $\chi(x)$ represents the deformation of the Yang-Mills coupling and $h(x)$ is a source for 
\begin{equation}\label{Eq:H}
\mathcal{H}(x)\equiv \frac{\delta W_{\rm CFT}[h(x)] }{\delta h(x)},
\end{equation}
a marginal operator in the holographic conformal gauge theory (CFT) with $W_{\rm CFT}$ being the logarithm of its partition function. The inter-system coupling, $\beta$, has mass dimension $-d$. The equations of motion for the auxiliary fields $\chi(x)$ and $h(x)$ lead to
\begin{equation}
\chi(x) = \beta \mathcal{H}(x), \quad h(x) = - \frac{\beta}{4 g_{YM}^2} {\rm tr}(F^2).
\end{equation}
Substituting the above back in the action \eqref{Eq:Action1}, we obtain
\begin{equation}\label{Eq:Action2}
S= -\frac{1}{4g_{YM}^2}\int{\rm d}^d x\;  {\rm tr}(F^2) + W_{\rm CFT}\left[h(x) = - \frac{\beta}{4 g_{YM}^2} {\rm tr}(F^2)\right].
\end{equation}
Finally, % we note that 
the holographic correspondence %allows the computation of 
defines $W_{\rm CFT}$ via 
\begin{equation}\label{Eq:dic1} 
W_{\rm CFT}\left[h(x) = - \frac{\beta}{4 g_{YM}^2} {\rm tr}(F^2)\right] = S_{\rm grav}\left[\phi^{(0)}(x) = - \frac{\beta}{4 g_{YM}^2} {\rm tr}(F^2) \right],
\end{equation}
where $ S_{\rm grav}$ is the renormalized on-shell action of the dual $(d+1)$-dimensional gravitational theory which can be taken to be simply Einstein's gravity coupled to a massless dilaton field $\Phi$ $dual$ to the CFT operator $\mathcal{H}(x)$. The non-normalizable mode of the dilaton,
\begin{equation}\label{Eq:BC-dilaton}
\phi^{(0)}(x)\equiv \lim_{r\rightarrow 0} \Phi(r,x), 
\end{equation}
which specifies its boundary value (the boundary of the bulk spacetime is at $r=0$), is identified with the source $h(x)$ that couples to the dual operator $\mathcal{H}(x)$.
%This completes our description of the dynamics of the full system. 

It is easy to see from the action \eqref{Eq:Action2} that the full system has a conserved energy-momentum tensor which takes the form
\begin{equation}\label{Eq:Tmn}
T^{\mu\nu} = t^{\mu\nu}_{\rm YM} + \mathcal{T}^{\mu\nu} + \frac{\beta}{4 g_{YM}^2} {\rm tr}(F^2)\mathcal{H} \eta^{\mu\nu},
\end{equation}
where $t^{\mu\nu}_{\rm YM}$ is the energy-momentum tensor of the Yang-Mills theory and $\mathcal{T}^{\mu\nu}$ is that of the holographic sector. The full dynamics of the system was solved in \cite{Ecker:2018ucc}. Convergence was reached typically in 4 iterations as demonstrated by $\partial_\mu T^{\mu\nu} =0$ being satisfied for all time to a very good accuracy.\footnote{The pile-up of numerical error breaks down the accuracy at very large time, however we could successfully extract the nature of the large time behavior. In particular, we were able to show the complete transfer of energy to the growing holographic black hole with both $t^{\mu\nu}_{YM}$ and the interaction term in the total energy $T^{\mu\nu}$ vanishing asymptotically at large time.} We will present a simpler version of this set-up below and also  the equations of motion explicitly. At this point, it could be mentioned that the most general democratic scalar couplings were found in \cite{Banerjee:2017ozx}.

%We now try to formulate the semi-holographic problem. 
The above model demonstrated that the energy in the Yang-Mills sector gets transferred completely to a growing black hole in the bulk holographic geometry for homogeneous initial conditions even if the bulk geometry is initially \textit{empty}, i.e. a vacuum anti-de Sitter space with a vanishing dilaton.\footnote{It was shown in \cite{Ecker:2018ucc} that we can start with such vacuum initial conditions by taking a suitable numerical limit in which the mass of an initial seed black hole is sent to zero.} At late times, both $t^{\mu\nu}_{YM} $ and the interaction term in the total energy-momentum tensor \eqref{Eq:Tmn} decayed. The rate of transfer of energy to the black hole, as mentioned in the Introduction, was controlled by the mutual coupling $\beta$ and was very slow for small $\beta$. It is to be noted that the model does not have any linear-coupling to the gauge field in the final thermal state. Since ${\rm tr}(F^2)$ vanishes at late times, the coupling to the bulk dilaton can only be quadratic in the gauge field. This does not allow us to relate the transfer of energy to the black hole to a quasi-normal mode of the full system easily. This motivates {the} simpler construction {below}.

{The democratic coupling allows flexibility of constructing phenomenological models that capture the low energy dynamics of the full system based on effective descriptions of the subsystems. This has enabled understanding of the hydrodynamics of the composite system based on effective metric coupling of two fluids \cite{Kurkela:2018dku} and a preliminary study of the hybrid hydrodynamic attractor \cite{Mitra:2020mei}. However, it is important to retain the dynamical black hole for capturing the infrared dynamics of the holographic sector in order to obtain the late time behavior and understand thermalization of the full system.}

The effective metric coupling, unlike the scalar coupling discussed above, leads to hybridization of the thermal fluctuations of the black hole and the perturbative system. To understand hydrodynamization and thermalization in semi-holography, we should study the hybrid system of a gas of gluons described by kinetic theory coupled to the black hole by an effective metric coupling. The linearized hybrid modes were studied in the simpler version in \cite{Kurkela:2018dku} in which the black hole was substituted by a fluid. Here, we retain the black hole, but replace the kinetic theory by a massless scalar field, and the effective metric coupling by a linear scalar coupling. We find that the resulting simplified model retains many characteristics of the more complex models explored so far 
%(including the possibility of reverse transfer of energy from the holographic to the weakly coupled sector along with irreversible transfer of energy back to the black hole in the final stage), 
and gives several new insights.

\subsection{{Novel scalar semiholography}}

The simplified model introduces %considers 
only a massless gauge-invariant scalar %{meson/baryon }
field at the boundary that couples to a black hole  in the bulk via a (massless) bulk dilaton. Since such a field is gauge-invariant, we can couple it linearly to the holographic system unlike the gauge field which can couple only via ${\rm tr}(F^2)$, the energy-momentum tensor, etc. Therefore, instead of \eqref{Eq:Action2}, we can consider the following action:
\begin{equation}\label{Eq:Action3}
S= -\frac{1}{2}\int{\rm d}^d x \,\, \partial_\mu\chi \partial^\mu\chi + W_{\rm CFT}\left[h(x) = - \beta \chi(x) \right],
\end{equation}
with $h(x)$ being the source of a marginal operator $\mathcal{H}(x)$ of the CFT. 
{Note that here the inter-system coupling $\beta$ has mass dimension $-(d-2)/2$, different
than the one introduced in (\ref{Eq:Action1}).}
Furthermore, $$ W_{\rm CFT}\left[h(x) = - \beta \chi(x) \right] = S_{\rm grav}\left[\phi^{(0)}(x) = - \beta \chi(x)  \right],$$
where $ S_{\rm grav}$ is the renormalized on-shell action of the dual $(d+1)$-dimensional gravitational theory with a dilaton $\Phi$ whose boundary condition is given by \eqref{Eq:BC-dilaton}. It is easy to see that the energy-momentum tensor of the full system is 
\begin{equation}\label{Eq:Full-T}
T^{\mu\nu} = t^{\mu\nu}_{\chi} + \mathcal{T}^{\mu\nu} ,
\end{equation}
where $$ t^{\mu\nu}_{\chi} = \partial^\mu\chi\partial^\nu\chi - \frac{1}{2}\eta^{\mu\nu} (\partial_\alpha\chi\partial^\alpha\chi),$$ is the energy-momentum tensor of the boundary scalar field and $\mathcal{T}^{\mu\nu}$ is that of the holographic CFT. It will be shown below that the explicit equations of motion of the full system directly implies the conservation of the full energy-momentum tensor {with respect to the physical background metric $\eta_{\mu\nu}$}. Remarkably, the full energy-momentum tensor, unlike \eqref{Eq:Tmn}, is simply the sum of those of the two subsystems without an explicit interaction term. This feature will be helpful for us to deduce the dynamical consequences of the hybrid quasinormal modes. Linear semi-holographic couplings leading to such an energy-momentum tensor of the full system have been explored in other contexts in \cite{Joshi:2019wgi,Kibe:2020gkx}.\footnote{The linear couplings can be easily motivated also for fermions in the context of applications to condensed matter physics, since the boundary fermion is an electron which can be considered to be a gauge-neutral hadron made out of the partons of the holographic theory \cite{Faulkner:2010tq,Mukhopadhyay:2013dqa,Doucot:2017bdm,Doucot:2020fvy}.}

%It is obvious that 
The equation of motion for the boundary scalar field from \eqref{Eq:Action3} is 
\begin{equation}\label{Eq:chi}
\partial_{\mu}\partial^{\mu}\chi = \beta \mathcal{H},
\end{equation}
where we have used \eqref{Eq:H}. The Ward identity (following from the diffeomorphism invariance) of the holographic theory implies that
\begin{equation}\label{Eq:WICFT}
\partial_\mu \mathcal{T}^{\mu\nu} = \mathcal{H}\,\partial^\nu h = - \beta \,\mathcal{H}\,\partial^\nu \chi.
\end{equation}
It is then easy to see from \eqref{Eq:chi} and \eqref{Eq:WICFT} that the full energy-momentum tensor given by \eqref{Eq:Full-T} is indeed conserved, because
\begin{equation}
    \partial_\mu t^{\mu\nu}_{\chi}=\beta \,\mathcal{H}\,\partial^\nu \chi
    +\partial^\mu\chi\,\partial_\mu \partial^\nu\chi-\frac12\partial^\nu(\partial_\mu\chi\partial^\mu\chi)=\beta \,\mathcal{H}\,\partial^\nu \chi.
\end{equation}

Explicitly, the equations of motion for the metric and dilaton in the $(d+1)$-dimensional gravitational theory dual to the holographic strongly coupled large $N$ CFT are
\begin{eqnarray}\label{Eq:GravEoms}
R_{MN} - \frac{1}{2} R G_{MN} - \frac{d(d-1)}{2L^2} G_{MN}  &=& \kappa\left(\nabla_M \Phi\nabla_N \Phi -\frac{1}{2}G_{MN}\nabla_P\Phi\nabla^P\Phi\right),\nonumber\\
G^{MN}\nabla_M\nabla_N \Phi &=& 0.
\end{eqnarray}
Since all physical quantities of the gravitational theory should be measured in units of the AdS radius,\footnote{e.g. the dimensionless Planck's constant is related to the rank of the gauge group via $M_{Pl}L^{d-1} \approx N^2$} $L$, we set $L= 1$ for convenience. The generic solutions of the equations of motion have the following expansion in the Fefferman-Graham coordinates in which the radial coordinate is $r$, $G_{r\mu} =0$, $G_{rr} = 1/r^2$ and the boundary is at $r=0$:
\begin{eqnarray}
G_{\mu\nu}&=& \frac{1}{r^2}\left(g^{(0)}_{\mu\nu} + \cdots +r^d g^{(d)}_{\mu\nu}+ \mathcal{O}(r^d \log r) \right), \\
\Phi &=& \phi^{(0)} + \cdots + r^d \phi^{(d)} +\mathcal{O}(r^d \log r).
\end{eqnarray}
The log terms above appear specifically for even $d$ and capture the conformal anomaly of the dual CFT. To specify a unique solution, we need to specify the sources $g^{(0)}_{\mu\nu}$ (a.k.a. the \textit{boundary metric}) and $\phi^{(0)}$ aside from providing the initial conditions. For the semi-holographic construction, these sources are determined by the gauge-invariant operators of the perturbative sector as discussed above. In the absence of effective metric couplings, %\st{we have simply} 
\begin{equation}
g^{(0)}_{\mu\nu} = \eta_{\mu\nu},
\end{equation}
so the boundary metric is %just 
the physical background metric. Furthermore, as implied by \eqref{Eq:Action3}, 
\begin{equation}\label{Eq:phi0chi}
\phi^{(0)} = - \beta \chi.
\end{equation}
The expectation values of the operators in the state of the CFT dual to the gravitational solution (determined now by the sources and the initial conditions) can be obtained from functional differentiation of the renormalized gravitational action \cite{Balasubramanian:1999re,deHaro:2000vlm,Skenderis:2002wp}. The results are
\begin{equation}
\mathcal{T}_{\mu\nu} = \frac{d}{\kappa} g^{(d)}_{\mu\nu} + \mathcal{X}_{\mu\nu}, \quad \mathcal{H} = \frac{d}{\kappa} \phi^{(d)} + \psi,
\end{equation}
where $\mathcal{X}_{\mu\nu}$ and $\psi$ are local functionals of the sources of the theory, namely $g^{(0)}_{\mu\nu}$ and $\phi^{(0)}$. The constraints of Einstein's equations imply two Ward identities, namely the conservation of $\mathcal{T}_{\mu\nu}$ given by \eqref{Eq:WICFT} and the trace condition,
\begin{equation}
g^{(0)}_{\mu\nu}\mathcal{T}^{\mu\nu}=\eta_{\mu\nu}\mathcal{T}^{\mu\nu} =0.
\end{equation}
%when the boundary metric $g^{(0)}_{\mu\nu}$ is flat. 
We will provide more explicit details in the ingoing Eddington-Finkelstein coordinates, which will be convenient for solving the dynamics numerically.

In what follows, we will explicitly analyze the hybrid quasinormal modes of this simplified semi-holographic system and study its non-linear dynamics. For numerical convenience, we will prefer to avoid the logarithmic terms in the radial expansion of the bulk fields. Therefore, we choose $d=3$ for which we obtain the coupled system of a three-dimensional massless scalar field and a dynamical black hole with a dilaton field in $AdS_4$.

 From \eqref{Eq:chi}, \eqref{Eq:GravEoms} and \eqref{Eq:phi0chi} 
{one can see} that our simplified model has %an ``axionic'' 
a global {shift} symmetry under which
\begin{equation}\label{Eq:Axionic}
\chi \rightarrow \chi + \chi_0, \quad \Phi \rightarrow \Phi - \beta\chi_0
\end{equation}
for a constant $\chi_0$. Note that {this} is a symmetry of the full non-linear theory. In the decoupling limit $\beta =0 $, the shifts of $\chi$ and $\Phi$ lead to two independent global symmetries. The coupling breaks these symmetries to the specific combination \eqref{Eq:Axionic}. Therefore, at finite but small $\beta$, the system has a \psg\ mode associated with a softly broken symmetry. It will be of fundamental interest to us as it will govern the relaxation dynamics of the system. 

{In what follows it is crucial that the diagonal shift-symmetry \eqref{Eq:Axionic} is exact. Our model
can therefore be interpreted also as a simple effective theory for a composite Goldstone boson interacting with a dissipative bath, where the
underlying spontaneous symmetry breaking in the full (boundary plus bulk) system is not part of
the model but takes place at a more fundamental level.}

% \AM{Our model can be applied to any shift-symmetric scalar but in what follows it is crucial that the diagonal shift-symmetry \eqref{Eq:Axionic} is not broken either explicitly or spontaneously. Therefore, it is natural to interpret our model as that of a Goldstone boson interacting with a dissipative bath. Note that the relevant spontaneous symmetry breaking (SSB) in the full (boundary plus bulk) system, which is responsible for the emergence of the (composite and hybrid) Goldstone, is itself not part of the model which retains only the relevant low energy modes. Thus our semi-holographic model provides a simple effective theory of a dissipative Goldstone mode.}

Generalized versions of our setup {may} help us to model the real time dynamics of the QCD axion and {may} be useful also for other phenomenological applications of holography. We will leave this for the future.

\section{%Results for the h
Hybrid quasinormal modes}\label{Sec:QNMresults}

\subsection{Quasinormal modes in semi-holography}\label{Sec:QNMMethod}

QNMs are eigenfunctions of linearized perturbations which characterize the relaxation/growth of perturbations that govern the system away from a thermal equilibrium state. The thermal equilibrium state of the semi-holographic model discussed in the previous section is that of the bulk geometry being a static black brane with a vanishing or constant dilaton field $\Phi$, while the boundary field $\chi$ also vanishes or is a constant. We will investigate {whether} this thermal background is stable against perturbations both linearly and non-linearly. In this subsection, we discuss how we can compute the hybrid quasinormal modes of this thermal equilibrium solution of the semi-holographic system, and save the nonlinear discussion for Sec.~\ref{Sec:NonlinearMethod}.
%we will discuss how we can compute the full non-linear dynamics with arbitrary (good) initial conditions in the following subsection. 
For reasons mentioned before, we will consider a $(2+1)$-dimensional system (the bulk dual to the holographic sector is therefore $(3+1)$-dimensional). The description of our method will be modelled on the pedagogical account presented in \cite{Yaffe} -- we will highlight the crucial modifications  brought in by the semi-holographic coupling. We also refer the reader to \cite{Berti:2009kk} for a comprehensive review of quasinormal modes of black branes.

The $AdS_4$-Schwarzschild black brane dual to the thermal holographic sector takes the following form in the ingoing Eddington-Finkelstein coordinates 
\begin{equation}\label{Eq:Sch-metric}
ds^2 =- 2 \frac{L^2}{r^2} dt dr -\frac{L^2}{r^2}(1 - M r^3) dt^2  + \frac{L^2}{r^2}(dx^2 + dy^2 ),
\end{equation}
where $M$ is the Arnowitt-Deser-Misner (ADM) mass of the black brane and $L$ is the AdS radius which we set to $1$%for the sake of convenience% as discussed before
. The dual thermal equilibrium state has a temperature 
\begin{equation}\label{Eq:HawkingT}
T = \frac{3 M^{\frac{1}{3}}}{4 \pi} = \frac{3}{4 \pi\, r_h},
\end{equation}
where $r = r_h = M^{-1/3}$ is the radial position of the horizon.
%QNM can be classified into different channels depending on which components of %the bulk metric
% \eqref{metric} are perturbed to generate them. In this paper, we will only work with the scalar channel. The equation of motion for a massless bulk scalar field is 

Here we will focus on the hybrid fluctuations of the bulk dilaton and the boundary scalar field. At the linearized level, the metric perturbations will be exactly the same as those in the purely holographic case because the coupling to the boundary scalar is quadratic. The massless bulk dilaton field obeys the Klein-Gordon equation %as shown in 
\eqref{Eq:GravEoms}. 
{With the background metric \eqref{Eq:Sch-metric},
a %The 
Fourier decomposition of the profile of the bulk dilaton %in terms of Fourier modes gives us
according to}
 \begin{equation}
\Phi(r, t, \vec{x}) = \int {\rm d}^2 k\int {\rm d}\omega \, \,e^{\mathrm{i}(\vec{k}\cdot \vec{x} - \omega t)} f(k,\omega,r)
\end{equation}
%Substituting this into the Klein-Gordon equation %, $\nabla_\mu \nabla^\mu \Phi=0$
% in the background metric \eqref{Eq:Sch-metric} 
yields
\begin{equation}\label{Eq:qnm-eq}
(M r^3 -1) f^{\prime \prime} (k,\omega,r) + \frac{M r^3 +2 - 2 \mathit{i} r \omega}{r} f^{\prime}(k,\omega,r) + \frac{k^2 r + 2 \mathit{i} \omega}{r}f(k,\omega,r) = 0.
\end{equation}
%is the massless bulk scalar field written in terms of its Fourier decomposition. 
Note that rotational symmetry of the black brane implies that the spectrum 
{only depends on}
%should be determined by 
$k =\sqrt{\vert\vec{k}\vert^2 }$.

%and satisfy the appropriate conditions at the boundary $r=0$ which will be determ. For a given value of wavevector q, such solutions only exist for a discrete set of frequencies $\omega = \omega_i$, which are the quasinormal mode frequencies. 

The bulk dilaton couples linearly to the boundary scalar $\chi$, so it cannot be solved in isolation. It is more convenient to write the boundary equation in terms of the non-normalizable mode $$\phi^{(0)}(\omega, k) = \lim_{r\rightarrow 0} f(k,\omega, r),$$ which according to \eqref{Eq:phi0chi} should equal to $-\beta \chi(\omega,k)$. The equation of motion for the boundary scalar field \eqref{Eq:chi} in Fourier space can then be rewritten as
\begin{equation}\label{Eq:qnm-eq2}
(\omega^2 - k^2)\phi^{(0)}(k,\omega) = - \beta^2 \mathcal{H}(k,\omega), 
\end{equation}
where $ \mathcal{H}(k,\omega)$ can be obtained from the renormalized on-shell action (see Sec.~\ref{Sec:NonlinearMethod} for more details) and is explicitly given by
\begin{equation}\label{Eq:H-lin}
\mathcal{H}(k,\omega) = 3 \phi^{(3)}(\omega, k) + i \omega^3 \phi^{(0)}(k,\omega) - \frac{3}{2}\mathit{i} \omega k^2\phi^{(0)}(k,\omega),
\end{equation}
where $\phi^{(3)}(\omega, k)$ is the $r^3$ term of the near-boundary radial expansion of any solution of \eqref{Eq:qnm-eq} that takes the form
\begin{eqnarray}\label{Eq:asymp-f}
f(k,\omega,r) &=& \phi^{(0)}(k,\omega)  - \mathit{i}\omega\phi^{(0)}(k,\omega) -\frac{k^2 r^2}{2}\phi^{(0)}(k,\omega) + r^3 \phi^{(3)}(k,\omega)\nonumber\\&&
+r^4 \left(\left(-\frac{1}{8}k^4-\frac{1}{4}\mathit{i}M\omega\right)\phi^{(0)}(k,\omega) -\mathit{i}\omega\phi^{(3)}(k,\omega) \right) + \mathcal{O}(r^5).
\end{eqnarray}

Physical solutions of the hybrid system corresponding to a causal response to perturbations must be ingoing at the horizon $r=r_h$. A generic solution of \eqref{Eq:qnm-eq} behaves as 
\begin{equation}\label{Eq:ingoing-bc}
f(k,\omega, r) \approx c_1  + c_2 (r - r_h)^{\frac{2\mathit{i}\omega}{3 M^{1/3}}},
\end{equation}
and the ingoing boundary condition means setting $c_2 =0$. The solution must also satisfy the semi-holographic boundary condition at $r = 0$ given by \eqref{Eq:qnm-eq2} and \eqref{Eq:H-lin} which amounts to specifying a relation between $\phi^{(0)}(k,\omega)$ and $\phi^{(3)}(k,\omega)$, the two independent coefficients of the near-boundary radial expansion \eqref{Eq:asymp-f} of $f(k,\omega, r)$. For a given value of wave vector $k$, such solutions satisfying the boundary conditions at both $r =0$ and $r = r_h$ can only exist for a discrete set of frequencies $\omega = \omega_i(k)$, which are the complex QNM frequencies. 

We readily note that when $\beta = 0$, the boundary condition at $r=0$ reduces simply to $\phi^{(0)} = 0$ as evident from \eqref{Eq:qnm-eq2} and \eqref{Eq:H-lin}. In this case, we get the usual conditions for the QNMs which require the linearized fluctuations to be normalizable. Since QNMs are intrinsic fluctuations of a system, we require them to exist source-free. This precisely implies $\phi^{(0)} = 0$ when the holographic system is decoupled from the boundary degrees of freedom. We thus reproduce the usual QNMs in the $\beta \rightarrow 0$ limit, along with the $\omega = \pm k$ modes of the decoupled boundary massless scalar field. Note that at finite mutual coupling $\beta$, the boundary conditions at $r=0$ given by \eqref{Eq:qnm-eq2} and \eqref{Eq:H-lin} still imply that the quasinormal mode fluctuations are intrinsic, i.e. they can exist without any external source. These equations impose the condition that the full hybrid system is not subjected to any external force.

{The numerical method of determining the quasinormal modes by imposing both the ingoing boundary condition \eqref{Eq:ingoing-bc} at the horizon and \eqref{Eq:qnm-eq2} at the boundary has been discussed in details in Appendix \ref{section:Appendix A}.}

\subsection{Homogeneous quasinormal modes}\label{Sec:QNMhom}
%We study the behavior of few of the lowest lying QNM frequencies in the complex frequency plane as we vary either k or $\beta$ keeping the other fixed. Note that we have set the temperature $T=1$ and the AdS radius $L=1$.
The homogeneous QNMs give us fundamental understanding of the relaxation dynamics of the system. In the decoupling limit, there exists two independent global symmetries, namely the constant shifts of the boundary and bulk scalar fields, which are broken to the specific combination \eqref{Eq:Axionic} at finite value of $\beta$. Therefore, in the decoupling limit, there are two poles at the origin at zero momentum. At small $\beta$ one of these is lifted but should be close to the origin. We call the latter the \psg\ mode $\wpg$ for reasons to be discussed in the following subsection. Nonlinear simulations to be presented in Section \ref{Sec:Nonlinear} confirm that $\wpg$ governs the homogeneous thermal relaxation of the full system. Explicitly, we find that
\begin{equation}\label{Eq:psg}
\wpg(k=0) \approx - i5.6\pi \beta^2 T^2
\end{equation} 
at small values of $\beta$. The other (unlifted) pole which stays at the origin at $k=0$ for any value of $\beta$ will behave as a diffusion pole $\wdiff \approx - i Dk^2$ at small $k$ and non-vanishing $\beta$. The diffusion constant $D$ is negative above a critical value of $\hatbeta
$ which is approximately $0.48
$ as discussed later.

\begin{figure}[t]
\centering
\begin{subfigure}[$\hatbeta$ = 0]{\includegraphics[scale=0.5]{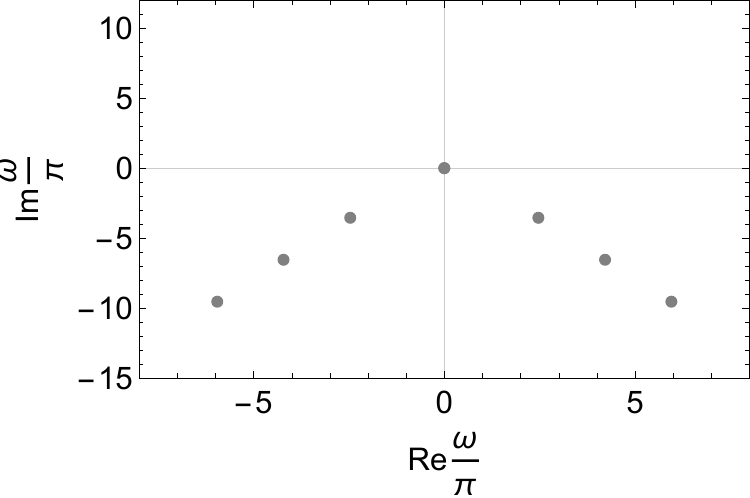}\label{Fig:QNMHomoa}}
\end{subfigure}
\hfill
\begin{subfigure}[$\hatbeta$ = 0.18]{\includegraphics[scale=0.5]{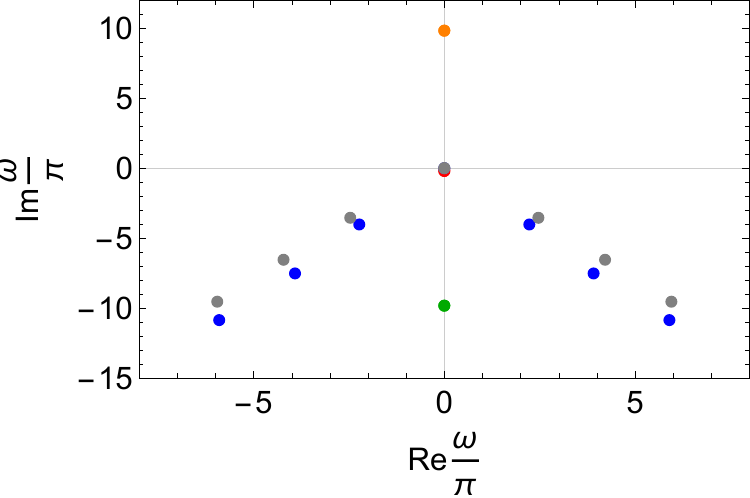}\label{Fig:QNMHomob}}\end{subfigure}
\subfigure[$\hatbeta$ = 0.4]{\includegraphics[scale=0.5]{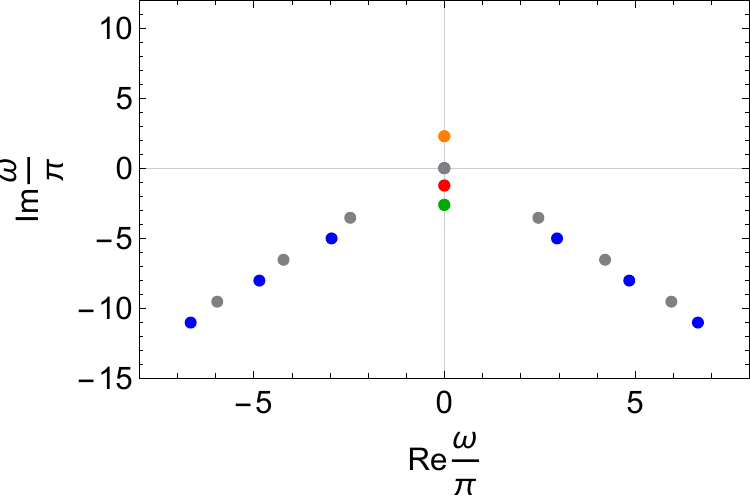}\label{Fig:QNMHomoc}}
\hfill
\subfigure[$\hatbeta$ = 0.45]{\includegraphics[scale=0.5]{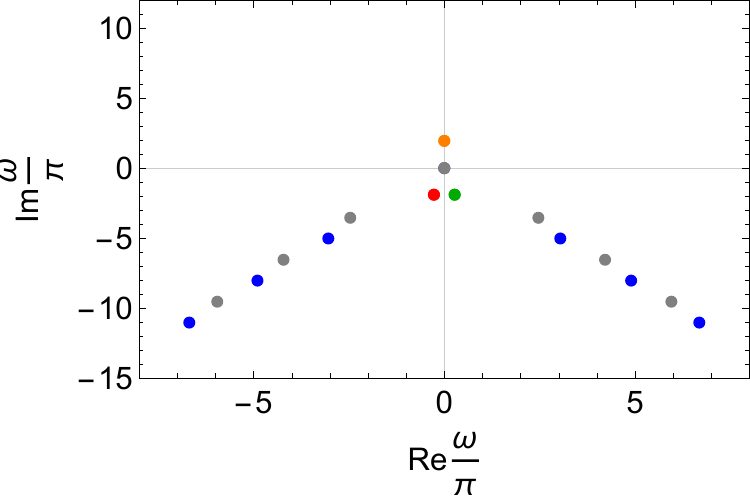}\label{Fig:QNMHomod}}
\subfigure[$\hatbeta$ = 10]{\includegraphics[scale=0.5]{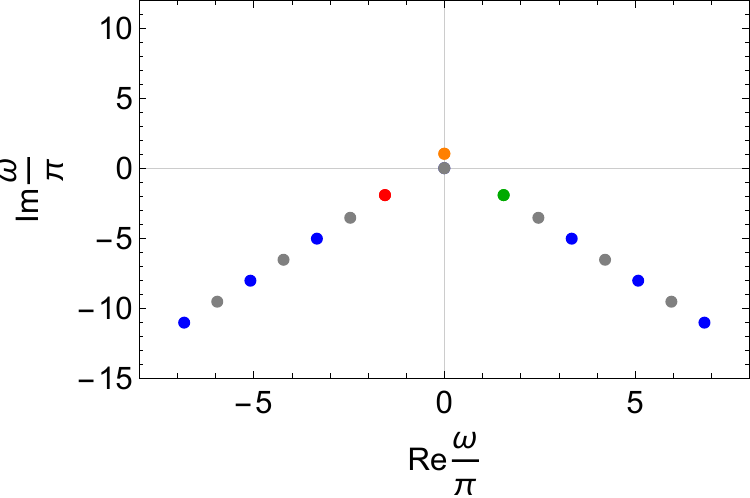}\label{Fig:QNMHomoe}}
\caption{Behavior of the first 8 QNMs in complex frequency plane with varying $\hatbeta$ at $k = 0$ are shown here. Gray dots in all figures are the QNM frequencies for $\hatbeta=0$. Fig. 1(a) shows the poles in the decoupling limit. Note that there are two poles at the origin corresponding to the independent constant shifts of $\chi$ and $\Phi$. Fig. 1(b) shows that at small values of $\hatbeta$ one of the poles, namely $\wpg$ (shown in red), is displaced slightly below the origin following \eqref{Eq:psg}, while two new poles $\wgl$ (shown in green) and $\wus$ (shown in orange) appear from $-i\infty$ and $+i\infty$ respectively on the imaginary axis. Fig.1(c) shows that with increasing $\hatbeta$, $\wgl$ moves upward, while $\wpg$ and $\wus$ move downward on the imaginary axis. Eventually $\wgl$ and $\wpg$ collide on the negative imaginary axis and transform into usual quasinormal modes as shown in Fig. 1(d). On the other hand $\wus$ attains a limiting value on the positive imaginary axis as $\hatbeta \rightarrow\infty$. As shown in Fig. 1(e), in the latter limit, all other QNMs (shown in blue) realign approximately on the same straight line on which the poles were located approximately in the decoupling limit but roughly at half-spacing.}
\end{figure}

{Already the homogeneous quasinormal modes show a complex behavior.} {For the sake of notational convenience, we define $\hatbeta \equiv \beta\sqrt{T}$ and use this dimensionless variable for the discussion. Note that the quasinormal modes will be of the general functional form $$\omega_{\rm QNM} = Tf(k/T, \hatbeta).$$}In Fig.~\ref{Fig:QNMHomoa}-~\ref{Fig:QNMHomoe}, we have plotted the eight homogeneous (complex) quasinormal modes of the full system with lowest absolute values for various values of $\hatbeta$.  For small and non-vanishing values of $\hatbeta$, there are three poles on the imaginary axis (aside from $\wdiff$ which is at the origin for all values of $\hatbeta$), namely
\begin{enumerate}
\item the \psg\ mode $\wpg$ {(plotted in red)}, which is parametrically close to the origin and is well approximated by \eqref{Eq:psg} at small $\hatbeta$,
\item a mode which we denote as $\wgl$ {(plotted in green)} that approaches the origin along the negative imaginary axis from $-i\infty$ as the value of $\hatbeta$ is increased from zero, and
\item a mode which we denote as $\wus$ {(plotted in orange)} that approaches the origin along the positive imaginary axis from $+i\infty$ as the value of $\hatbeta$ is increased from zero.
\end{enumerate}
In Fig.~\ref{Fig:QNMHomob}, corresponding to $\hatbeta = 0.18$, %we readily spot 
the red $\wpg$ is slightly below the origin, whereas the green $\wgl$ and orange $\wus$ are well separated from the origin on the negative and positive imaginary axes, respectively. Here, the poles in the decoupling limit \cite{Starinets:2002br}, which have been plotted in Fig.~\ref{Fig:QNMHomoa}, are shown again in gray color. The twin poles $\wgl$ and $\wus$ clearly have no analogues in the decoupling limit. The remaining blue poles in Fig.~\ref{Fig:QNMHomob} are on the lower half plane, and are only slightly displaced from their values in the decoupling limit shown in gray.

As evident from Fig.~\ref{Fig:QNMHomob}-~\ref{Fig:QNMHomoe}, the unstable pole $\wus$ stays on the positive imaginary axis for all non-vanishing values of $\hatbeta$. It moves closer to the origin and attains a limiting value $1.03\times i  \pi T$ as $\hatbeta \rightarrow \infty$. This mode apparently implies an instability of the thermal state (corresponding to constant $\chi$ and $\Phi$ on a black brane geometry). In actuality, this only implies an instability over a short time scale as will be evident from our non-linear simulations presented in Section \ref{Sec:Nonlinear}. {Note, unlike the case of a closed system, a mode with ${\rm Im}\,\omega>0$ may or may not imply instability in an open system. In our case, we do not have any instability in the homogeneous situation because of two reasons.} The total conserved energy of the system shown in \eqref{Eq:E-tot} is a sum of two non-negative terms, namely the boundary scalar kinetic energy and the black hole mass. Thus none of these can grow without bound in magnitude as they are bounded from both below and above. Furthermore, Birkhoff's theorem\footnote{{The massless dilaton in our case has to be constant at the horizon for regularity, which implies it is constant everywhere for a static configuration. A constant dilaton has vanishing stress tensor and hence we obtain the AdS-Schwarzschild black brane solution. In the case of a holographic superconductor \cite{Hartnoll:2008kx}, there is a non-trivial potential for the scalar field and/or a non-trivial radial mass profile.  The analysis of stable stationary configuration is more complicated but can be done via the method of Hollands and Wald \cite{Hollands:2012sf}.}} guarantees that the homogeneous thermal state is the unique static solution {where of course entropy cannot be produced. As the entropy given by the area of the apparent horizon grows monotonically (as explicitly verified in Sec.~\ref{Sec:Nonlinear}), the endpoint of evolution in the homogeneous case should be the static black brane.} 

So we can anticipate what is borne out by our non-linear solutions: for an arbitrary homogeneous perturbation about the thermal state, the unstable pole $\wus$ governs the rapid transfer of {some} energy from the {holographic sector} to the boundary scalar field, which is followed by a slow, complete and irreversible transfer of energy back to the black hole over a timescale governed by ${\rm Im}\,\wpg$ at small $\beta$. {The pole $\wus$ thus does not signal an imminent transition to another phase (unlike e.g.\ the case of holographic superconductors \cite{Amado:2009ts}), only the propensity to process a perturbation in this particular way.}

The pole $\wgl$ is associated with a Gregory-Laflamme type of instability at finite $k$ as discussed later.\footnote{{As we shall see, the fate of this additional pole when $k$ is increased depends on the value of $\beta$. For sufficiently small $\beta$ it always gives rise to a Gregory-Laflamme type instability, while at larger $\beta$ the diffusion pole can take over this role. Then the label ``G'' simply stands for the color ``green'' in the plots. By contrast, $\wus$ always refers to an unstable mode.}} However, it remains on the lower half plane for all values of $\hatbeta$ at $k=0$.

In Fig.~\ref{Fig:QNMHomoc}, corresponding to $\hatbeta = 0.4$, we see that $\wpg$ has moved down while $\wgl$ has moved up along the negative imaginary axis. At a sightly higher value of $\hatbeta$, these poles collide on the negative imaginary axis, after which they move almost horizontally keeping the imaginary part almost unchanged as evident from Fig.~\ref{Fig:QNMHomod} corresponding to $\hatbeta = 0.45$. Thus $\wpg$ and $\wgl$ are transformed to usual quasinormal mode poles for higher values of $\hatbeta$.

As $\hatbeta$ is increased towards infinity, all poles at $k= 0$, except for $\wdiff$ (which stays at the origin) and $\wus$ (which goes to the limiting value on the positive imaginary axis), realign approximately on the same straight lines on the lower half plane along which the decoupled quasinormal poles were placed. Furthermore, in this limit $\hatbeta\rightarrow\infty$, the poles are almost halfway in between the quasinormal mode poles of the decoupling limit. This is evident from Fig.~\ref{Fig:QNMHomod} corresponding to $\hatbeta = 10$.

\begin{figure}[t]
\centering
\includegraphics[scale=1]{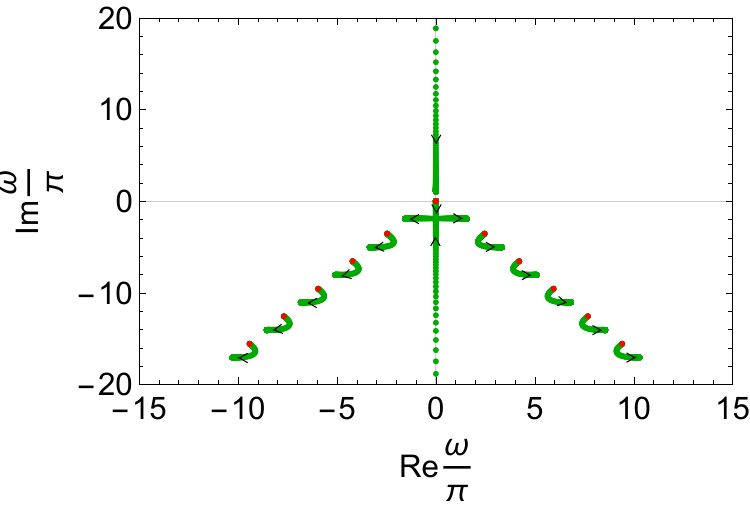}
\caption{Evolution of the homogeneous QNM poles with varying $\hatbeta$ on the complex $\omega$ plane. The red dots are the QNM frequencies for $\hatbeta=0$. As $\hatbeta$ is increased from zero, the QNM frequencies follow the green trajectories and the black arrows denote direction of motion of QNM frequencies with increasing $\beta$ until eventual saturation.}\label{Fig:qnm-semi}
\end{figure}

{We provide a summary of the above discussion as a snapshot in Fig.~\ref{Fig:qnm-semi}. 
}
%\st{In Fig.%~\ref{Fig:qnm-semi}
%we have plotted how all the $k=0$ hybrid QNM poles evolve with $\beta$ at fixed $T =1$.}

%Fig.~\ref{qnm-semi} and \ref{Fig:QNMHomo} show the plot of QNM frequencies of the semi-holographic system in the complex $\omega$ plane for $k=0$. The blue dots are the QNM frequencies for $\beta=0$, which corresponds to the decoupled case and agrees with \cite{}. As $\beta$ is increased from zero, the QNM frequencies follow the red trajectories and eventually saturate. We observe that one of the poles from the origin starts moving down along the negative imaginary axis, which we call the \emph{Drude-like pole}. At the same time, another pole (which starts from $-i\infty$%negative imaginary infinity
%) starts moving up along the negative imaginary axis. These two poles collide with each other at $\beta \approx 0.44$. After collision, both of these poles acquire real parts and start moving away from the imaginary axis. All the poles eventually align themselves along the original christmas tree with their new positions midway between the poles of the original christmas tree. There is one pole, however, which travels from $+i\infty$ down along the positive imaginary axis with increasing $\beta$, before saturating to a point on the positive imaginary axis close to the origin. 

%QNM frequencies for fixed non-zero $k$ with varying $\beta$ show qualitatively similar behavior. Fig.~\ref{qnm-k0p1} illustrates this for $k=0.1$. 

\subsection{Quasinormal modes at finite momentum}\label{Sec:QNMinhom}
\begin{figure}[t]
\centering
\subfigure[k = 0]{\includegraphics[scale=0.5]{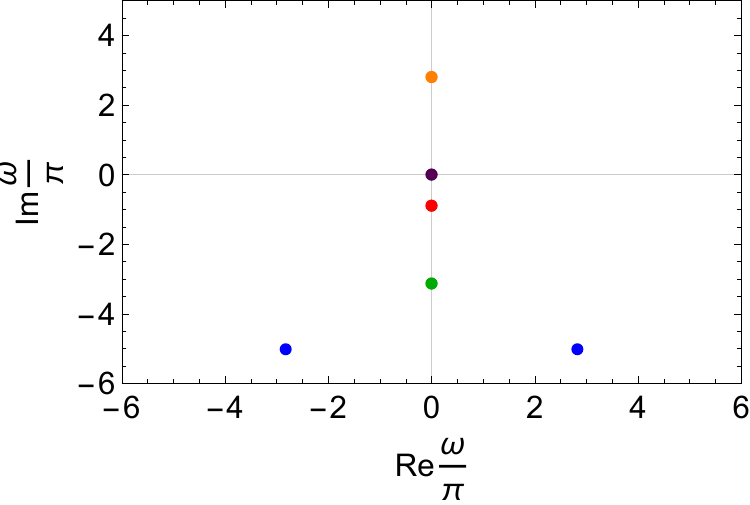}}
\hfill
\subfigure[k = 0.554$\pi$]{\includegraphics[scale=0.5]{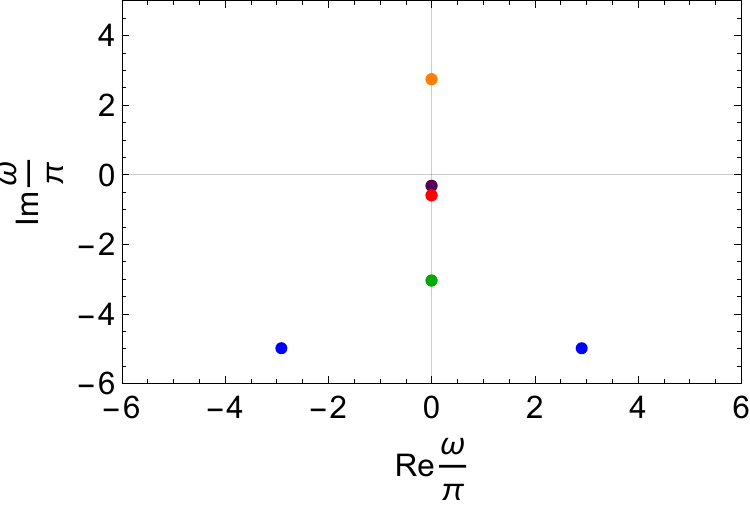}}
\subfigure[k = 2.4$\pi$]{\includegraphics[scale=0.5]{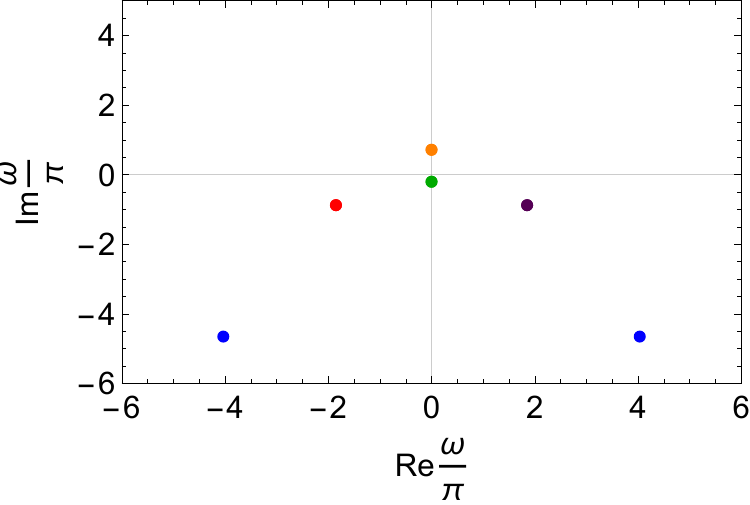}}
\hfill
\subfigure[k = 2.4334$\pi$]{\includegraphics[scale=0.5]{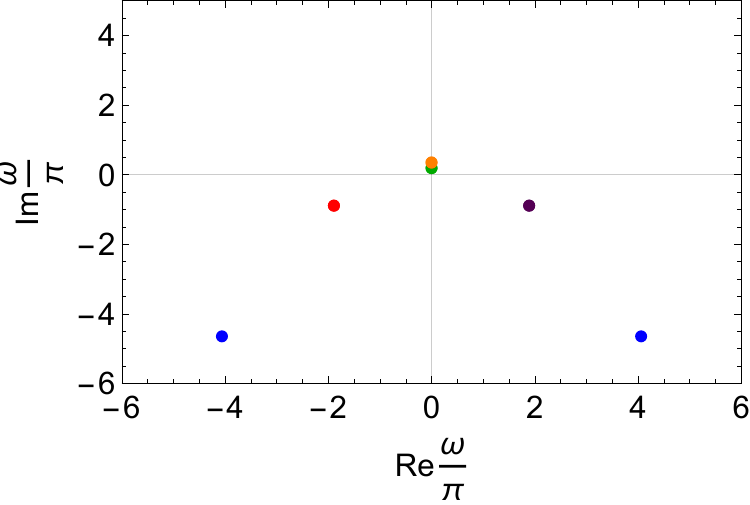}}
\subfigure[k = 2.54$\pi$]{\includegraphics[scale=0.5]{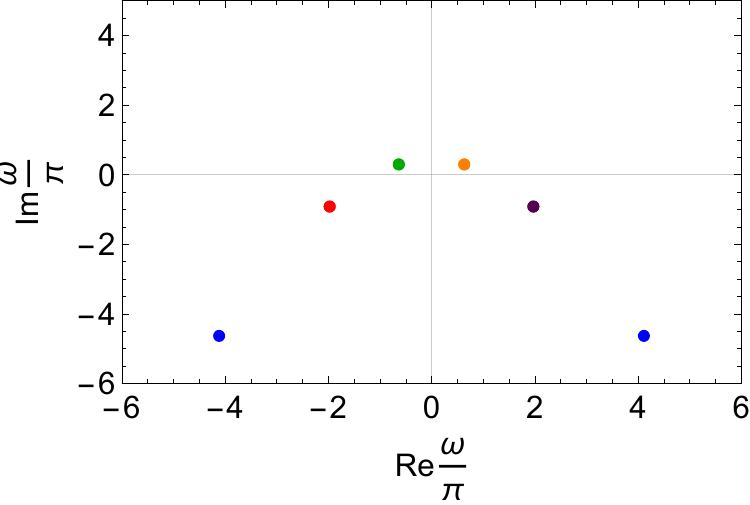}}
\caption{Behavior of the first six QNMs in complex frequency plane with varying $k$ at $\hatbeta = 0.35$. As $k$ evolves, the diffusion pole $\wdiff$ shown in purple (and is at the origin at $k=0$) collides with the \psg\ mode $\wpg$ shown in red on the negative imaginary axis, and subsequently both of them transform to a pair of quasinormal modes with almost $k$-independent negative imaginary parts. The other mode $\wgl$ shown in green moves upwards on the imaginary axis, crosses the origin and collides with the downward moving $\wus$ shown in orange above the origin. Subsequently both of them transform into a pair of unstable quasi-normal modes with almost $k$-independent positive imaginary parts.}\label{Fig:qnm-b0p35}
\end{figure}

\begin{figure}[t]
\centering
\subfigure[${\rm Im}\, \wdiff(k)$]{\includegraphics[scale=0.5]{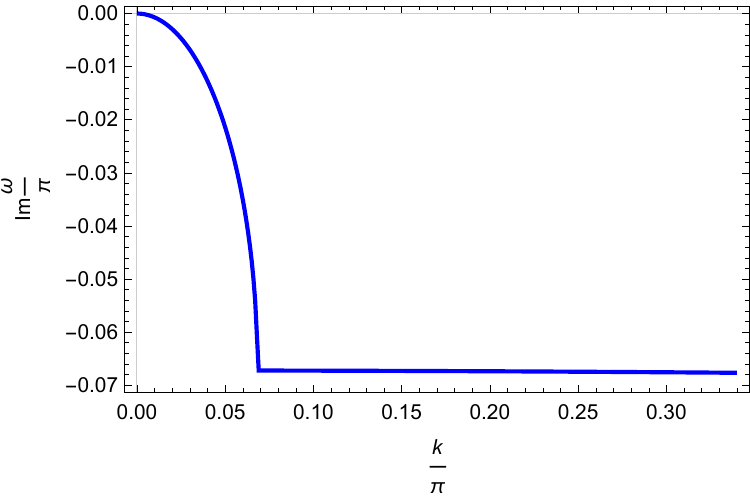}}
\hfill
\subfigure[${\rm Re}\,\wdiff(k) $]{\includegraphics[scale=0.5]{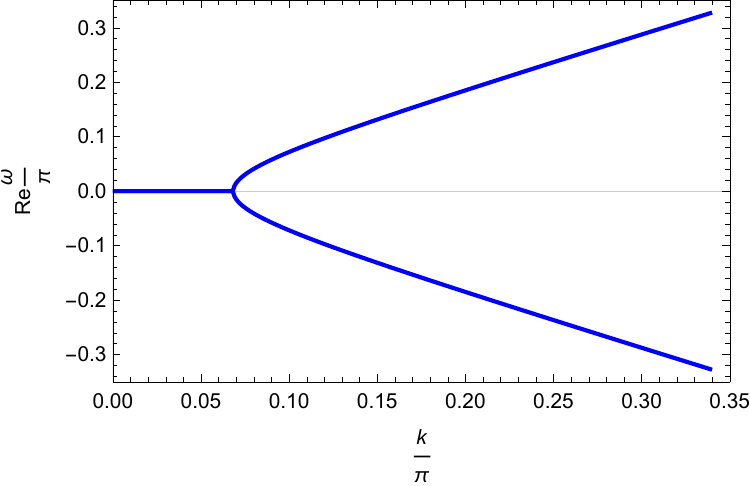}}
\caption{${\rm Re}\,\wdiff(k)$ and ${\rm Im}\,\wdiff(k)$ at $\hatbeta=0.15$. As the value of $k$ is increased, the diffusive behavior of $\wdiff(k)$ with ${\rm Re}\,\wdiff(k) =0$ stops at $k = k_* \approx 0.07 \pi$ when the collision with the $\wpg$ pole takes place and a pair of complex quasinormal modes are produced. For $k > k_*$, ${\rm Im}\,\wdiff(k)$ remains almost constant while ${\rm Re}\,\wdiff(k)\approx \vert k- k_*\vert^{1/2}$ when $k$ is close to $k_*$. At higher values of $k$, ${\rm Re}\,\wdiff(k)$ is linear in $k$ with slope very close to $1$.}\label{Fig:crossover}
\end{figure}

\begin{figure}
\centering
\subfigure[k = 0]{\includegraphics[scale=0.5]{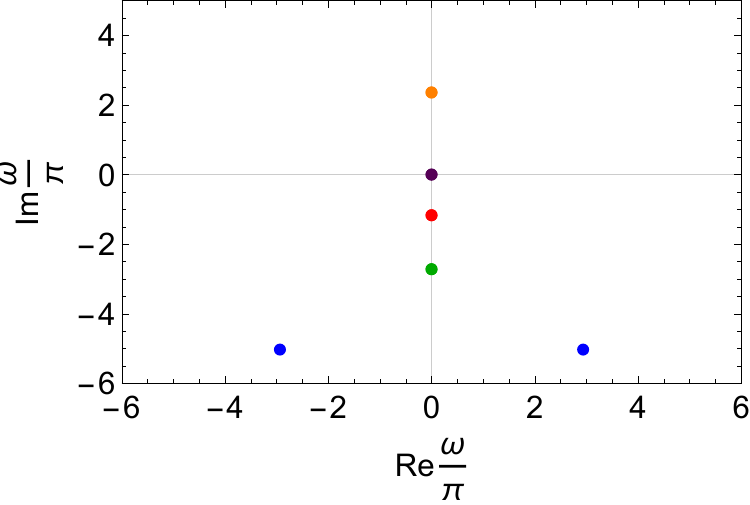}}
\hfill
\subfigure[k = $\pi$]{\includegraphics[scale=0.5]{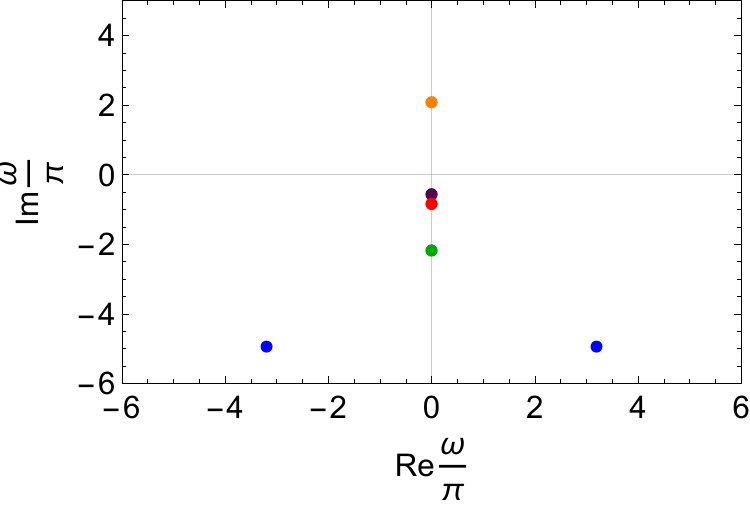}}
\subfigure[k = 1.1 $\pi$]{\includegraphics[scale=0.5]{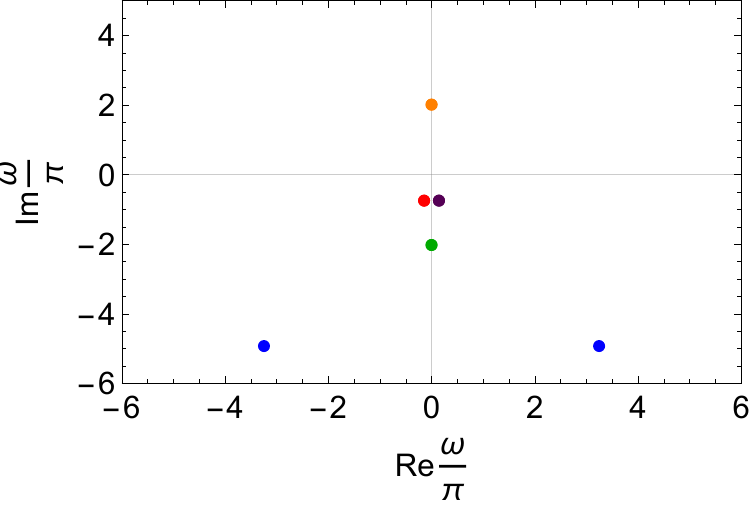}}
\hfill
\subfigure[k = 1.285 $\pi$]{\includegraphics[scale=0.5]{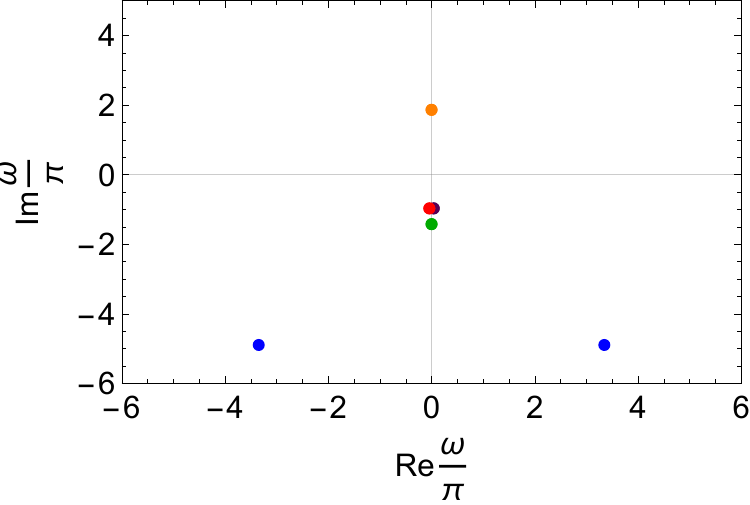}}
\subfigure[k = 1.29 $\pi$]{\includegraphics[scale=0.5]{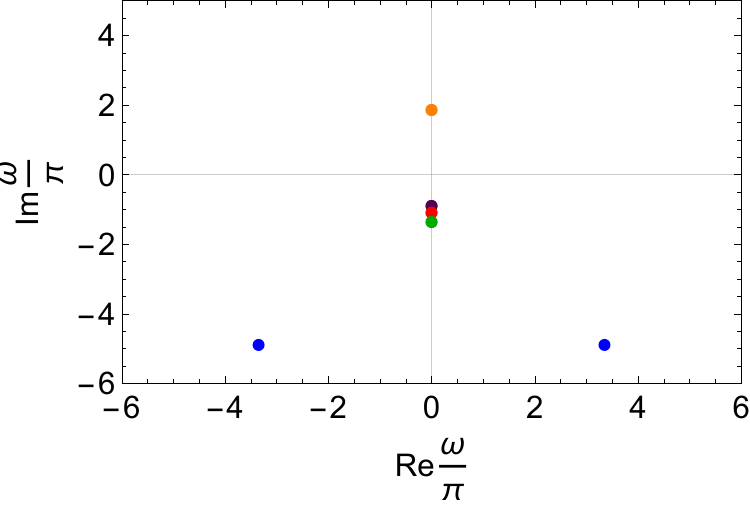}}
\hfill
%\subfigure[k = 1.294]{\includegraphics[scale=0.32]{lowest_eight_QNMs_beta=0p391_k=1p294.pdf}}
\subfigure[k = 1.3 $\pi$]{\includegraphics[scale=0.5]{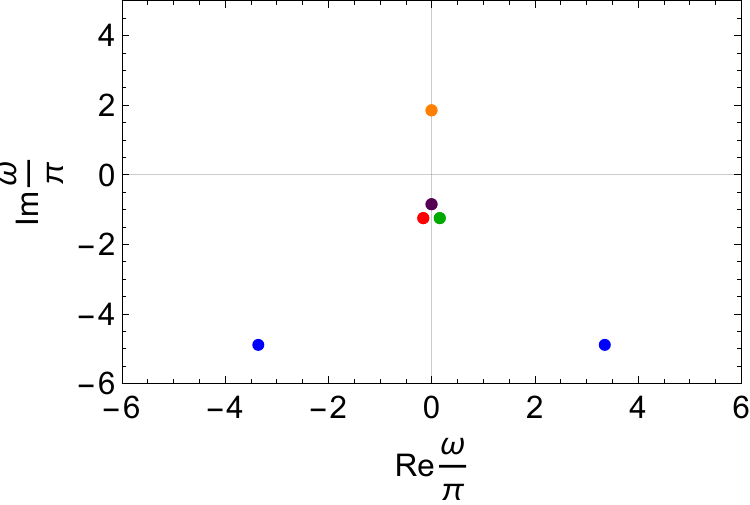}}
\subfigure[k = 1.86 $\pi$]{\includegraphics[scale=0.5]{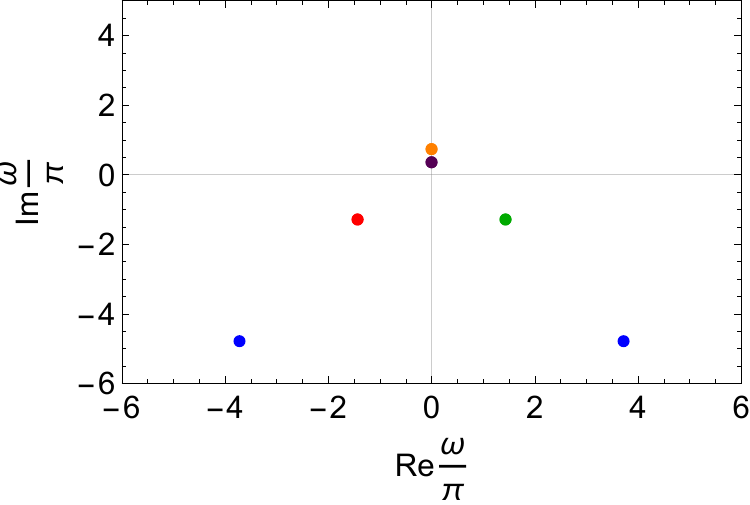}}
\hfill
\subfigure[k = 2 $\pi$]{\includegraphics[scale=0.5]{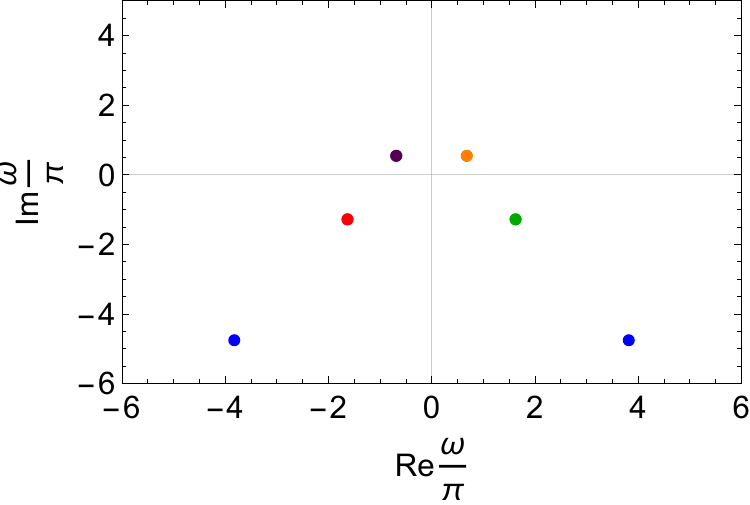}}
\caption{Behavior of the first six QNMs in complex frequency plane with varying $k$ at $\hatbeta = 0.391$. Unlike the case of $\hatbeta = 0.35$ in Fig.~\ref{Fig:qnm-b0p35}, the diffusion pole (shown in purple) and the \psg\ mode (shown in red) do not move permanently away from the imaginary axis after collision. They return back to collide again on the negative real axis, after which one of them moves down colliding with the upward moving $\wgl$ (shown in green) and producing a pair of horizontally moving stable quasinormal modes. The other one moves upwards and crosses the origin, etc., as in the case of $\beta = 0.35$.}\label{Fig:qnm-b0p391}
\end{figure}

\begin{figure}
\centering
\includegraphics[scale=0.8]{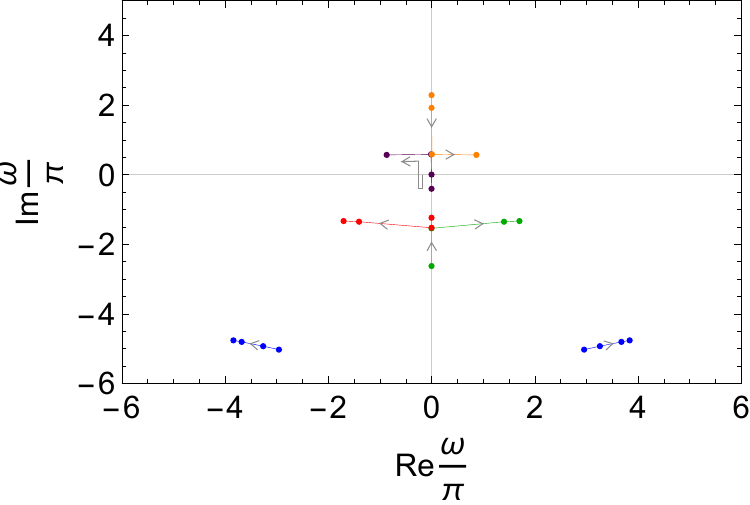}
\caption{Behavior of the first six QNMs in complex frequency plane with varying $k$ at $\hatbeta = 0.4$. The arrows indicate the movement of the poles in increasing $k$, with each colored dot representing $k=0$, $k = 1.0939 \pi$, $k= 1.77525 \pi$  and $k=2 \pi$ respectively. Unlike the cases of $\hatbeta = 0.35$ shown in Fig.~\ref{Fig:qnm-b0p35} and $\hatbeta = 0.391$ shown in Fig.~\ref{Fig:qnm-b0p391}, the {diffusive} pole (shown in purple) returns back to the origin producing the Gregory-Laflamme type instability -- there exists a finite value of $k$, namely $k_0$, at which $\wdiff(k_0) =0$ and ${\rm Im}\wdiff(k) > 0$ for $k> k_0$. The collision on the negative imaginary axis happens between the \psg\ pole shown in red and $\wgl$ pole shown in green at $k\approx 1.1\pi$.}\label{Fig:qnm-b0p4}
\end{figure}

\begin{figure}
\centering
\includegraphics[scale=0.8]{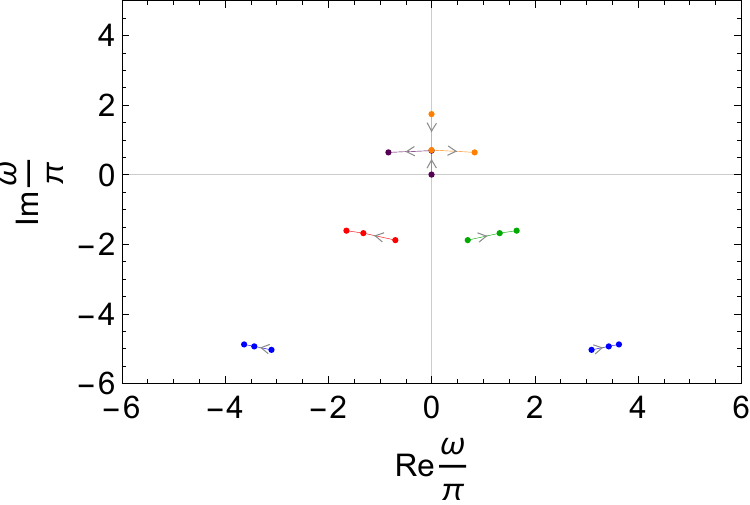}
\caption{Behavior of the first six QNMs in complex frequency plane with varying $k$ at $\hatbeta = 0.5$. The arrows indicate the movement of the poles in increasing $k$, with each colored dot representing $k=0$, $k = 1.1572 \pi$ and  $k=1.5\pi$ respectively. Unlike the cases of $\hatbeta = 0.4$ shown in Fig.~\ref{Fig:qnm-b0p4}, the diffusive pole shown in purple has a negative diffusion constant, moves along the positive imaginary axis and collides with $\wus$ around $k\approx 1.1\pi,$. Furthermore, all poles on the lower half plane have finite real parts for all values of $k$ -- no collision between poles take place on the lower half plane at any value of $k$. }\label{Fig:qnm-b0p5}
\end{figure}

The behavior of QNM frequencies with varying $k$ at fixed $\beta$ and $T$ is {even} richer. Qualitatively different behaviors are observed for 
\begin{enumerate}
\item $\hatbeta \lessapprox  0.391$ 
\item $\hatbeta \approx 0.391$
\item $0.391 \lessapprox \hatbeta\lessapprox 0.4425$
\item $\hatbeta \gtrapprox 0.4425$
\end{enumerate}
These are illustrated in Figs.~\ref{Fig:qnm-b0p35}-\ref{Fig:qnm-b0p5}. {In the decoupling limit $\beta =0$, we recover the propagating Goldstone modes $\omega = \pm k$ of the boundary scalar and the usual complex quasi-normal modes of the bulk scalar at any value of $k$ as mentioned before. As described below, even a small value of $\beta$ changes the character of the Goldstone modes while two other non-trivial modes emerge as in the homogeneous case described previously.}

\paragraph{The case of $\hatbeta
\lessapprox  0.391$:} The representative case of $\hatbeta = 0.35$ in this category is illustrated in Fig.~\ref{Fig:qnm-b0p35}. The pole $\wdiff$ (shown in purple), which is at the origin at $k=0$, becomes diffusive, i.e. it behaves as $$\wdiff \approx - i Dk^2$$at small $k$. We will discuss the dependence of the diffusion constant $D$ on $\hatbeta$ later. The \psg\ mode (shown in red) moves upwards along the negative imaginary axis with increasing $k$ and eventually collides with $\wdiff$. We denote the value of $k$ where this collision happens as $k_*$, which is $\approx 0.555 \pi$ for the chosen value of $\hatbeta$. For $k > k_*$, these two poles $\wdiff$ and $\wpg$ transform into a pair of QNMs which evolve almost horizontally (with constant imaginary parts). The momentum $k_*$ is {critical} because the dispersion relation of $\wdiff$ giving the effective diffusive dynamics of the boundary massless field $\chi$ becomes non-analytic (with discontinuous first derivatives) at $k=k_*$ signalling the breakdown of an effective hydrodynamic description.

A representative characterisation of ${\rm Re}\,\wdiff(k)$ and ${\rm Im}\,\wdiff(k)$ for $\hatbeta 
= 0.15$ is presented in Fig.~\ref{Fig:crossover}. The value of $k_*$ at which $\wdiff$ and $\wpg$ collide turns out to be $0.07 \pi \,T$, producing a pair of complex quasinormal poles which have almost $k$-independent imaginary part and with non-vanishing real parts ${\rm Re}\,\wdiff(k)\approx \pm \vert k- k_*\vert^\delta$ as $k\rightarrow k_* + \epsilon$ (with $\epsilon > 0$). We find that $\delta \approx 1/2$ and is independent of the choice of $\hatbeta$  in this range to a very good accuracy. As one would expect, ${\text{Re} \wdiff(k)}/{k } \approx 1$  as $k \rightarrow \infty$. It is to be noted that we can get $k_*$ arbitrarily close to the origin by tuning $\beta$ to smaller values. {This feature that propagating modes with non-vanishing $\mathrm{Re}\, \omega(k)$ exists for $k> k_*$ is called the $k$-gap \cite{Baggioli:2019jcm}.} {In the context of phonon-like modes, this has been observed in \cite{Baggioli:2018vfc,Baggioli:2018nnp}.}

Such phenomena of breakdown of hydrodynamics due to collision between a hydrodynamic and a quasi-hydrodynamic mode at a real and parametrically small value of momentum {leading to a $k$-gap} is  a characteristic property of the collective modes of liquids \cite{PhysRevB.101.214312}. In fact, our model interpolates between various qualitatively different behaviors about which we will have more to say in the following subsection. 

The transformation of massless propagating Goldstone modes of the type $\omega = \pm k$ (which is exactly how the modes of $\chi$ behave in the decoupling limit) to a pair of diffusion and a \psg\ modes has been observed before in other models of holography \cite{Davison:2014lua,Baggioli:2020loj} where such modes are also obtained from an explicit and soft symmetry breaking \cite{Davison:2014lua,Grozdanov:2018fic,Ammon:2019wci}\footnote{Remarkably, it has been shown in \cite{Grozdanov:2018ewh} that the $k$-gap is produced naturally also in holographic models with higher form fields.} and have been also discussed from an effective field theory point of view \cite{Hayata:2014yga,Hidaka:2019irz}.\footnote{In \cite{Davison:2014lua}, actually the reverse transformation of a pair of complex poles to a pair of purely imaginary poles at higher values of $k$ was reported.} For a recent review, see \cite{Baggioli:2021xuv}. The independent %axionic 
shift symmetries are broken to the diagonal explicitly via the semi-holographic coupling in our model, while translation symmetry is explicitly broken in the models discussed in \cite{Baggioli:2021xuv}.\footnote{A similar phenomenon of emergence of a diffusive pole in presence of time-translation symmetry breaking has been discussed in \cite{Hayata:2018qgt}.} However, in contrast to the latter models, ours can be thought of as an open quantum system {with a finite total conserved energy} if we consider the boundary scalar field as the system and the black hole with the dilaton field as the bath. {Indeed our model has an additional feature, namely the presence of an additional modes $\wgl$ mode on the negative imaginary axis that also contributes to low energy dynamics. This is absent in usual setups.} As evident from Fig.~\ref{Fig:qnm-b0p35}, the $\wgl$ mode moves upwards along the imaginary axis, and crosses the origin at a finite momentum $k_0$. At $k=k_0$, the system therefore has a Gregory-Laflamme type instability! Also when $k$ is near $k_0$, $\wgl$ cannot be excluded from the low energy description of the system.  

As $k$ is increased above $k_0$, $\wgl$ collides with $\wus$ (which moves downwards with increasing $k$) on the positive imaginary axis close to the origin, and both transform into a pair of unstable quasinormal modes with small imaginary parts as shown in Fig.~\ref{Fig:qnm-b0p35}. Subsequently, they move almost horizontally with almost $k$-independent positive imaginary parts. 

The Gregory-Laflamme type instability \cite{Gregory:1993vy} can have profound consequences for the dynamics of this system. Since the total conserved energy given by \eqref{Eq:E-tot} is a sum of two non-negative terms, a repetition of our argument in the previous subsection, namely that the poles on the upper half plane could only lead to initial instabilities involving reverse transfer of energy from the black hole to the boundary scalar field, could have got through had there been no zero modes at finite $k$. The presence of a zero mode at $k=k_0$ may imply that the system may not be able to evolve to the static thermal configuration eventually and the final end point could be turbulent or glassy \cite{Lehner:2011wc,Emparan:2015gva}. Unfortunately, non-linear simulation of this system in presence of inhomogeneities is difficult and we postpone this to a future work. In fact, unlike the usual Gregory-Laflamme instability of the black string, ours is intrinsically a $(3+1)$-dimensional gravitational problem.

\paragraph{The case of $\hatbeta
\approx  0.391$:} This is illustrated in Fig.~\ref{Fig:qnm-b0p391}. The collision of the diffusion pole $\wdiff$ with the \psg\ pole proceeds as in the previous case. However, after these transform into a pair of stable quasinormal modes moving horizontally away from the imaginary axis with increasing $k$, they reverse back to the imaginary axis and once again collide there.\footnote{This second collision is closer to the phenomenon described in \cite{Davison:2014lua}.} Subsequently one of these poles moves downwards on the negative real axis, colliding with the pole $\wgl$ that moves upwards with increasing $k$, and thus producing two almost horizontally moving quasinormal modes. The other one moves upwards and produces the Gregory-Laflamme type instability as before. It is to be noted that for these values of $\hatbeta$, there exists a region of value of $k$ in which the three poles $\wdiff$, $\wpg$ and $\wgl$ are on the negative imaginary axis and almost degenerate.\footnote{It is possible that there exists a value of $\hatbeta$ around $0.391$ on the negative real axis where the three poles $\wdiff$, $\wpg$ and $\wgl$ coincide on the negative imaginary axis at a specific value $k$.} 

\paragraph{The case of $\,\,0.391 \lessapprox \hatbeta 
\lessapprox 0.4425$:}The illustrative case of $\hatbeta = 0.4$ is shown in Fig.~\ref{Fig:qnm-b0p4}. This is very distinct from the previous cases because the diffusive pole $\wdiff$ never collides with the \psg\ pole. It initially behaves as a diffusion pole, but it starts moving upwards along the negative imaginary axis as the value of $k$ is increased and eventually crosses the origin producing a Gregory-Laflamme type instability  -- there exists a finite value of $k$, namely $k_0$, at which $\wdiff(k_0) =0$ and ${\rm Im}\,\wdiff(k) > 0$ for $k> k_0$. The \psg\ pole $\wpg$ moves downwards with increasing $k$, in contrast to the previous cases, and collides with the upward moving $\wgl$ pole, producing a pair of stable horizontally moving QNM poles. 

%\begin{figure}[t]
%\centering
%\subfigure[k = 0]{\includegraphics[scale=0.5]{lowest_eight_QNMs_beta=0p4_k=0.pdf}}
%\hfill
%\subfigure[k = 1.08 $\pi$]{\includegraphics[scale=0.5]{lowest_eight_QNMs_beta=0p4_k=1p08.pdf}}
%\subfigure[k = 1.1]{\includegraphics[scale=0.5]{lowest_eight_QNMs_beta=0p4_k=1p1.pdf}}
%\hfill
%\subfigure[k = 1.75 $\pi$]{\includegraphics[scale=0.5]{lowest_eight_QNMs_beta=0p4_k=1p75.pdf}}
%\subfigure[k = 2 $\pi$]{\includegraphics[scale=0.5]{lowest_eight_QNMs_beta=0p4_k=2.pdf}}
%\caption{Behavior of the first six QNMs in complex frequency plane with varying $k$ at $\hatbeta = 0.4$. Unlike the cases of $\hatbeta = 0.35$ shown in Fig~\ref{Fig:qnm-b0p35} and the case of $\hatbeta = 0.391$ shown in Fig.~\ref{Fig:qnm-b0p391}, the diffusion pole (shown in purple) returns back to the origin producing the Gregory-Laflamme type instability -- there exists a finite value of $k$, namely $k_0$, at which $\wdiff(k_0) =0$ and ${\rm Im}\wdiff(k) > 0$ for $k> k_0$. The collision on the negative imaginary axis happens between the \psg\ pole shown in red and $\wgl$ pole shown in green.}\label{Fig:qnm-b0p4}
%\end{figure}

\paragraph{The case of $\hatbeta
\gtrapprox 0.4425
$:} The representative case of $\hatbeta = 0.5$ is shown in Fig.~\ref{Fig:qnm-b0p5}. Firstly, even at $k=0$, there exists no pole on the negative imaginary axis. We recall from the previous subsection that the $\wpg$ and $\wgl$ poles are complex for $\hatbeta
\gtrapprox 0.4425$. Secondly, the diffusive pole $\wdiff$ has a negative diffusion constant at small $k$ for $\hatbeta
\gtrapprox 0.48$ and moves upwards on the positive imaginary axis to collide with the downward moving $\wus$ pole. Consequently, for $\hatbeta
\gtrapprox 0.48$, there exists no finite value of $k$ for which there is a quasinormal mode pole at the origin, and hence no Gregory-Laflamme type phenomena. {We find that as we approach $\hatbeta_c\approx 0.48$ from below,   $k_0$ scales like $ \vert\hatbeta-\hatbeta_c\vert^\rho$ with $\rho\approx 1/2$.} The negative diffusion constant could lead to clumping instabilities, which should be investigated via a numerical simulation of the {inhomogeneous} non-linear dynamics in future work. In the narrow range %of beta values, namely 
$0.4425 \lessapprox 
\hatbeta
\lessapprox 0.48$, the diffusion constant is positive as in the previous case, but the value of $k_0$ at which $\wdiff$ crosses the origin again moves towards zero as $\hatbeta
$ gets closer to $\hatbeta_c \approx 0.48$. It seems likely that the non-linear dynamics of the system is qualitatively different for $\hatbeta
\geq \hatbeta_c$. 

%\begin{figure}[t]
%\centering
%\subfigure[k = 0]{\includegraphics[scale=0.5]{lowest_eight_QNMs_beta=0p5_k=0.pdf}}
%\hfill
%\subfigure[k = 1.15 $\pi$]{\includegraphics[scale=0.5]{lowest_eight_QNMs_beta=0p5_k=1p15.pdf}}
%\subfigure[k = 1.5 $\pi$]{\includegraphics[scale=0.5]{lowest_eight_QNMs_beta=0p5_k=1p5.pdf}}
%\caption{Behavior of the first six QNMs in complex frequency plane with varying $k$ at $\hatbeta = 0.5$. Unlike the cases of $\hatbeta = 0.4$ shown in Fig~\ref{Fig:qnm-b0p4}, the diffusive pole shown in purple has a negative diffusion constant and it collides with $\wus$ as in the case of $\beta = 0.4$ illustrated in Fig.~\ref{Fig:qnm-b0p4}. However, $k_0 =0$. Furthermore, all poles on the lower half plane have finite real parts for all values of $k$ -- no collision between poles take place on the lower half plane at any value of $k$.}\label{Fig:qnm-b0p5}
%\end{figure}

\subsection{On the diffusion constant $D$ and the Gregory-Laflamme momentum $k_0$}\label{Sec:Diffusion}
\begin{figure}[t]
\centering
\centering
\subfigure[Diffusion constant as a function of the mutual coupling]{\includegraphics[scale=0.5]{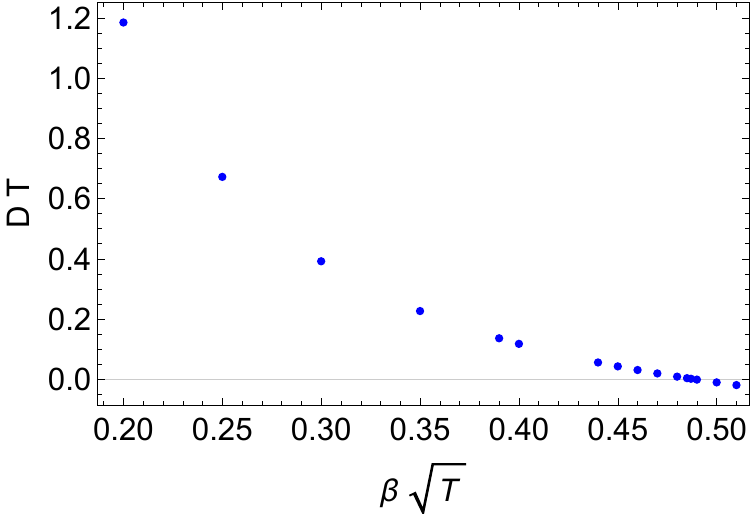}}
\hfill
\subfigure[Checking bounds on the diffusion constant]{\includegraphics[scale=0.5]{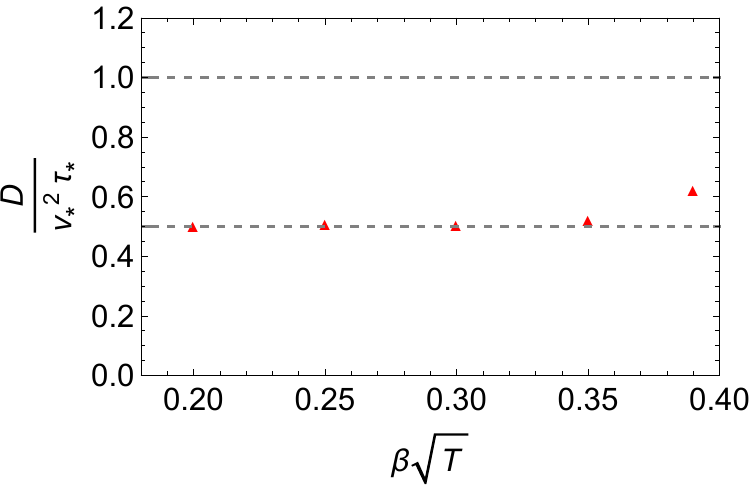}}
\caption{The diffusion constant $D$ as a function of $\hatbeta$ has been plotted above on the left. Note that the diffusion constant becomes negative around $\hatbeta\approx 0.48$. The right plot shows that the upper bound in \eqref{Eq:Dspec} is half-saturated at small $\hatbeta$.}\label{Fig:D}
\end{figure}

The mode $\wdiff$ behaves as a diffusive mode at small $k$ as discussed above. The plot of the dimensionless product of the diffusion constant times the temperature ($DT$) as a function of the dimensionless mutual coupling ($\hatbeta$) is presented in Fig.~\ref{Fig:D}. We find that the diffusion constant $D$ decreases monotonically with $\hatbeta$ and changes sign at $\hatbeta_c\approx 0.48$ as discussed previously. Since the diffusive behavior exists for $k \ll k_*$ (where $k_*$ is the momentum at which $\wdiff$ collides with $\wpg$), and $k_*$ can be arbitrarily close to the origin for small $\hatbeta$, it is very difficult to determine $D$ at small values of $\hatbeta$ numerically.

It has been argued that the diffusion constant should satisfy an upper bound \cite{Hartman:2017hhp,Lucas:2017ibu,Arean:2020eus,Grozdanov:2020koi,Wu:2021mkk,Jeong:2021zsv} in a wide class of many-body systems, i.e.\footnote{See \cite{Baggioli:2020ljz} for a discussion in the context of the Goldstone diffusivity.}
\begin{equation}\label{Eq:Dspec}
 D \lessapprox  v_*^2 \tau_*, \quad {\rm with } \quad v_*= \vert\omega_*\vert/\vert k_*\vert, \quad {\rm and} \quad\tau_* = \vert\omega_*\vert^{-1}.
\end{equation}
Above $k_*$ and $\omega_*$ are the values of the momentum and frequency respectively at which the hydrodynamic description breaks down, i.e. they set the limits of the convergence of the gradient expansion. To be precise, $k_*$ is the value of momentum at which the hydrodynamic mode collides with a gapped mode or a branch point, and $\omega_*$ is the value of the complex frequency at that point \cite{Grozdanov:2017ajz,Blake:2017ris,Blake:2018leo,Grozdanov:2019uhi}. Interestingly, $k_*$ can be complex and should be then determined by the analytic continuation of the hydrodynamic and non-hydrodynamic modes. {It is expected that $v_*$ is essentially an effective state-dependent Lieb-Robinson velocity governing the ballistic growth of the operators at late time (see \cite{Roberts:2016wdl} and \cite{Hartman:2017hhp}).} The inequality \eqref{Eq:Dspec} is saturated typically in holographic theories and in other models such as SYK chains \cite{Blake:2016sud,Gu:2016oyy}.

A lower bound on the diffusion constant has also been conjectured \cite{Hartnoll:2014lpa,Blake:2016wvh,Blake:2016sud,Grozdanov:2020koi,Wu:2021mkk}\footnote{See also \cite{Jeong:2021zhz}.} to hold for many-body systems primarily inspired by the KSS bound \cite{Kovtun:2004de} on $\eta/s$ (which should be stated in terms of the product of the diffusion constant and the temperature more generally). In case of fermionic systems, $v_*$ has been identified with the Fermi velocity in fermionic systems \cite{Hartnoll:2014lpa} while the corresponding $\tau$ is the Planckian scattering time \cite{Hartnoll:2014lpa}. For holographic systems, the velocity is identified with the butterfly velocity $v_B$ and $\tau$ with the corresponding Lyapunov time $\tau_L$ \cite{Blake:2016wvh,Blake:2016sud}. In the case that the dispersion relation of the diffusive mode is univalent over the entire complex $z$-plane with $z\equiv k^2$ except for a branch point and at $z =\infty$, then according to \cite{Grozdanov:2019uhi}, {$v_*$} should indeed be the butterfly velocity. We will not have much to say about the lower bound in our model because we believe that we need an independent computation to establish the butterfly velocity and the Lyapunov time in our model (see below).

For $\hatbeta
\lessapprox 0.391$, the value of $k_*$ in our model is simply the (real) momentum at which the $\wdiff$ collides with $\wpg$, and $\omega_* = \vert \wdiff(k_*)\vert = \vert {\rm Im}\,\wdiff(k_*)\vert$. We find that indeed for $\hatbeta
\lessapprox 0.35$, 
\begin{equation}
\frac{D}{v_*^2\tau_*} \approx {0.51 \pm 0.01},
\end{equation}
as shown on the right in Fig. \ref{Fig:D}. Thus the upper bound in \eqref{Eq:Dspec} is half-saturated. In the regime $0.35\lessapprox \hatbeta
\lessapprox 0.391$,  the stricter inequality \eqref{Eq:Dspec} holds, i.e. $$\frac{D}{v_*^2\tau_*} > 0.5.$$
For instance, when $\beta = 0.39/\sqrt{T}$, we find that $$ \frac{D}{v_*^2\tau_*}\approx {0.62}.$$
For $0.391\lessapprox
\hatbeta
\lessapprox0.48$, it is unclear what should be the value of $k_*$. Three possibilities exist, namely
\begin{enumerate}
\item $k_*$ is the inflexion point at which $\wdiff(k)$ reverses its motion and moves back towards the origin if $\wdiff(k)$ is non-analytic here (see the case $\hatbeta = 0.4$ in Fig.~\ref{Fig:qnm-b0p4}),
\item $k_*$ is the value of the momentum at which $\wdiff$ collides with $\wus$ in the upper half plane (see the case $\hatbeta = 0.4$ in Fig.~\ref{Fig:qnm-b0p4} again), and
\item $k_*$ is the complex momentum at which $\wdiff(k)$ collides with $\wpg$ or another pole after analytic continuation.
\end{enumerate}
Actually $k_*$ would be the smallest of these three possibilities. 

In absence of an understanding of the analytic properties of $\wdiff$ as a function of $z\equiv k^2$, it is unclear how we can identify the butterfly velocity and Lyapunov exponent in our model{;} {an independent computation following \cite{Shenker:2013pqa} could be necessary to settle this}. At present, we cannot comment on the validity of the lower bound on the diffusion constant in our model.  
It is worth mentioning though that as shown in the section \ref{Sec:Nonlinear}, the energy relaxation time is $$\tau_{\rm eq} = \frac{1}{2 \,{\rm Im}\,\wpg(k=0)} < \frac{1}{2\omega_*} = \frac{\tau_*}{2}$$ at small $\hatbeta$. 
The inequality above follows from our previous discussion that for $k < k_*$ and small $\hatbeta$, $\wpg$ is purely imaginary and its imaginary part decreases with increasing $k$. This would imply that a lower bound similar to \eqref{Eq:Dspec} but with the inequality reversed would be almost saturated if $v_B \approx v_*$ and $\tau_{\rm eq} \approx \tau_L$ at small $\hatbeta$. 
 
It is not clear how a negative diffusion constant or even the vanishing of the diffusion constant could be reconciled with \eqref{Eq:Dspec}.\footnote{The vanishing of the diffusion constant could be compatible with the $\wgl$ pole coming close to the origin at finite momentum for higher values of $\hatbeta$. Also since open systems can have poles in the complex upper half frequency plane without implying instability of the thermal equilibrium state, one needs to reevaluate the bounds on transport coefficients arising from the analytic properties of the poles in such systems.} In the future, we would like to investigate this further.

\begin{figure}
\centering
\includegraphics[scale=0.7]{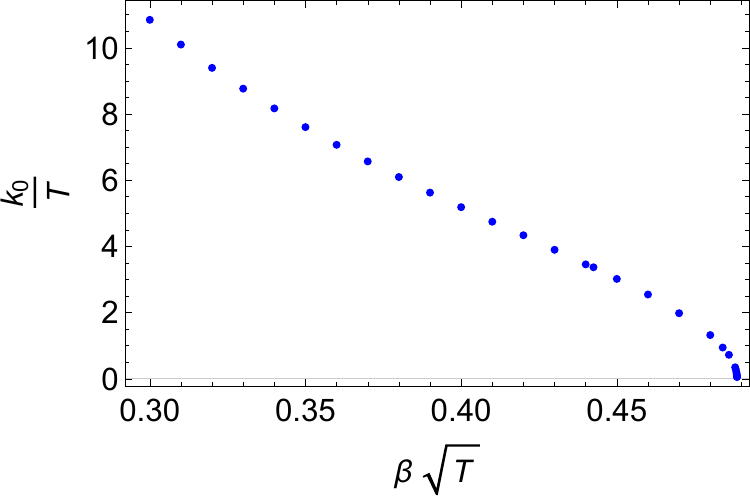}
\caption{The Gregory-Laflamme momentum $k_0$ as a function of $\hatbeta$. The diffusion constant changes sign precisely when $k_0$ vanishes.}\label{Fig:GLbeta}
\end{figure}

The Gregory-Laflamme momentum $k_0$ at which $\wgl$ or $\wdiff$ crosses the origin from the lower half plane also monotonically decreases with $\hatbeta$ as shown in Fig.~\ref{Fig:GLbeta}. As discussed in the previous subsection, $k_0$ goes to zero when the diffusion constant changes sign. Note that $k_0$ diverges in the limit $\hatbeta \rightarrow 0$. Since in the latter limit $\wgl$ moves towards $-i\infty$ at $k=0$, it crosses the origin at higher and higher values of $k_0$.

{Another type of instabilities is known to appear in a weakly coupled plasma (and also glasma \cite{Romatschke:2005pm}) out of equilibrium:
plasma instabilities involving gauge fields such as Weibel instabilities
\cite{Mrowczynski:1993qm,Romatschke:2003ms,Arnold:2003rq,Romatschke:2006wg,Rebhan:2009ku}. Those are clearly
left out by our simplified model for the ultraviolet degrees of freedom. A full-fledged semi-holographic description of large-$N_c$ Yang-Mills plasmas should in principle also
contain those. However, in the context of heavy-ion collisions it has been found that
they tend to be too slow in their on-set to play a crucial role \cite{Romatschke:2006wg,Berges:2013eia}. In particular, the
bottom-up scenario of Ref.~\cite{Baier:2000sb} may be qualitatively right even though it ignores
plasma instabilities.}

{Let us also point out that plasma instabilities are qualitatively
different from the instabilities we have found in the present simplified semi-holographic model.
The former are present for a certain range $0<k<k_\mathrm{max}$ with vanishing
growth rate at $k=0$, unlike the mode $\wus$. By contrast, the Gregory-Laflamme instabilities
set in only above a nonzero value $k_0$.}

Before concluding this subsection, we would like to refer the reader to \cite{Amoretti:2018tzw,Ammon:2019apj,Donos:2019txg,Baggioli:2020nay,Ghosh:2020lel} for the computation of quasinormal modes with an examination of the diffusive behavior in \textit{purely} holographic systems with a broken global symmetry. An analytic expression for the diffusion constant was obtained in \cite{Donos:2019txg} particularly in terms of thermodynamic data.

\subsection{Emergence of conformality at infinite mutual coupling}\label{Sec:Conformal}
In all our previous semi-holographic models, we have found emergence of conformality when the mutual coupling between the subsectors becomes infinite. At the level of quasinormal modes, the latter should imply that the quasinormal frequencies should behave as
\begin{equation}
\omega_{\rm QNM} = T\, f(k/T), \quad {\rm when} \quad \beta\sqrt{T}\rightarrow \infty.
\end{equation}
At fixed $T$ and $k$, the above implies that all $\omega_{\rm QNM}$ should saturate to finite values in the limit $\beta \rightarrow\infty$. We have indeed seen this feature in the case of $k=0$ as shown in Fig.~\ref{Fig:qnm-semi}. 

\begin{figure}[t]
\centering
\subfigure[$\hatbeta$ = 0]{\includegraphics[scale=0.5]{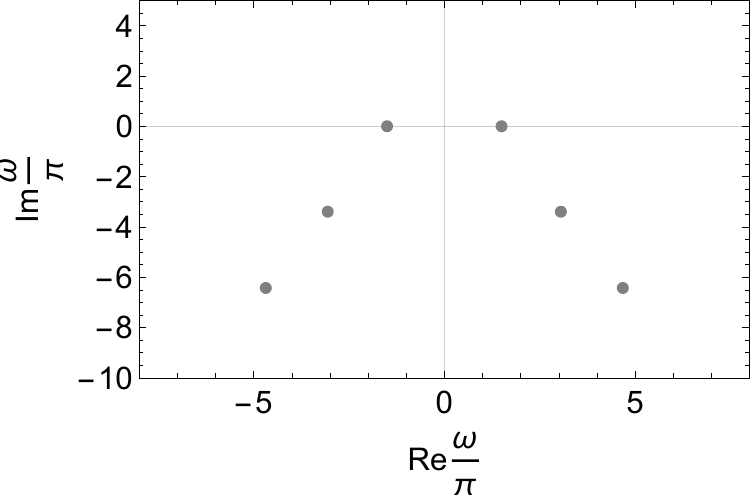}}
\hfill
\subfigure[$\hatbeta$ = 0.3]{\includegraphics[scale=0.5]{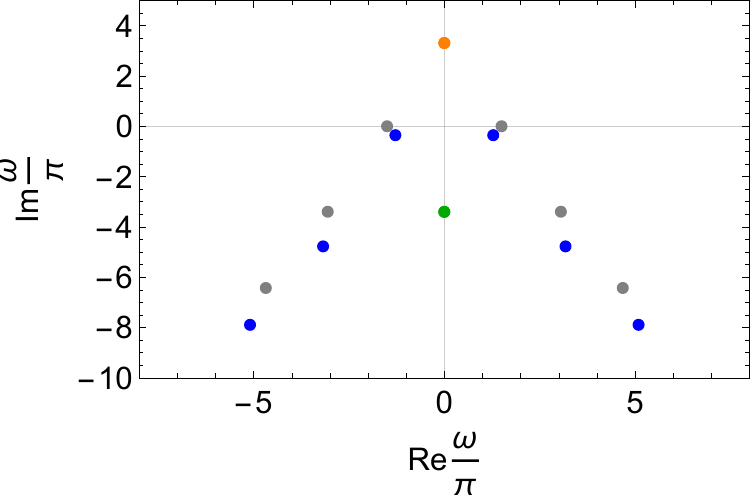}}
\subfigure[$\hatbeta$ = 0.4]{\includegraphics[scale=0.5]{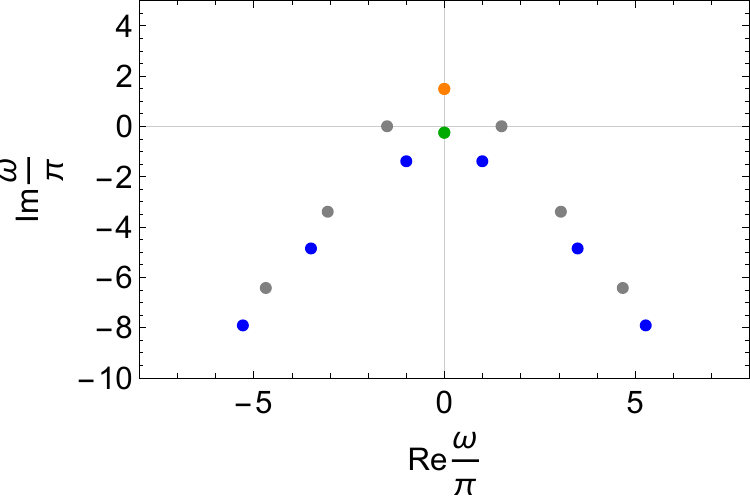}}
\hfill
\subfigure[$\hatbeta$ = 0.43]{\includegraphics[scale=0.5]{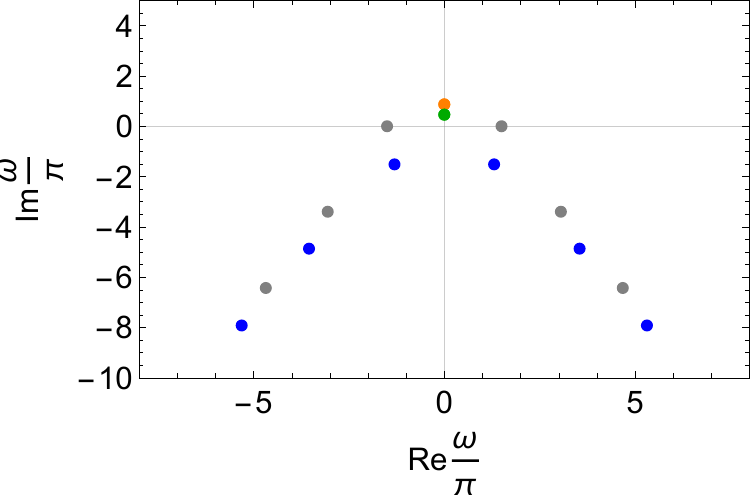}}
\subfigure[$\hatbeta$ = 0.44]{\includegraphics[scale=0.5]{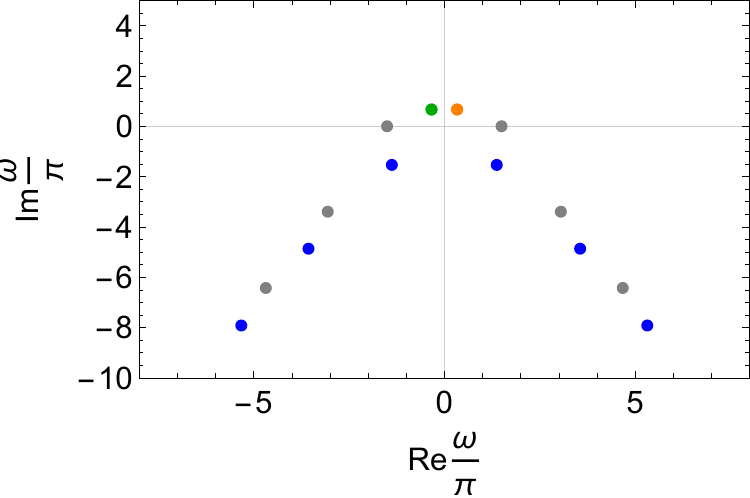}}
\subfigure[$\hatbeta$ = 10]{\includegraphics[scale=0.51]{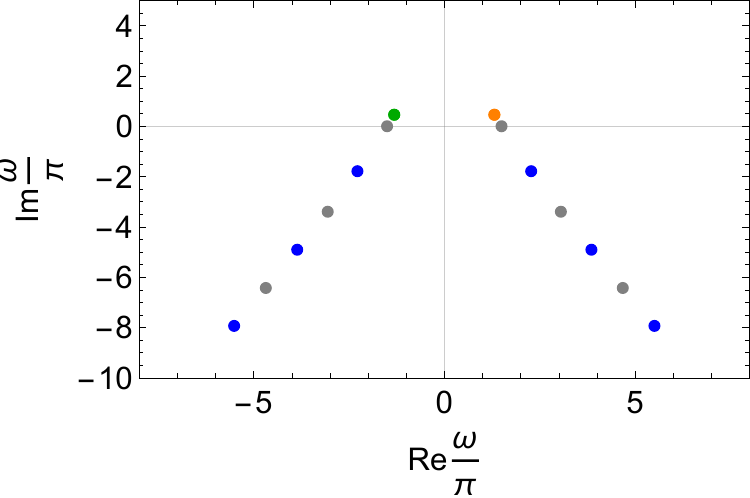}}
\caption{Behavior of first 8 QNMs in complex frequency plane with varying $\hatbeta$ for $k = 1.5$. The dots in gray represent the $k=0$ poles in all the figures. The pair of poles just below the horizontal axis on the lower half plane are the diffusion and \psg\ poles. It is obvious that these are the smooth deformations of the $\omega = \pm k$ poles of $\chi$ in the decoupling limit. The orange pole is $\wus$ and the green pole is $\wgl$.  We readily observe that just as in the case of $k=0$, the poles attain finite values as $\hatbeta \rightarrow\infty$ and realign approximately on the same straight line, and (except for the two unstable poles in the upper half plane) are halfway in between the poles in the decoupling limit.}\label{Fig:Saturate-k}
\end{figure}

In Fig.~\ref{Fig:Saturate-k}, we have plotted the QNM poles at various values of $\beta$ at $T=1$ and $k=1.5$. We readily notice that in the limit $\beta\rightarrow\infty$
\begin{enumerate}
\item all QNMs become independent of $\hatbeta$ saturating to finite values, and
\item they align themselves approximately on the same straight line as in the case of the decoupling limit, but placed approximately halfway between the latter poles (except for the two poles in the upper half plane).
\end{enumerate}
We have verified that the above features hold at any value of $k$ %\st{implying}
{indicating} the emergence of conformality at infinite mutual coupling.

\section{Non-linear evolution}\label{Sec:Nonlinear}

\subsection{%Non-linear simulation:
Methodology of non-linear simulations}\label{Sec:NonlinearMethod}

The full non-linear dynamics of the semi-holographic model can be readily computed based on the iterative procedure proposed in \cite{Iancu:2014ava}, and successfully demonstrated in the more complex case of classical Yang-Mills and dilaton plus black brane system in \cite{Ecker:2018ucc}. {Here we implement a simpler version of \cite{Ecker:2018ucc} with more general initial conditions in the case of homogeneous non-linear dynamics.}

The asymptotically $AdS_4$ metric representing the holographic sector can be assumed to be of the following form
\begin{equation}
ds^2 = - \frac{2}{r^2} {\rm d}t {\rm d}r -A(r,t) {\rm d}t^2 + S^2(r,t) ({\rm d}x^2 + {\rm d}y^2),
\end{equation}
in the in-going Eddington-Finkelstein coordinates with the boundary being at $r =0$. We can consistently set the anisotropy to zero, i.e. assume $G_{xx} = G_{yy}$. The bulk dilaton profile takes the form $\Phi(r,t)$ while the boundary scalar field $\chi(t)$ is only a function of time. The gravitational equations of motion \eqref{Eq:GravEoms} reduce to the following nested set of partial differential equations:
\begin{eqnarray}\label{Eq:ASPhi1}
\partial_r^2 S +\frac{2}{r}\partial_r S &=& -\frac{\kappa}{4}\,S (\partial_r\Phi)^2, \\\label{Eq:ASPhi2}
\partial_r(d_+{S}) &=& -\frac{3}{2r^2}S - \frac{d_+ S\partial_r S}{S},\\\label{Eq:ASPhi3}
\partial_r(d_+\Phi) &=&- \frac{d_+ S\partial_r \Phi}{S}- \frac{d_+ \Phi\partial_r S}{S},\\\label{Eq:ASPhi4}
\partial_r^2 A -\frac{1}{r}\partial_r A  &=& -4\frac{d_+ S\partial_r S}{S^2} + \kappa\, d_+\Phi \partial_r \Phi,\\\label{Eq:ASPhi5}
d_+^2 S  &=& -\frac{r^2}{2}d_+ S\,\partial_r \frac{A}{r^2} - \frac{1}{4}\kappa \, S(d_+ \Phi)^2,
\end{eqnarray}
where $d_+$ is the directional derivative along the outgoing null radial geodesic, i.e. 
\begin{equation}\label{Eq:d+}
d_+ \equiv\partial_t - \frac{1}{2}A(r,t) \partial_r.
\end{equation}

%The functions $A(r,t)$, $S(r,t)$, and $\phi(r,t)$ can be expressed  as following power series in $r$ near the boundary ($r \rightarrow \infty$):
%\begin{subequations}
%\begin{equation}
%A(r,t) = r^2 \sum_{n=0}^{\infty}a_n r^{-n}
%\end{equation}
%\begin{equation}
%S(r,t) = r \sum_{n=0}^{\infty}s_n r^{-n}
%\end{equation}
%\begin{equation}
%\phi(r,t) = \sum_{n=0}^{\infty}\phi_n r^{-n}
%\end{equation}
%\end{subequations}
Setting the boundary metric to be $\eta_{\mu\nu}$ and solving the equations of gravity \eqref{Eq:GravEoms} order by order in $r$ near the boundary gives
\begin{eqnarray}\label{Eq:A-exp}
A(r,t) &=&  \frac{1}{r^2} -\frac{3}{4} 
{{\phi^{(0)}}'}^2 
+ r a^{(3)}+ r^2 \left(\frac{1}{2} \phi^{(3)} {\phi^{(0)}}'-\frac{1}{24}
 {{\phi^{(0)}}'}^4\right)\nonumber \\&&
+r^3 \left(\frac{1}{4} \phi^{(3)} {\phi^{(0)}}''+\frac{1}{4}
  {\phi^{(3)}}' {\phi^{(0)}}'-\frac{1}{12} {{\phi^{(0)}}'}^3
 {\phi^{(0)}}''\right) + \mathcal{O}(r^4),\\\label{Eq:S-exp}
S(r,t) &=& \frac{1}{r} -\frac{r}{8}{{\phi^{(0)}}'}^2 +\frac{1}{384}r^3 
   \left({{\phi^{(0)}}'}^4-48 \phi^{(3)} {\phi^{(0)}}'\right)
   \nonumber\\ &&
   + \frac{1}{40}r^4 {\phi^{(0)}}'\left( a^{(3)} {\phi^{(0)}}' - 4
  {\phi^{(3)}}'+{{\phi^{(0)}}'}^2 {\phi^{(0)}}''\right)+ \mathcal{O}(r^{5}),
  \\\label{Eq:Phi-exp}
\Phi(r,t)&=& \phi^{(0)}+r
  {\phi^{(0)}}'+ r^3 \phi^{(3)} 
  \nonumber\\ && 
  +r^4 \left(-\frac{1}{4} a^{(3)} {\phi^{(0)}}'+ {\phi^{(3)}}'-\frac{1}{4} {\phi^{(0)}}'^2\right)
+\mathcal{O}(r^{5}),
\end{eqnarray}
where $'$ denotes the time-derivative. Above, the residual gauge freedom $r \rightarrow r + f(t)$ of the Eddington-Finkelstein gauge {is fixed} by setting the coefficient of $r$ in the expansion of $A(r,t)$ to zero. The normalizable modes, namely $a^{(3)}(t)$ and $\phi^{(3)}(t)$, remain undetermined in this procedure and need to be extracted from the full bulk solution that is determined by the initial conditions. The constraint of Einstein's equation implies that
\begin{equation}\label{Eq:constraint}
{a^{(3)}}' = \frac{3}{2} \phi^{(3)} {\phi^{(0)}}'+\frac{1}{2} {\phi^{(0)}}^{'''} {\phi^{(0)}}'-\frac{3}{8} {{\phi^{(0)}}'}^4 .
\end{equation}
Performing holographic renormalization of the on-shell action and {taking the functional derivative} with respect to the sources, we obtain the expectation values of the energy momentum tensor and the scalar operator in the dual CFT state \cite{Ecker:2018ucc}. These turn out to be
\begin{subequations}
\begin{equation}
\mathcal{T}^{\mu \nu} = \text{diag} \left( -2a^{(3)}, a^{(3)}, a^{(3)} \right)
\end{equation}
and
\begin{equation}\label{vev}
\mathcal{H} =  3\phi^{(3)}+ {\phi^{(0)}}'''-\frac{3}{4} {{\phi^{(0)}}'}^3,
\end{equation}
\end{subequations}
respectively. Considering terms which are linear in $\phi^{(0)}$, the above reproduces the result for the linear fluctuations \eqref{Eq:H-lin} when it is homogeneous. As a consistency check, we readily note that \eqref{Eq:constraint} reproduces the Ward identity of the CFT 
\begin{equation}\label{Eq:WICFT-hom}
\partial_\mu \mathcal{T}^{\mu \nu} = \mathcal{H}\partial^\nu\phi^{(0)}, \quad {\rm i.e.}\quad -2 {a^{(3)}}' =-\mathcal{H}{\phi^{(0)}}' = \beta \mathcal{H} 
\chi',
\end{equation}
where we have used the key relation $\phi^{(0)} = -\beta \chi$ which sets the value of the non-normalizable mode in terms of the boundary scalar field. The equation of motion for the boundary field $\chi$ given by \eqref{Eq:chi} reduces to
\begin{equation}\label{Eq:chi-hom}
\chi'' = -\beta \mathcal{H} =- \beta\left(3\phi^{(3)}- \beta^3{\chi}'''+\frac{3}{4} \beta^3{\chi'}^3\right).
\end{equation}
The conservation of the energy-momentum tensor of the total system \eqref{Eq:Full-T} amounts to the total energy $E_{\rm tot}$, which is simply the sum of the boundary scalar field's kinetic energy and the ADM mass of the black hole, {remaining constant}. Indeed it is easy to verify using \eqref{Eq:WICFT-hom} and \eqref{Eq:chi-hom} that
\begin{equation}\label{Eq:E-tot}
E_{\rm tot}'=0,\quad E_{\rm tot} = E_{\rm kin} + E_{\rm BH}, \quad {\rm with}\quad E_{\rm kin} = \frac{1}{2} {\chi'}^2 \quad{\rm and}\quad \quad E_{\rm BH} = - 2 a^{(3)}.
\end{equation}

{The non-linear dynamics of the full system is determined uniquely by the initial conditions for $\chi(t_0)$, $\chi'(t_0)$, $a^{(3)}(t_0)$ and the initial profile  of the bulk dilaton $\Phi(r, t_0)$ at the initial time $t_0$. The iterative method of computing this numerically %\st{has been} 
{is} discussed in Appendix \ref{section:Appendix B}.}

\subsection{Results %of the non-linear simulations
for the non-linear evolution of the homogeneous case}

In this section we present our results for the non-linear simulations of generic {but} homogeneous initial conditions. As discussed before, each evolution is uniquely specified by the initial configuration of the bulk dilaton field $\Phi$ and the initial values of $a^{(3)} = - (1/2) E_{\rm bh} = - M$ (the ADM mass of the black hole), and $\chi$ and $\chi'$, i.e. the values of the boundary scalar field and its time-derivative. Note that due to the presence of the symmetry given by \eqref{Eq:Axionic} we can set the initial value of $\chi$ (and therefore the boundary mode of the bulk dilaton field given by $\phi^{(0)} = -\beta\chi$) to zero without loss of generality.

For the purpose of illustration, we %\st{can} 
consider two different types of initial conditions (ICs) at the initial time $t= 0$: 
\begin{align}
%&\text{
&{\rm IC1}: \quad \chi =0, \quad \chi' =0, \quad \Phi(r) = r^5 e^{-r^2},\quad a^{(3)} = -1;
%($\phi$,$\phi_0$,$\phi_0^\prime$,$a_3$)=(${e^{-r^{-2}}}{r^{-5}},  0,  0,  -1$)}
\label{Eq:IC1}\\
&{\rm IC2}: \quad \chi =0, \quad \chi' =0.1, \quad \Phi(r) = -0.1 \beta r \, e^{-r^4},\quad a^{(3)} = -1.\label{Eq:IC2}
%&\text{IC2: ($\phi$,$\phi_0$,$\phi_0^\prime$,$a_3$)=(${0.01e^{-r^{-2}}}{r^{-5}},  0,  0.01,  -1$)}
\end{align}
%\begin{equation}
%\phi(r,t=0) =  \frac{e^{-r^{-2}}}{r^{5}} \qquad \phi_0 (t=0) = 0 \quad \phi_0'(t=0) = 0 \quad a_3 (t=0) = -1
%\end{equation}
%IC2:
%\begin{equation}
%\phi(r,t=0) =  \frac{0.01}{r}e^{-r^{-4}} \qquad \phi_0 (t=0) = 0 \quad \phi_0'(t=0) = 0.01 \quad a_3 (t=0) = -1.
%\end{equation}
These two initial conditions represent cases where the initial kinetic energy of the boundary scalar field is zero and non-zero, respectively. In both cases, the iterative method discussed in Section \ref{Sec:NonlinearMethod} converges to a very good accuracy after four iterations, and the total energy \eqref{Eq:E-tot} is conserved, i.e. is time-independent for a sufficiently long time which allows us to reliably draw our conclusions. A more detailed discussion about the numerical accuracy is provided in Appendix \ref{section:Appendix B}. 

\begin{figure}[t]
\centering
\includegraphics[scale=0.7]{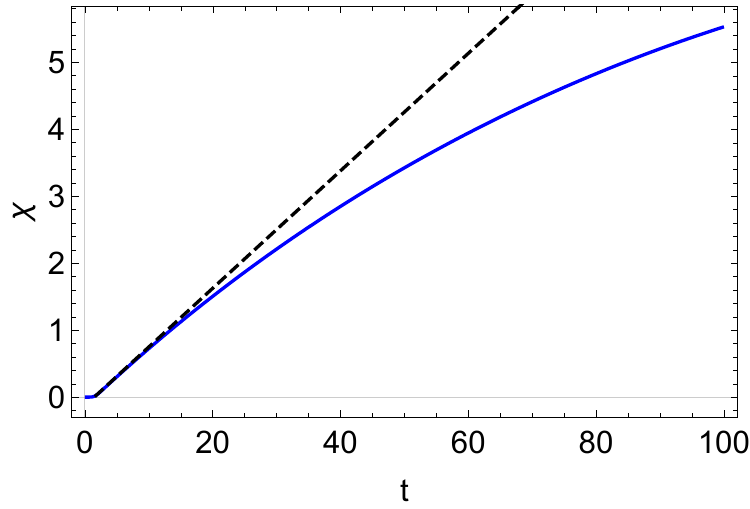}
\caption{Boundary scalar field $\chi$ as a function of time for $\beta =0.1$ and initial conditions given by \eqref{Eq:IC1}. We see that $\chi$ slowly saturates to a constant value.}\label{Fig:chi-t}
\end{figure}

\begin{figure}[t]
\centering
\subfigure[$E_{\rm kin}$, the kinetic energy of the boundary scalar field as a function of time for various values of $\beta$ and initial conditions set by \eqref{Eq:IC1}. ]{\includegraphics[scale=0.55]{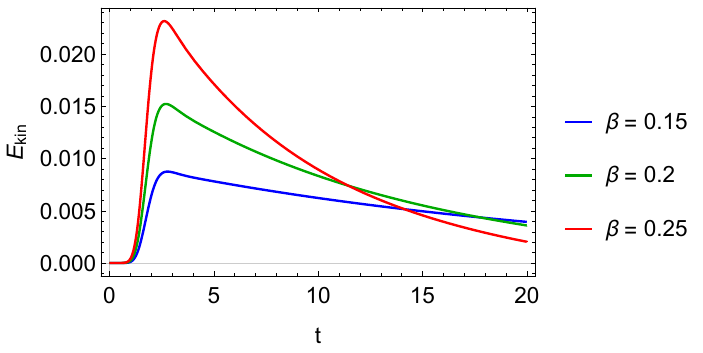}}
\hfill
\subfigure[$E_{\rm BH}$, the energy of the holographic sector as a function of time  for various values of $\beta$ and initial conditions set by \eqref{Eq:IC1}.]{\includegraphics[scale=0.55]{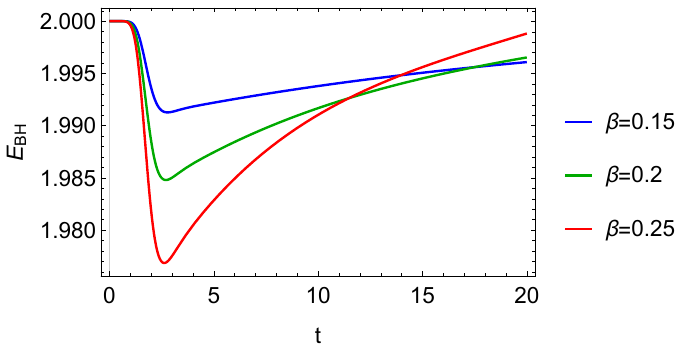}}
\caption{The kinetic energy of the boundary scalar field and that of the holographic sector as a function of time for various values of $\beta$ and initial conditions set by \eqref{Eq:IC1}. The sum is conserved. Higher values of $\beta$ lead to larger energy extraction from the black brane by the boundary scalar field but then the energy is returned back to the black brane irreversibly and completely at a higher rate. The time $t_{max}$ at which the boundary scalar field attains its maximum energy $E_{max}$ is independent of $\beta$ and is $\approx 2.9$ as shown in the Table \ref{Table:IC1}.}\label{Fig:Energies-IC1-t}
\end{figure}

For both initial conditions, as we have anticipated in Section \ref{Sec:QNMhom}, initially there is transfer of  energy from the black hole to the boundary scalar field. This is followed by complete and irreversible transfer of energy to the black hole. When $\beta$ is small, we also anticipated that the initial transfer of energy to the boundary should be rapid while the subsequent reverse transfer of energy back to the black hole should be slow. The final state is just the thermal state given by the black hole with a constant mass and with constant values of the boundary scalar and bulk dilaton fields. A plot of the boundary scalar field $\chi(t)$ with time is provided in Fig.~\ref{Fig:chi-t} for $\beta = 0.1$ and initial conditions set by \eqref{Eq:IC1}. The kinetic energy of the scalar field $E_{\rm kin}(t)$ and that of the holographic sector $E_{\rm BH}(t) = 2 - E_{\rm kin}(t)$ are provided in Fig.~\ref{Fig:Energies-IC1-t} with the same initial conditions  {note} the total energy, $E_{\rm kin}(t)+E_{\rm BH}(t)$, is $2$). We find that the kinetic energy of the boundary scalar $E_{\rm kin}$ fits perfectly to an exponentially decaying function, i.e. 
$$E_{\rm kin}(t)\approx \alpha e^{-\gamma t} \quad {\rm with} \quad \gamma>0$$at late time. 

From the QNM analysis, we can also anticipate $\gamma$ quantitatively. At late time, we expect that $$ \chi(t) \approx \chi_f + \tilde\chi \, e^{-\gamma_{\rm Q}t}$$where $\chi_f$ is the final value of $\chi$ and $\gamma_{\rm Q}$ is determined by the homogeneous (purely imaginary) \psg\ mode with the identification $\wpg = - i \gamma_{\rm Q}$, since $\wpg$ is the QNM pole that is closest to the origin and is on the lower half plane. The kinetic energy of the scalar field should then behave as $$E_{\rm kin}\approx e^{-2\gamma_{Q}t}$$at late time and therefore $\gamma = 2 \gamma_{\rm Q}$. Finally, since $\wpg$ is given by \eqref{Eq:psg} at small $\beta$, we obtain that
\begin{equation}\label{Eq:gamma-pred}
\gamma \approx 11.2\pi\beta^2 T^2
\end{equation}
with $T$ determined by the final mass of the black hole (which is $1/2 E_{BH}$) via \eqref{Eq:HawkingT}. 

\begin{figure}[t]
\centering
\subfigure[$\alpha$ scales as $\beta^2$ for fixed initial conditions. Blue dots are the values of $\alpha$ from Table \ref{Table:IC1} for the corresponding $\beta$ values in the table. Red dashed line is fitted function $a \beta^2 + b$ with $a = 0.406424$ and $b= -0.0000136$.% giving the best fit.
]{\includegraphics[scale=0.5]{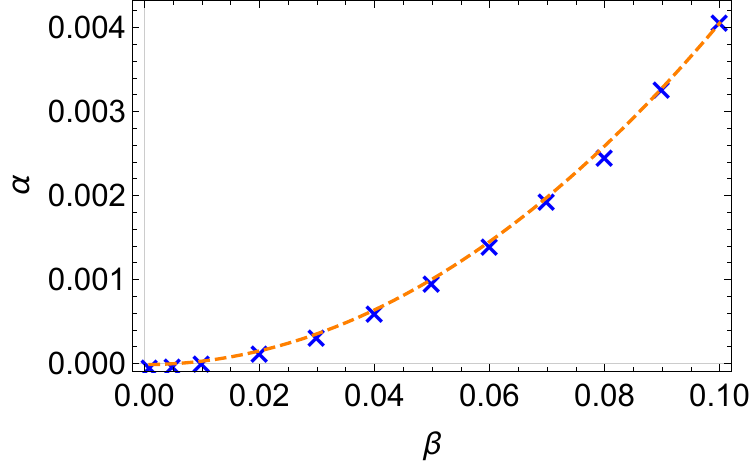}}
\hfill
\subfigure[$\gamma$ scales as $\beta^2$ for fixed initial conditions. Blue dots are the values of $\gamma$ from Table \ref{Table:IC1} for the corresponding $\beta$ values in the table. Red dashed line is fitted function $a \beta^2 + b$ with $a = 2.00568$ and $b= -6.056 \times 10^{-6}$.% giving the best fit.
]{\includegraphics[scale=0.5]{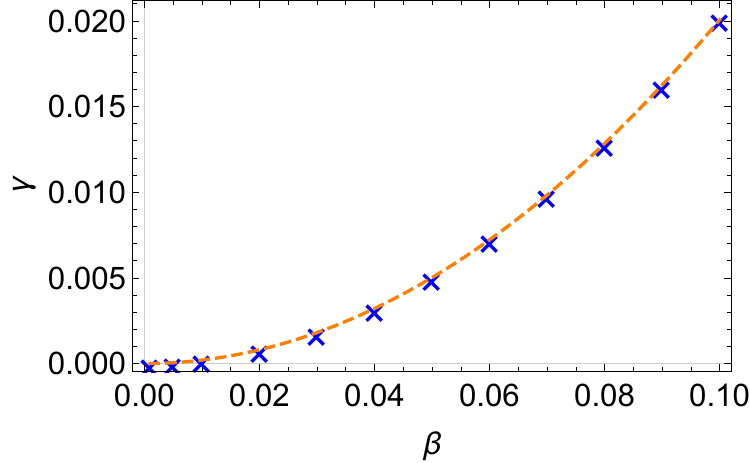}}
\hfill
\subfigure[$E_{max}$ scales as $\beta^2$ for fixed initial conditions. Blue dots are the values of $E_{max}$ from Table \ref{Table:IC1} for the corresponding $\beta$ values in the table. Red dashed line is fitted function $a \beta^2 + b$ with $a = 0.39445$ and $b= 3.791 \times 10^{-6}$% giving the best fit.
]{\includegraphics[scale=0.5]{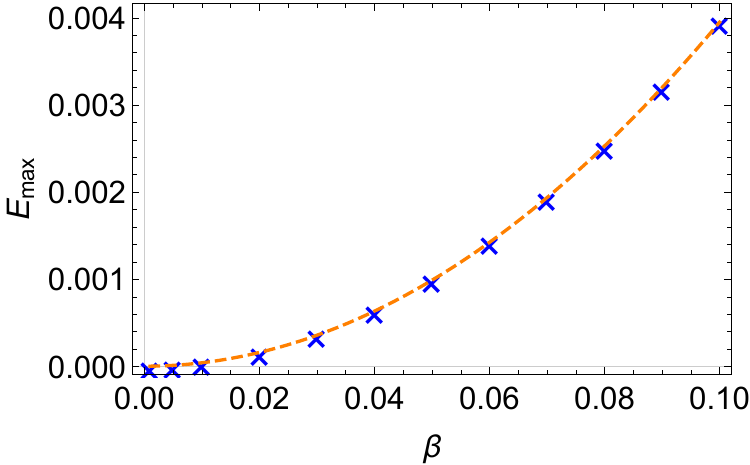}}
\caption{Scaling of $\alpha$, $\gamma$ and $E_{max}$ with semi-holographic coupling $\beta$ for initial conditions set by \eqref{Eq:IC1}. {We are using units in which the initial black hole mass is unity.}}\label{Fig:alphagamabeta}
\end{figure}

{In Table }\ref{Table:IC1}, the values of $\alpha$ and $\gamma$ have been computed for $\beta$ {ranging} between $0.001$ and $0.1$ and initial conditions set by \eqref{Eq:IC1}. First, we find that the value of $\gamma$ satisfy \eqref{Eq:gamma-pred} {with remarkable} accuracy. For the initial conditions set by \eqref{Eq:IC1}, the final mass of the black hole should equal {to} its initial mass (which is unity) because the initial kinetic energy of $\chi$ is zero. Therefore, we obtain from \eqref{Eq:HawkingT} that the final temperature is $3/(4\pi) \approx 0.239$. From \eqref{Eq:gamma-pred} with $\beta =0.1$, we find $$\gamma %\approx 11.2\times \pi\times 0.01 \times 0.239^2
\approx 0.0201$$which is precisely the value calculated in Table \ref{Table:IC1}. 
The other values of $\gamma$ %in Table \ref{Table:IC1} 
are reproduced to the same degree of accuracy by \eqref{Eq:gamma-pred}. 
Fig.~\ref{Fig:alphagamabeta} confirms that $\gamma$ and \eqref{Eq:gamma-pred} scale like $\beta^2$ for fixed initial conditions. %As evident from Table \ref{Table:IC1}, 
The time $t_{max}$ at which the kinetic energy of the boundary scalar attains its maximum value, {$E_{max}$}, is independent of $\beta$. {In} Fig.~\ref{Fig:alphagamabeta} {we see that} $E_{max}$
%\ref{Table:IC1} 
also scales like $\beta^2$ for fixed initial conditions. We also note from Table \ref{Table:IC1}  that 
\begin{align}
    \alpha \approx E_{max}.
\end{align}

\begin{table}[t]
\centering
\begin{tabular}{|c|c|c|c|c|}
\hline
$\beta$ & $\alpha$ & $\gamma$ & $E_{max}$ & $t_{max}$ %& Area   
\\
\hline
0.001 & 3.928 $\times$ $10^{-7}$ & 2 $\times$ $10^{-6}$ & 3.981 $\times$ $10^{-7}$ & 2.9 %& 3.228 $\times$ $10^{-6}$ 
\\
\hline
0.005 & 9.821 $\times$ $10^{-6}$ & 0.00005 & 9.983 $\times$ $10^{-6}$ & 2.9% & 0.0000834
\\
\hline
0.01 & 0.0000393 & 0.0002 & 0.00004 & 2.9 
%& 0.00033
\\
\hline
0.02 & 0.000157 & 0.0008 & 0.000161 & 2.9 
%& 0.00132
\\
\hline
0.03 & 0.000355 & 0.0018 & 0.00036 & 2.9  
% &0.00294 
\\
\hline
0.04 & 0.000633 & 0.0032 & 0.000644 & 2.9
%&0.00514 
\\
\hline
0.05 & 0.000993 & 0.005 &0.000994  & 2.9 
%& 0.00786
\\
\hline
0.06 & 0.001437 & 0.007216 & 0.0014298 & 2.9 
%& 0.0110299
\\
\hline
0.07 & 0.001969 & 0.00983 & 0.001934 & 2.9 %& 0.0145534
\\
\hline
0.08 & 0.0025 & 0.0128 &  0.00252 & 2.9 %& 0.0183
\\
\hline
0.09 & 0.0033 & 0.0162 & 0.003199 & 2.9 %& 0.0222
\\
\hline
0.1 & 0.0041 & 0.0201 & 0.00394& 2.9 %& 0.0262
 \\
\hline
\end{tabular}
\caption{$E_{\rm kin}$, the kinetic energy of the boundary scalar, is fitted to function $\alpha e^{-\gamma t}$ from $t=10$ to $t=18.5$ for different values of $\beta$ with the initial conditions set by \eqref{Eq:IC1}. $t_{max}$ is the time when $E_{\rm kin}$ attains its maximum value $E_{max}$. The values of $\gamma$ match with those predicted by \eqref{Eq:gamma-pred} to a remarkably good accuracy. 
 We note that $t_{max}$ is independent of $\beta$ while $\alpha\approx E_{max}$ and both scale as $\beta^2$. 
For fits see Fig.~\ref{Fig:alphagamabeta}. {We are using units in which the initial black hole mass is unity.}}\label{Table:IC1}
\end{table}

The above scaling properties lead to a rather interesting result. For small $\beta$ and initial conditions set by \eqref{Eq:IC1}, we should have
\begin{equation}
\mathcal{A}\equiv \int_0^\infty E_{\rm kin} dt \approx \alpha \int_{t_{max}}^\infty e^{-\gamma t} dt\approx \frac{\alpha}{\gamma}e^{-\gamma t_{max}}
\end{equation}
Since $\alpha$ and $\gamma$ both scale as $\beta^2$ and $t_{max}$ is independent of $\beta$, we obtain that
\begin{equation}\label{Eq:uncertainty}
\lim_{\beta \rightarrow 0}\mathcal{A} \approx 0.2.
\end{equation}
Thus the limit $\beta \rightarrow 0$ is non-trivial. Furthermore, $\mathcal{A}$ is independent of $\beta$ for small $\beta$ to a very good approximation. This result implies that if the boundary system draws more energy from the black hole (bath), then it has to give it back to the black hole (bath) at a higher rate. It would be interesting to see if  \eqref{Eq:uncertainty} could be a generic feature of out-of-equilibrium open quantum systems with the value of the limit possibly depending on the initial conditions. 

%Because of this $\beta^2$ scaling of $\alpha$ and $\gamma$, $\beta \rightarrow 0$ limit is highly non-trivial. Area under the scalar energy curve goes to a finite constant (instead of zero) as $\beta \rightarrow 0$. Because, Area $\sim$ $\int_{\gamma_0}^{\infty} \alpha \: e^{- \gamma t} dt$ this implies Area $\sim$ $\frac{\beta^2 e^{-\beta^2 \gamma_0}}{\beta^2}$ $\sim$ $ e^{-\beta^2 \gamma_0}$, a non-zero constant as $\beta \rightarrow 0$. ($\gamma_0 \approx 2.9$ is the time from when the scalar energy curve starts decaying  exponentially as $\alpha e^{-\gamma t}$) 

\begin{figure}[t]
\centering
\subfigure[Position of apparent horizon for $\beta =0.1$ and initial conditions set by \eqref{Eq:IC1}]{\includegraphics[scale=0.5]{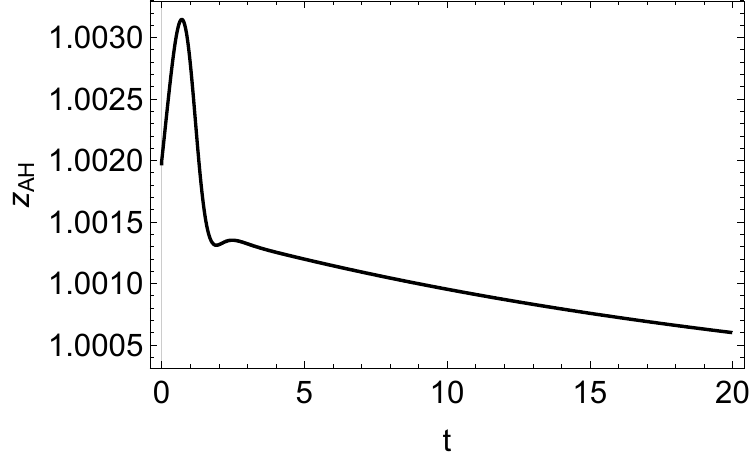}}
\hfill
\subfigure[The entropy of the system for different values of $\beta$ and initial conditions set by \eqref{Eq:IC1}.]{\includegraphics[scale=0.7]{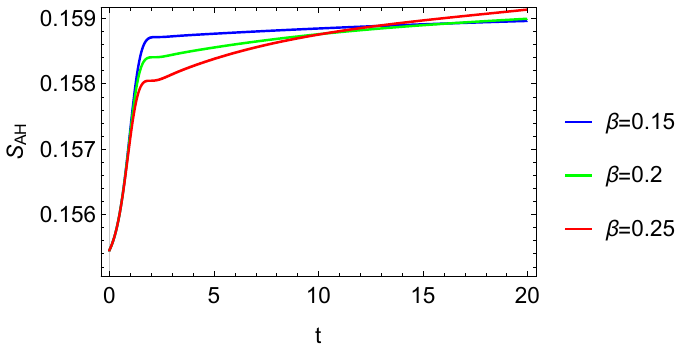}}
\caption{The plots of the radial position of the apparent horizon and its area (the entropy of the full system) as a function of time for different values of $\beta$ and initial conditions set by \eqref{Eq:IC1}. Although the apparent horizon first moves away from the boundary (at $z =0$) and then towards it, the entropy grows monotonically.}\label{Fig:entropy}
\end{figure}

The area of the apparent horizon acts as a proxy for the entropy of an out-of-equilibrium semi-holographic system as noted in \cite{Ecker:2018ucc}. We indeed find that the entropy grows monotonically although the black hole mass is non-monotonic as a function of time. In Fig.~\ref{Fig:entropy}, we have plotted the radial position and area of the apparent horizon as a function of time for various values of $\beta$ and initial conditions set by \eqref{Eq:IC1}. {We observe that the entropy saturates to its final thermal value more quickly for smaller values of the coupling $\beta$.}

\begin{figure}
\centering
\subfigure[$\dot{E}_{\rm kin}/E_{\rm kin}$ as a function of time for $\beta = 0.1$ and initial conditions set by \eqref{Eq:IC1}. Note that the final value $0.0201$ is consistent with Table \ref{Table:IC1}.]{\includegraphics[scale=0.5]{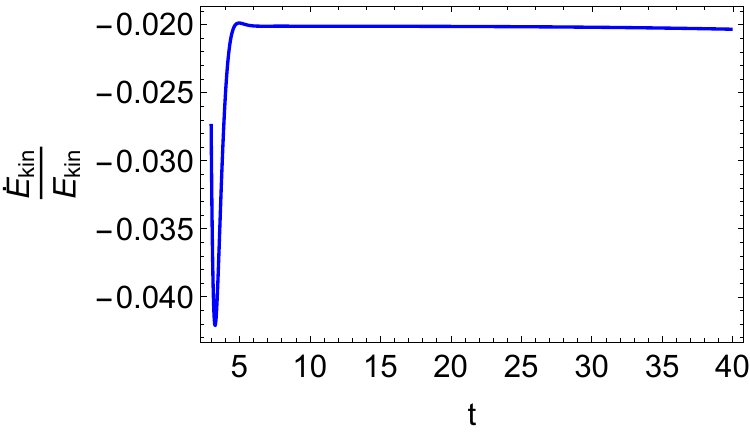}}
\hfill
\subfigure[$\dot{S}_{AH}/S_{AH}$ as a function of time for $\beta = 0.1$ and initial conditions set by \eqref{Eq:IC1}.]{\includegraphics[scale=0.55]{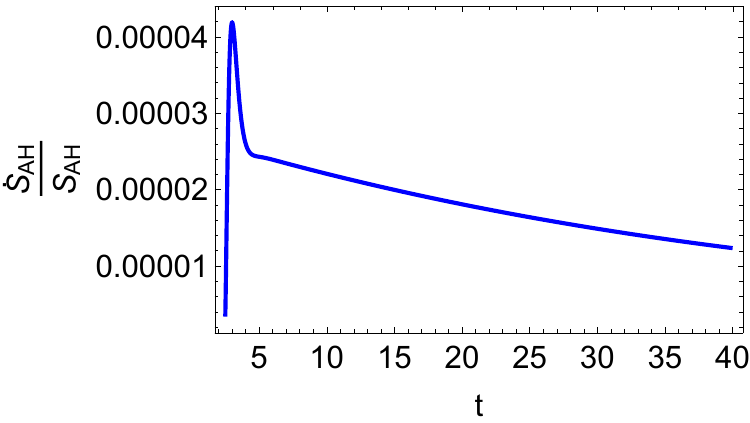}}
\caption{$\dot{E}_{\rm kin}/E_{\rm kin}$ and $\dot{S}_{AH}/S_{AH}$  for $\beta = 0.1$ and initial conditions set by \eqref{Eq:IC1}. While the energy transfer to the bath occurs at a constant rate at late time, the rate of growth of entropy vanishes as it should if the system equilibrates.}\label{Fig:isentropy}
\end{figure}

Fig.~\ref{Fig:isentropy} indicates that although ${\dot{E}_{\rm kin}}/{E_{\rm kin}} \approx \gamma$ is setting the rate of energy exchange between the subsystems, the rate of growth of entropy ${\dot{S}_{AH}}/{S_{AH}}$  decays to zero at late times. This suggests that asymptotically there is an isentropic transfer of energy between the boundary scalar and the black hole. In fact, this is a proof of thermal equilibration because the latter implies that the rate of growth of entropy should vanish at late time.

\begin{figure}[t]
\centering
\subfigure[$E_{\rm kin}$, the kinetic energy of the boundary scalar field as a function of time for various values of $\beta$ and initial conditions set by \eqref{Eq:IC2}. ]{\includegraphics[scale=0.45]{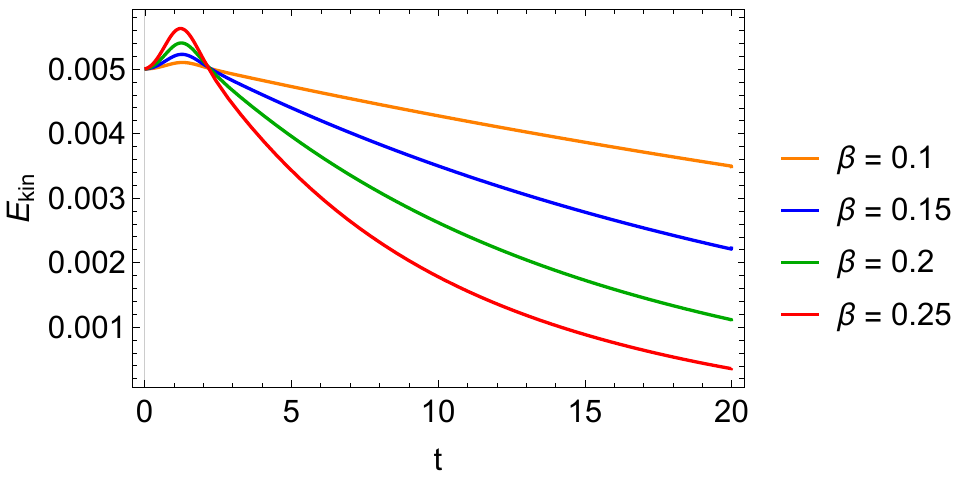}}
\hfill
\subfigure[$E_{\rm BH}$, the energy of the holographic sector as a function of time  for various values of $\beta$ and initial conditions set by \eqref{Eq:IC2}.]{\includegraphics[scale=0.45]{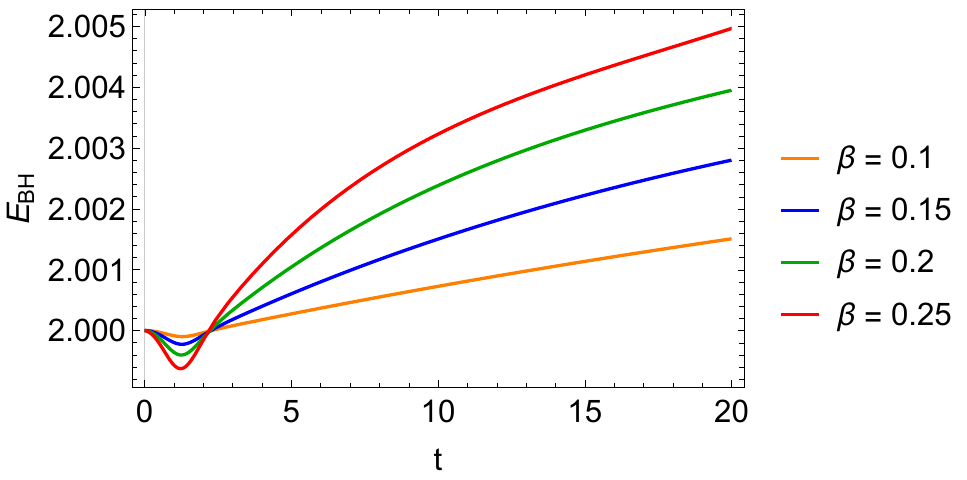}}
\caption{The kinetic energy of the boundary scalar field and that of the holographic sector as a function of time for various values of $\beta$ and initial conditions set by \eqref{Eq:IC2}. The sum is conserved. Higher values of $\beta$ lead to larger energy extraction from the black brane by the boundary scalar field but then the energy is returned back to the black brane irreversibly and completely at a higher rate. The time $t_{max}$ at which the boundary scalar field attains its maximum energy $E_{max}$ is independent of $\beta$. Thus the qualitative features are exactly like in the case of the initial condition \eqref{Eq:IC1} shown in Fig. \ref{Fig:Energies-IC1-t}.}\label{Fig:Energies-IC2-t}
\end{figure}

We observe the same qualitative features for simulations of the system with initial conditions set by \eqref{Eq:IC2} in which the initial energy in the scalar field is non-vanishing (see Fig. \ref{Fig:Energies-IC2-t}). Of course $\gamma$, the rate of decay of the scalar kinetic energy at late time, is quantitatively the same as well. Furthermore, the maximum transfer of energy to the scalar sector, i.e. $ E_{max} - E_{kin}(t_0)$ with $t_0$ being the initial time, scales as $\beta^2$ and $t_{max}$ is almost independent of $\beta$. The rate of the monotonic growth of the entropy of the system goes to zero at late time confirming thermal equilibration. We have found that indeed that these features are present for generic initial conditions while the final mass of the black hole is simply determined by the fact that it is equal to the total initial energy of the system. Only the transient behavior at initial time depends on the initial conditions.

Parametrically slow thermalization in non-conformal holography had been observed in \cite{Janik:2016btb,Gursoy:2016ggq}. However the mechanism discussed in these works was not related to a \psg\ mode but rather to the presence of two intersecting branches of black brane solutions.

\section{Conclusions and outlook}\label{Sec:Conclude}

Our simple semi-holographic model is fundamentally an open quantum system involving one preserved and one weakly broken global symmetry. For generic homogeneous initial conditions and weak inter-system coupling, an unstable mode implies rapid transfer of energy from the holographic sector to the massless field at the boundary, while the purely imaginary \psg\ mode governs a slow, irreversible and complete transfer of energy to the black brane at later stages. The entropy of the system, represented by the area of the apparent horizon of the black brane, grows monotonically although the black brane mass behaves non-monotonically. Higher values of the inter-system coupling leads to more extraction of energy from the black brane by the boundary scalar field, but then a quicker irreversible and complete transfer of energy back to the black brane. Furthermore, the integral of the kinetic energy of the boundary scalar field with time for a vanishing initial value, remains finite even in the limit when the inter-system coupling vanishes. This feature deserves a more basic understanding with insights from non-equilibrium statistical mechanics.

The inhomogeneous dynamics is even richer. We find that for any value of the inter-system coupling, the mode at the origin becomes diffusive at finite momentum. At weak inter-system coupling, the \psg\ mode also moves up on the negative imaginary axis and collides with the diffusion pole producing a pair of complex poles as the momentum is increased. {This results in the system having low energy propagating modes for momentum $k> k_c$ where $k_c$ is close to zero for small inter-system coupling. This feature is quite ubiquitous in dissipative systems with a softly broken symmetry and is called the $k$-gap~\cite{Baggioli:2019jcm,PhysRevB.101.214312}.} 

{Additionally, our system} has a third pole which also moves from negative imaginary infinity along the negative imaginary axis with increasing momentum and produces an instability as it crosses the origin. This is similar to the Gregory-Laflamme instability \cite{Gregory:1993vy}. At intermediate values of the coupling, these three poles are approximately degenerate for an intermediate range of momentum. At higher values of the inter-system coupling, the diffusion pole (instead of the third pole) reverses back and crosses the origin again as the momentum is increased producing the Gregory-Laflamme type instability, while the \psg\ mode collides with the third pole on the negative imaginary axis. At even higher values of the inter-system coupling, the diffusion constant of the diffusive mode becomes negative, and no pole crosses the real axis from the lower half plane at any value of the momentum. The Gregory-Laflamme momentum, at which one of the three modes has zero energy, exists only below a critical value of the inter-system coupling.

The model thus exhibits diverse behavior for different values of the inter-system coupling. Since the total conserved energy of the system is simply the sum of two non-negative terms, namely the kinetic energy of the massless scalar field at the boundary and the black brane energy, we have argued that in absence of zero modes with finite momenta, the unstable poles only imply short-term instability involving inverse transfer of energy from the holographic sector to the boundary scalar field. The dynamics is constrained by the facts that the energies cannot grow indefinitely and that the entropy represented by the total area of the apparent horizon should increase monotonically. The rate of growth of entropy becomes zero typically when the system reaches the thermal state represented by the static black brane geometry. However, there is no guarantee that the system has an entropy current generically although there exists a global monotonic entropy function. It is likely that the presence of zero modes at finite momenta can lead the system to turbulent or glassy final states even in the presence of an entropy current and the fate of the instability is  also describable by a quasihydrodynamic theory.\footnote{If the infrared behavior of the system can be described by hydrodynamics, there is a generic expectation of the existence of an entropy current especially in holographic theories (see \cite{Hubeny:2011hd,Romatschke:2009kr} for a review). A more general understanding of the existence of the entropy current has been led by the construction of an equilibrium partition function \cite{Banerjee:2012iz}. It is therefore of importance to understand if a quasihydrodynamic theory can describe the dynamics of the black brane horizon and thus the infrared behavior of the full system even in the presence of instabilities -- see \cite{Emparan:2015gva} for a hydrodynamic description of the fate of the Gregory-Laflamme instability.}  In the former case, one may achieve equipartition of energy between the perturbative and holographic sectors at least for lower values of the inter-system coupling. We leave this issue for future investigation.

It would also be of interest to find an appropriate causal and consistent quasihydrodynamic effective theory\footnote{{In the context of holography, an early attempt has been made in \cite{Iyer:2009in,Iyer:2011qc} by adopting a consistent truncation of the full dynamics to that of the evolution of the energy-momentum tensor operator.}} which can describe the interplay of the three modes that play a role in the low energy (macroscopic) dynamics of the system. Furthermore, we would like to have a better understanding of whether the diffusion constant saturates or satisfies conjectured bounds beyond those values of the inter-system coupling reported here. The latter would require us to study the applicability of quasihydrodynamics \cite{Grozdanov:2018fic,Hayata:2014yga} in this system in more detail, and furthermore study the Lyapunov exponent and the butterfly velocity. This discussion needs to be reconciled with the change of sign of the diffusion constant with increased inter-system coupling. We leave this for the future.

We would also like to investigate whether the usual effective metric and scalar couplings of semi-holography \cite{Banerjee:2017ozx,Kurkela:2018dku} have similar features to the simple linear inter-system coupling described here. The numerical simulations done earlier demonstrate similar slow irreversible transfer of energy to the black brane \cite{Ecker:2018ucc} and could be indeed related to weak breaking of global symmetries via non-linearities. 

\acknowledgments{
It is a pleasure to thank Matteo Baggioli and Christian Ecker for helpful discussions, and Matteo Baggioli and Blaise Gout\'{e}raux for comments on the manuscript. 
% {We are also grateful to Christian Ecker for providing his helpful insights on the numerics.} 
{The numerical simulation used in Section \ref{Sec:Nonlinear}
for the full non-linear dynamics of the semi-holographic model was building
on code developed by Christian Ecker in our previous joint work \cite{Ecker:2018ucc}
and we are grateful to him for providing helpful insights.}
AM was supported by the Ramanujan Fellowship grant and the Early Career Research award of the Science and Engineering Board (SERB) of the Department of Science and Technology (DST) of India, IFCPAR/CEFIPRA grant no 6304-3 and also the new faculty seed grant and the Center of Excellence initiative of IIT Madras. AS is supported by the Austrian Science Fund (FWF), project no. J4406. This research was supported in part by the International Centre for Theoretical Sciences (ICTS) for the online program - Extreme Nonequilibrium QCD (code: ICTS/ExNeqQCD2020/10).
}
\appendix

\section{Computation of the quasinormal modes}\label{section:Appendix A}

We can solve for the quasinormal frequencies of the semi-holographic system numerically by modification of the usual procedure described in \cite{Yaffe}. In the usual case, one can reduce the problem of finding the quasinormal mode spectrum to a linear eigenvalue problem. In the semi-holographic case, it will be a cubic eigenvalue problem that smoothly reduces to the usual linear problem in the decoupling limit. 

Following \cite{Yaffe}, let us define the dimensionless radial coordinate $u = {r}/{r_h}$ so that the boundary is at $u=0$ and the horizon is at $u=1$ (note from \eqref{Eq:HawkingT} that $r_h = 3/(4\pi T)$). The domain of interest is $u \in [0,1]$. As evident from the near boundary expansion \eqref{Eq:asymp-f}, we can define a function $g_{k,\omega}(u)$ via
\begin{equation}\label{Eq:g-def}
g(u) = \frac{64}{27\,u^3}\left(f(k,\omega,u) -\left(\phi^{(0)}(k,\omega)  - \mathit{i}\omega\phi^{(0)}(k,\omega) -\frac{k^2 r^2}{2}\phi^{(0)}(k,\omega) \right)\right).
\end{equation}
%\begin{equation}\label{Eq:g-def}
%g(u) = \frac{1}{r^3 (\pi T)^3}\left(f(k,\omega,r) -\left(\phi^{(0)}(k,\omega)  - \mathit{i}\omega\phi^{(0)}(k,\omega) -\frac{k^2 r^2}{2}\phi^{(0)}(k,\omega) \right)\right).
%\end{equation}
We also drop the $k,\omega$ subscripts in $g_{k,\omega}(u)$ for notational simplicity. Crucially, $g(u)$ is analytic at $u=1$ owing to the ingoing boundary condition. The above definition of $g(u)$ ensures a smooth decoupling limit in which $\beta\rightarrow 0$. It is also convenient to define the dimensionless wave number $q = k/(\pi T)$ and the dimensionless frequency $\varpi = \omega/(\pi T)$. We can readily obtain the differential equation for $g(u)$ by substituting the above in \eqref{Eq:qnm-eq}.

%The unknown function $g$ will be expanded in series of Chebyshev polynomials (of the first kind). Chebyshev polynomials are completely regular (analytic) and thus any sum of Chebyshev will be regular and thus all terms will automatically satisfy both horizon and boundary regularity conditions. 

%Note that the relevant range of $u := \frac{r_h}{r}$ is $[0,1]$. The point $u=0$ is the spacetime boundary, while $u=1$ is the horizon. For later convenience, we introduce a dimensionless rescaled momentum $k = \frac{q}{\pi T}$ and decay rate $\lambda = \frac{\mathit{i} \omega}{\pi T}$. 
%The QNM equation in terms of $g(u)$ is
%\begin{equation}
%\frac{27}{64}u \left(u^3-1\right) g''(u)+\frac{1}{2}\left(14 u^3-3 \lambda  u-8\right) g'(u)+ \frac{3}{16}\left(3 k^2 u-16 \lambda +48 u^2\right)g(u) = 0
%\end{equation}
We numerically approximate $g(u)$ as a linear combination of the first $M + 1$ Chebyshev polynomials which are linearly mapped onto the domain $[0,1]$. While working on the interval $[-1,1]$, the \textit{Gauss-Lobatto grid points} for a series of $M+1$ Chebyshev polynomials are taken to be $\cos(\frac{n \pi}{M})$ for $n=0,1,...,M$. A linear map of these grid points to the domain $[0,1]$ is given by $\frac{1}{2} \left( 1 - \cos(\frac{n \pi}{M}) \right)$%with $n=0,1,...,M$
. Next, we plug the Chebyshev expansion of $g(u)$ into the QNM equation for $g(u)$ and evaluate the resulting (truncated) QNM equation at each grid point in the domain $[0,1]$. The resulting equation, at each grid point, is a linear combination of the unknown expansion coefficients. We extract the coefficient $q_{ij}$ which multiplies the $j^{\text{th}}$ expansion coefficient in the equation for the $i^{\text{th}}$ grid point. We assemble a matrix $Q = ||q_{ij}||$, such that the complete set of equations takes the schematic form 
\begin{equation}\label{Eq:Sch1}
Q\cdot\text{coeffs} +  \phi^{(0)} P= 0
\end{equation}
where $P$ is a vector. 

Since the equation for $g(u)$ involves $\phi^{(0)}(k,\omega)$, we need to give an additional input, namely the boundary condition at $r=0$ given by \eqref{Eq:qnm-eq2} and \eqref{Eq:H-lin} which can be rewritten in the form
\begin{equation}\label{Eq:bdyr0}
(\varpi^2 - q^2)\phi^{(0)} + \tilde\beta^2 \left(
  i \varpi^3 \phi^{(0)}(k,\omega) - \frac{3}{2}\mathit{i} \varpi q^2\phi^{(0)}+3 \, g(u=0) \right)= 0,
\end{equation}
where we have used $\phi^{(3)} = (\pi T)^3 g(u=0)$ which follows from the defining equation \eqref{Eq:g-def} for $g(u)$ and the asymptotic expansion \eqref{Eq:asymp-f}, and $\tilde\beta \equiv \beta \sqrt{\pi T}$ is the dimensionless mutual coupling.\footnote{Note that in $d=3$ the boundary scalar field $\chi$ has mass dimension $1/2$. Also $\beta\chi$ is dimensionless since $\phi^{(0)}$ is the source of a marginal CFT operator.} Obviously, $g(u=0)$ is also a linear sum of the Chebyshev coefficients. We can then combine these coefficients and $\phi^{(0)}(k,\omega)$ into a column vector $V$, and also \eqref{Eq:Sch1} and \eqref{Eq:bdyr0} together into a matrix equation of the form
\begin{equation}\label{Eq:Sch2}
\widetilde{Q}\cdot {V} = 0.
\end{equation}
The above system of linear equations for the elements of $V$ will have solutions only for certain values of $\varpi$ (the dimensionless frequency) for which ${\rm det}\,\widetilde{Q} =0$. These values of $\varpi$ will constitute the quasinormal spectrum of the full system. However, one can follow a strategy which is better than solving for ${\rm det}\,\widetilde{Q} =0$  for determining the quasinormal mode spectrum.

We note that each element of $\widetilde{Q}$ is at most \emph{cubic} in $\varpi$.\footnote{The cubic term arises from the dependence of $\mathcal{H}$ on the third time-derivative $\phi^{(0)}$ as explicit in \eqref{Eq:H-lin}. In absence of semi-holographic coupling, $Q$ depends only linearly on $\varpi$.} Therefore, we can write $\widetilde{Q} = Q_0  + \varpi Q_1 + \varpi^2 Q_2 + \varpi^3 Q_3$ where $Q_0$, $Q_1$, $Q_2$, $Q_3$ are independent of $\varpi$. Thus \eqref{Eq:Sch2} is simply a cubic eigenvalue problem whose solutions give us the desired QNM for a given value of $q$, the dimensionless wave number. 

We can readily solve a cubic eigenvalue problem by converting it into a generalized eigenvalue problem. To see this, we simply note that we can rewrite \eqref{Eq:Sch2} in the form
\begin{equation}\label{Eq:Sch3}
\alpha\cdot \begin{pmatrix} V \\ \varpi V \\\varpi^2 V \end{pmatrix} = \varpi \,\,\delta\cdot \begin{pmatrix} V \\ \varpi V \\\varpi^2 V \end{pmatrix}
\end{equation}
where 
\begin{subequations}
\begin{equation}
\alpha =
\begin{pmatrix}
O & I & O\\
O & O & I\\
Q_0 & Q_1 & Q_2
\end{pmatrix},
\end{equation}
and
\begin{equation}
\delta =
\begin{pmatrix}
I & O & O\\
O & I & O\\
O & O & -Q_3
\end{pmatrix},
\end{equation}
\end{subequations}
{where $O$ is the null matrix and} $I$ is {the} identity matrix ({with the} same rank as $Q$). We can obtain the full QNM spectrum by using standard routines for solving the generalized eigenvalue problem of the type \eqref{Eq:Sch3}.

\section{Iterative procedure for computing the
non-linear dynamics}\label{section:Appendix B}

The non-linear dynamics of the full system can be solved by the following iterative procedure with the input initial conditions for $\chi(t_0)$, $\chi'(t_0)$, $a^{(3)}(t_0)$ and the initial profile  of the bulk dilaton $\Phi(r, t_0)$. Note that it is not necessary to specify higher order time derivatives of $\chi$ at initial time although the equation for time-evolution of $\chi$ given by \eqref{Eq:chi-hom} is third order. Nevertheless the initial profile of the bulk dilaton $\Phi(r,t_0)$ needs to be consistent with the near-boundary radial expansion of $\chi$, i.e. we need $$\Phi(r,t_0) = -\beta \chi(t_0)- \beta r
  \chi'(t_0) + f(r)$$ with $f(r) = \mathcal{O}(r^2)$ at $r = 0$ so that it is consistent with \eqref{Eq:Phi-exp} after setting $\phi^{(0)} = -\beta\chi$. 
  
The dynamics of the full system can be solved by the following iterative procedure. The initial conditions described above should be held fixed at each stage of the iteration.
 
\begin{figure}[t]
\centering
\includegraphics[scale=0.5]{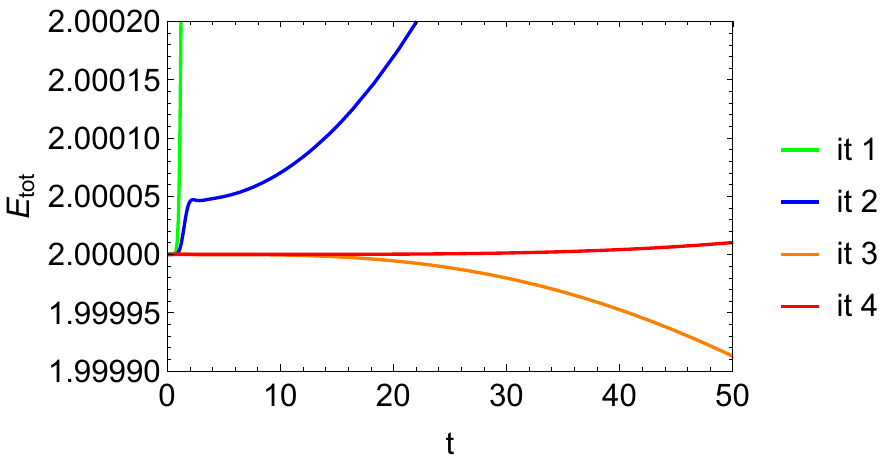}
\caption{The total energy is plotted for various iterations with initial conditions given by \eqref{Eq:IC1} with \textit{it 1} standing for first iteration, etc. We get convergence in 4 iterations as evident from the total energy being constant in the fourth iteration to a very good accuracy. }\label{Fig:iterations-IC1}
\end{figure}

\begin{enumerate}
\item We first solve the boundary scalar field equation \eqref{Eq:chi-hom} by setting its right hand side to zero. This implies that at the first step of the iteration $\chi(t) = \chi(t_0) + \chi'(t_0) t$ which gives the input to the holographic system by specifying the source $$\phi^{(0)} = -\beta(\chi(t_0) + \chi'(t_0) t)$$ for all time.

\item We proceed to solve the gravitational system of equations with the input source $\phi^{(0)}$ as determined above, and the initial bulk profile of $\Phi$ and the initial value of $a^{(3)}$. This can be achieved by utilizing the nested structure of the gravitational equations given by Eqs. \eqref{Eq:ASPhi1}-\eqref{Eq:ASPhi5} and employing the spectral method discussed in \cite{Chesler:2008hg,Chesler:2010bi}. We use 30 Chebyshev grid points to capture the dependence on the holographic radial coordinate and a fourth order Adam-Bashforth time-stepping to evolve in time. We also choose a suitable radial cutoff. Solving the full non-linear system of equations, we can extract $\phi^{(3)}(t)$ for from $\Phi(r,t)$.

\item Taking $\phi^{(3)}(t)$ and $\chi(t)$ as inputs from the previous iteration we now compute the right hand side of the boundary scalar field equation \eqref{Eq:chi-hom}. We solve this linear equation again with the source determined by the previous iteration and with the same initial conditions as before.

\item We again solve the gravitational system of equations with the source $\phi^{(0)}$ specified by the above solution of $\chi$. With the same initial radial profile of $\Phi$ and the initial value of $a^{(3)}$ as in the previous iteration, we obtain the new solutions for $S$, $A$ and $\Phi$. From the latter, we can readily update $\phi^{(3)}(t)$.

\item We again solve the boundary scalar field equation \eqref{Eq:chi-hom} with the specified initial conditions but now the right hand side determined by the inputs of the second iteration. This solution, which specify the boundary source, together with the initial conditions then uniquely determine the bulk metric and the bulk dilaton in the third iteration. We repeat our iterations until we get convergence to a desired numerical accuracy.
\end{enumerate}
For typical initial conditions, we achieve convergence in about four iterations. Once convergence is reached, we find that the total energy $E_{\rm tot}$ given by \eqref{Eq:E-tot} is also conserved to an excellent numerical accuracy. It is to be noted that formation of caustics limit the choices of initial profile of the bulk dilaton that can be simulated in this method of characteristics \cite{Chesler:2013lia}. We do not find any issue if the initial profile of the bulk dilaton is sufficiently localized in the radial direction. 
{In Fig. \ref{Fig:iterations-IC1}, we show how the total energy conservation improves with successive iterations. For a typical initial condition, we obtain convergence to a very good accuracy within just four iterations.}
 %We begin the simulation by specifying the initial profile for massless bulk scalar $\phi(r,t=0)$ along with initial values of $\phi^{(0)}(t)$, $\phi'_0(t)$and $a_3(t)$.
%The energy in each sector is given by
%\begin{equation}
%E_{scal} = \frac{1}{2} \chi'(t)^2 \qquad E_{bh} = -2 a^{3}(t) 
%\end{equation}
%where $a_3(t)$ is the normalizable mode term in the near-boundary expansion of $A(r,t)$. 
%The total energy is just the sum of the energies of two sectors.
%\begin{equation}\label{total-energy}
%E_{tot}  = \frac{1}{2}\chi'(t)^2-2 a_3(t) 
%\end{equation}
%We have used metric signature $(-,+,+,+)$. \\ \\
%The equation of motion for scalar field from the scalar sector $\chi(t)$ is 
%\begin{equation}
%\eta^{\mu \nu}\partial_{\mu}\partial_{\nu}\chi = \beta <\mathcal{O}>
%\end{equation}
%where $\beta$ is the semi-holograhic coupling. Since $\chi(t)$ is only function of $t$ in our case, eom of $\chi(t)$ reduces to 
%\begin{equation}
%\chi''(t) = - \beta <\mathcal{O}>
%\end{equation}
%Using the relation $\phi^{(0)} = \beta \chi$ and equations \eqref{vev}and \eqref{scalareom}, one can easily show that the total energy \eqref{total-energy} is conserved:
%\begin{equation}
%E_{tot}'(t) = \chi'(t)\chi''(t) - 2 a_3'(t) = 0
%\end{equation}

% \subsection{Numerical accuracy of the iterative procedure}\label{Sec:Appendix}

%\AR{\textbf{More on: Numerical accuracy of the iterative procedure?}}

\bibliographystyle{JHEP.bst}
\bibliography{semi_qnm_draft.bib}

\providecommand{\href}[2]{#2}\begingroup\raggedright\begin{thebibliography}{10}

\bibitem{Faulkner:2010tq}
T.~Faulkner and J.~Polchinski, \emph{{Semi-Holographic Fermi Liquids}},
  \href{https://doi.org/10.1007/JHEP06(2011)012}{\emph{JHEP} {\bfseries 06}
  (2011) 012}, [\href{https://arxiv.org/abs/1001.5049}{{\ttfamily 1001.5049}}].

\bibitem{Mukhopadhyay:2013dqa}
A.~Mukhopadhyay and G.~Policastro, \emph{{Phenomenological Characterization of
  Semiholographic Non-Fermi Liquids}},
  \href{https://doi.org/10.1103/PhysRevLett.111.221602}{\emph{Phys. Rev. Lett.}
  {\bfseries 111} (2013) 221602},
  [\href{https://arxiv.org/abs/1306.3941}{{\ttfamily 1306.3941}}].

\bibitem{Doucot:2020fvy}
B.~Dou\c{c}ot, A.~Mukhopadhyay, G.~Policastro and S.~Samanta,
  \emph{{Linear-in-T resistivity from semi-holographic non-Fermi liquid
  models}},  \href{https://arxiv.org/abs/2012.15679}{{\ttfamily 2012.15679}}.

\bibitem{Banerjee:2017ozx}
S.~Banerjee, N.~Gaddam and A.~Mukhopadhyay, \emph{{Illustrated study of the
  semiholographic nonperturbative framework}},
  \href{https://doi.org/10.1103/PhysRevD.95.066017}{\emph{Phys. Rev. D}
  {\bfseries 95} (2017) 066017},
  [\href{https://arxiv.org/abs/1701.01229}{{\ttfamily 1701.01229}}].

\bibitem{Kurkela:2018dku}
A.~Kurkela, A.~Mukhopadhyay, F.~Preis, A.~Rebhan and A.~Soloviev, \emph{{Hybrid
  Fluid Models from Mutual Effective Metric Couplings}},
  \href{https://doi.org/10.1007/JHEP08(2018)054}{\emph{JHEP} {\bfseries 08}
  (2018) 054}, [\href{https://arxiv.org/abs/1805.05213}{{\ttfamily
  1805.05213}}].

\bibitem{Iancu:2014ava}
E.~Iancu and A.~Mukhopadhyay, \emph{{A semi-holographic model for heavy-ion
  collisions}}, \href{https://doi.org/10.1007/JHEP06(2015)003}{\emph{JHEP}
  {\bfseries 06} (2015) 003},
  [\href{https://arxiv.org/abs/1410.6448}{{\ttfamily 1410.6448}}].

\bibitem{Mukhopadhyay:2015smb}
A.~Mukhopadhyay, F.~Preis, A.~Rebhan and S.~A. Stricker, \emph{{Semi-Holography
  for Heavy Ion Collisions: Self-Consistency and First Numerical Tests}},
  \href{https://doi.org/10.1007/JHEP05(2016)141}{\emph{JHEP} {\bfseries 05}
  (2016) 141}, [\href{https://arxiv.org/abs/1512.06445}{{\ttfamily
  1512.06445}}].

\bibitem{Ecker:2018ucc}
C.~Ecker, A.~Mukhopadhyay, F.~Preis, A.~Rebhan and A.~Soloviev, \emph{{Time
  evolution of a toy semiholographic glasma}},
  \href{https://doi.org/10.1007/JHEP08(2018)074}{\emph{JHEP} {\bfseries 08}
  (2018) 074}, [\href{https://arxiv.org/abs/1806.01850}{{\ttfamily
  1806.01850}}].

\bibitem{Mitra:2020mei}
T.~Mitra, S.~Mondkar, A.~Mukhopadhyay, A.~Rebhan and A.~Soloviev,
  \emph{{Hydrodynamic attractor of a hybrid viscous fluid in Bjorken flow}},
  \href{https://doi.org/10.1103/PhysRevResearch.2.043320}{\emph{Phys. Rev.
  Res.} {\bfseries 2} (2020) 043320},
  [\href{https://arxiv.org/abs/2006.09383}{{\ttfamily 2006.09383}}].

\bibitem{Baier:2000sb}
R.~Baier, A.~H. Mueller, D.~Schiff and D.~T. Son, \emph{{'Bottom up'
  thermalization in heavy ion collisions}},
  \href{https://doi.org/10.1016/S0370-2693(01)00191-5}{\emph{Phys. Lett. B}
  {\bfseries 502} (2001) 51--58},
  [\href{https://arxiv.org/abs/hep-ph/0009237}{{\ttfamily hep-ph/0009237}}].

\bibitem{Kovtun:2012rj}
P.~Kovtun, \emph{{Lectures on hydrodynamic fluctuations in relativistic
  theories}}, \href{https://doi.org/10.1088/1751-8113/45/47/473001}{\emph{J.
  Phys. A} {\bfseries 45} (2012) 473001},
  [\href{https://arxiv.org/abs/1205.5040}{{\ttfamily 1205.5040}}].

\bibitem{Grozdanov:2017ajz}
S.~Grozdanov, K.~Schalm and V.~Scopelliti, \emph{{Black hole scrambling from
  hydrodynamics}},
  \href{https://doi.org/10.1103/PhysRevLett.120.231601}{\emph{Phys. Rev. Lett.}
  {\bfseries 120} (2018) 231601},
  [\href{https://arxiv.org/abs/1710.00921}{{\ttfamily 1710.00921}}].

\bibitem{Grozdanov:2018fic}
S.~Grozdanov, A.~Lucas and N.~Poovuttikul, \emph{{Holography and hydrodynamics
  with weakly broken symmetries}},
  \href{https://doi.org/10.1103/PhysRevD.99.086012}{\emph{Phys. Rev. D}
  {\bfseries 99} (2019) 086012},
  [\href{https://arxiv.org/abs/1810.10016}{{\ttfamily 1810.10016}}].

\bibitem{Hartnoll:2014lpa}
S.~A. Hartnoll, \emph{{Theory of universal incoherent metallic transport}},
  \href{https://doi.org/10.1038/nphys3174}{\emph{Nature Phys.} {\bfseries 11}
  (2015) 54}, [\href{https://arxiv.org/abs/1405.3651}{{\ttfamily 1405.3651}}].

\bibitem{Blake:2016wvh}
M.~Blake, \emph{{Universal Charge Diffusion and the Butterfly Effect in
  Holographic Theories}},
  \href{https://doi.org/10.1103/PhysRevLett.117.091601}{\emph{Phys. Rev. Lett.}
  {\bfseries 117} (2016) 091601},
  [\href{https://arxiv.org/abs/1603.08510}{{\ttfamily 1603.08510}}].

\bibitem{Blake:2016sud}
M.~Blake, \emph{{Universal Diffusion in Incoherent Black Holes}},
  \href{https://doi.org/10.1103/PhysRevD.94.086014}{\emph{Phys. Rev. D}
  {\bfseries 94} (2016) 086014},
  [\href{https://arxiv.org/abs/1604.01754}{{\ttfamily 1604.01754}}].

\bibitem{Hartman:2017hhp}
T.~Hartman, S.~A. Hartnoll and R.~Mahajan, \emph{{Upper Bound on Diffusivity}},
  \href{https://doi.org/10.1103/PhysRevLett.119.141601}{\emph{Phys. Rev. Lett.}
  {\bfseries 119} (2017) 141601},
  [\href{https://arxiv.org/abs/1706.00019}{{\ttfamily 1706.00019}}].

\bibitem{Lucas:2017ibu}
A.~Lucas, \emph{{Constraints on hydrodynamics from many-body quantum chaos}},
  \href{https://arxiv.org/abs/1710.01005}{{\ttfamily 1710.01005}}.

\bibitem{Grozdanov:2020koi}
S.~Grozdanov, \emph{{Bounds on transport from univalence and pole-skipping}},
  \href{https://doi.org/10.1103/PhysRevLett.126.051601}{\emph{Phys. Rev. Lett.}
  {\bfseries 126} (2021) 051601},
  [\href{https://arxiv.org/abs/2008.00888}{{\ttfamily 2008.00888}}].

\bibitem{Joshi:2019wgi}
L.~K. Joshi, A.~Mukhopadhyay and A.~Soloviev, \emph{{Time-dependent $NAdS_2$
  holography with applications}},
  \href{https://doi.org/10.1103/PhysRevD.101.066001}{\emph{Phys. Rev. D}
  {\bfseries 101} (2020) 066001},
  [\href{https://arxiv.org/abs/1901.08877}{{\ttfamily 1901.08877}}].

\bibitem{Kibe:2020gkx}
T.~Kibe, A.~Mukhopadhyay, A.~Soloviev and H.~Swain, \emph{{$SL(2,R)$ lattices
  as information processors}},
  \href{https://doi.org/10.1103/PhysRevD.102.086008}{\emph{Phys. Rev. D}
  {\bfseries 102} (2020) 086008},
  [\href{https://arxiv.org/abs/2006.08644}{{\ttfamily 2006.08644}}].

\bibitem{Doucot:2017bdm}
B.~Dou\c{c}ot, C.~Ecker, A.~Mukhopadhyay and G.~Policastro, \emph{{Density
  response and collective modes of semiholographic non-Fermi liquids}},
  \href{https://doi.org/10.1103/PhysRevD.96.106011}{\emph{Phys. Rev. D}
  {\bfseries 96} (2017) 106011},
  [\href{https://arxiv.org/abs/1706.04975}{{\ttfamily 1706.04975}}].

\bibitem{Balasubramanian:1999re}
V.~Balasubramanian and P.~Kraus, \emph{{A Stress tensor for Anti-de Sitter
  gravity}}, \href{https://doi.org/10.1007/s002200050764}{\emph{Commun. Math.
  Phys.} {\bfseries 208} (1999) 413--428},
  [\href{https://arxiv.org/abs/hep-th/9902121}{{\ttfamily hep-th/9902121}}].

\bibitem{deHaro:2000vlm}
S.~de~Haro, S.~N. Solodukhin and K.~Skenderis, \emph{{Holographic
  reconstruction of space-time and renormalization in the AdS / CFT
  correspondence}}, \href{https://doi.org/10.1007/s002200100381}{\emph{Commun.
  Math. Phys.} {\bfseries 217} (2001) 595--622},
  [\href{https://arxiv.org/abs/hep-th/0002230}{{\ttfamily hep-th/0002230}}].

\bibitem{Skenderis:2002wp}
K.~Skenderis, \emph{{Lecture notes on holographic renormalization}},
  \href{https://doi.org/10.1088/0264-9381/19/22/306}{\emph{Class. Quant. Grav.}
  {\bfseries 19} (2002) 5849--5876},
  [\href{https://arxiv.org/abs/hep-th/0209067}{{\ttfamily hep-th/0209067}}].

\bibitem{Yaffe}
L.~Yaffe, ``{Mathematica Summer School for Theoretical Physics - Numerical
  holography using Mathematica}.'' http://msstp.org/?q=node/289, 2020.

\bibitem{Berti:2009kk}
E.~Berti, V.~Cardoso and A.~O. Starinets, \emph{{Quasinormal modes of black
  holes and black branes}},
  \href{https://doi.org/10.1088/0264-9381/26/16/163001}{\emph{Class. Quant.
  Grav.} {\bfseries 26} (2009) 163001},
  [\href{https://arxiv.org/abs/0905.2975}{{\ttfamily 0905.2975}}].

\bibitem{Starinets:2002br}
A.~O. Starinets, \emph{{Quasinormal modes of near extremal black branes}},
  \href{https://doi.org/10.1103/PhysRevD.66.124013}{\emph{Phys. Rev. D}
  {\bfseries 66} (2002) 124013},
  [\href{https://arxiv.org/abs/hep-th/0207133}{{\ttfamily hep-th/0207133}}].

\bibitem{Hartnoll:2008kx}
S.~A. Hartnoll, C.~P. Herzog and G.~T. Horowitz, \emph{{Holographic
  Superconductors}},
  \href{https://doi.org/10.1088/1126-6708/2008/12/015}{\emph{JHEP} {\bfseries
  12} (2008) 015}, [\href{https://arxiv.org/abs/0810.1563}{{\ttfamily
  0810.1563}}].

\bibitem{Hollands:2012sf}
S.~Hollands and R.~M. Wald, \emph{{Stability of Black Holes and Black Branes}},
  \href{https://doi.org/10.1007/s00220-012-1638-1}{\emph{Commun. Math. Phys.}
  {\bfseries 321} (2013) 629--680},
  [\href{https://arxiv.org/abs/1201.0463}{{\ttfamily 1201.0463}}].

\bibitem{Amado:2009ts}
I.~Amado, M.~Kaminski and K.~Landsteiner, \emph{{Hydrodynamics of Holographic
  Superconductors}},
  \href{https://doi.org/10.1088/1126-6708/2009/05/021}{\emph{JHEP} {\bfseries
  05} (2009) 021}, [\href{https://arxiv.org/abs/0903.2209}{{\ttfamily
  0903.2209}}].

\bibitem{Baggioli:2019jcm}
M.~Baggioli, M.~Vasin, V.~V. Brazhkin and K.~Trachenko, \emph{{Gapped momentum
  states}}, \href{https://doi.org/10.1016/j.physrep.2020.04.002}{\emph{Phys.
  Rept.} {\bfseries 865} (2020) 1--44},
  [\href{https://arxiv.org/abs/1904.01419}{{\ttfamily 1904.01419}}].

\bibitem{Baggioli:2018vfc}
M.~Baggioli and K.~Trachenko, \emph{{Low frequency propagating shear waves in
  holographic liquids}},
  \href{https://doi.org/10.1007/JHEP03(2019)093}{\emph{JHEP} {\bfseries 03}
  (2019) 093}, [\href{https://arxiv.org/abs/1807.10530}{{\ttfamily
  1807.10530}}].

\bibitem{Baggioli:2018nnp}
M.~Baggioli and K.~Trachenko, \emph{{Maxwell interpolation and close
  similarities between liquids and holographic models}},
  \href{https://doi.org/10.1103/PhysRevD.99.106002}{\emph{Phys. Rev. D}
  {\bfseries 99} (2019) 106002},
  [\href{https://arxiv.org/abs/1808.05391}{{\ttfamily 1808.05391}}].

\bibitem{PhysRevB.101.214312}
R.~M. Khusnutdinoff, C.~Cockrell, O.~A. Dicks, A.~C.~S. Jensen, M.~D. Le,
  L.~Wang et~al., \emph{Collective modes and gapped momentum states in liquid
  {Ga}: Experiment, theory, and simulation},
  \href{https://doi.org/10.1103/PhysRevB.101.214312}{\emph{Phys. Rev. B}
  {\bfseries 101} (Jun, 2020) 214312}.

\bibitem{Davison:2014lua}
R.~A. Davison and B.~Gout\'eraux, \emph{{Momentum dissipation and effective
  theories of coherent and incoherent transport}},
  \href{https://doi.org/10.1007/JHEP01(2015)039}{\emph{JHEP} {\bfseries 01}
  (2015) 039}, [\href{https://arxiv.org/abs/1411.1062}{{\ttfamily 1411.1062}}].

\bibitem{Baggioli:2020loj}
M.~Baggioli, \emph{{How small hydrodynamics can go}},
  \href{https://doi.org/10.1103/PhysRevD.103.086001}{\emph{Phys. Rev. D}
  {\bfseries 103} (2021) 086001},
  [\href{https://arxiv.org/abs/2010.05916}{{\ttfamily 2010.05916}}].

\bibitem{Ammon:2019wci}
M.~Ammon, M.~Baggioli and A.~Jim\'enez-Alba, \emph{{A Unified Description of
  Translational Symmetry Breaking in Holography}},
  \href{https://doi.org/10.1007/JHEP09(2019)124}{\emph{JHEP} {\bfseries 09}
  (2019) 124}, [\href{https://arxiv.org/abs/1904.05785}{{\ttfamily
  1904.05785}}].

\bibitem{Grozdanov:2018ewh}
S.~Grozdanov and N.~Poovuttikul, \emph{{Generalized global symmetries in states
  with dynamical defects: The case of the transverse sound in field theory and
  holography}}, \href{https://doi.org/10.1103/PhysRevD.97.106005}{\emph{Phys.
  Rev. D} {\bfseries 97} (2018) 106005},
  [\href{https://arxiv.org/abs/1801.03199}{{\ttfamily 1801.03199}}].

\bibitem{Hayata:2014yga}
T.~Hayata and Y.~Hidaka, \emph{{Dispersion relations of Nambu-Goldstone modes
  at finite temperature and density}},
  \href{https://doi.org/10.1103/PhysRevD.91.056006}{\emph{Phys. Rev. D}
  {\bfseries 91} (2015) 056006},
  [\href{https://arxiv.org/abs/1406.6271}{{\ttfamily 1406.6271}}].

\bibitem{Hidaka:2019irz}
Y.~Hidaka and Y.~Minami, \emph{{Spontaneous symmetry breaking and
  Nambu\textendash{}Goldstone modes in open classical and quantum systems}},
  \href{https://doi.org/10.1093/ptep/ptaa005}{\emph{PTEP} {\bfseries 2020}
  (2020) 033A01}, [\href{https://arxiv.org/abs/1907.08241}{{\ttfamily
  1907.08241}}].

\bibitem{Baggioli:2021xuv}
M.~Baggioli, K.-Y. Kim, L.~Li and W.-J. Li, \emph{{Holographic Axion Model: a
  simple gravitational tool for quantum matter}},
  \href{https://doi.org/10.1007/s11433-021-1681-8}{\emph{Sci. China Phys. Mech.
  Astron.} {\bfseries 64} (2021) 270001},
  [\href{https://arxiv.org/abs/2101.01892}{{\ttfamily 2101.01892}}].

\bibitem{Hayata:2018qgt}
T.~Hayata and Y.~Hidaka, \emph{{Diffusive Nambu-Goldstone modes in quantum
  time-crystals}},  \href{https://arxiv.org/abs/1808.07636}{{\ttfamily
  1808.07636}}.

\bibitem{Gregory:1993vy}
R.~Gregory and R.~Laflamme, \emph{{Black strings and p-branes are unstable}},
  \href{https://doi.org/10.1103/PhysRevLett.70.2837}{\emph{Phys. Rev. Lett.}
  {\bfseries 70} (1993) 2837--2840},
  [\href{https://arxiv.org/abs/hep-th/9301052}{{\ttfamily hep-th/9301052}}].

\bibitem{Lehner:2011wc}
L.~Lehner and F.~Pretorius, \emph{{Final state of Gregory\textendash{}Laflamme
  instability}}.
\newblock 2012.
\newblock \href{https://arxiv.org/abs/1106.5184}{{\ttfamily 1106.5184}}.

\bibitem{Emparan:2015gva}
R.~Emparan, R.~Suzuki and K.~Tanabe, \emph{{Evolution and End Point of the
  Black String Instability: Large D Solution}},
  \href{https://doi.org/10.1103/PhysRevLett.115.091102}{\emph{Phys. Rev. Lett.}
  {\bfseries 115} (2015) 091102},
  [\href{https://arxiv.org/abs/1506.06772}{{\ttfamily 1506.06772}}].

\bibitem{Arean:2020eus}
D.~Arean, R.~A. Davison, B.~Gout\'eraux and K.~Suzuki, \emph{{Hydrodynamic
  Diffusion and Its Breakdown near AdS2 Quantum Critical Points}},
  \href{https://doi.org/10.1103/PhysRevX.11.031024}{\emph{Phys. Rev. X}
  {\bfseries 11} (2021) 031024},
  [\href{https://arxiv.org/abs/2011.12301}{{\ttfamily 2011.12301}}].

\bibitem{Wu:2021mkk}
N.~Wu, M.~Baggioli and W.-J. Li, \emph{{On the universality of AdS$_{2}$
  diffusion bounds and the breakdown of linearized hydrodynamics}},
  \href{https://doi.org/10.1007/JHEP05(2021)014}{\emph{JHEP} {\bfseries 05}
  (2021) 014}, [\href{https://arxiv.org/abs/2102.05810}{{\ttfamily
  2102.05810}}].

\bibitem{Jeong:2021zsv}
H.-S. Jeong, K.-Y. Kim and Y.-W. Sun, \emph{{The breakdown of
  magneto-hydrodynamics near AdS$_2$ fixed point and energy diffusion bound}},
  \href{https://arxiv.org/abs/2105.03882}{{\ttfamily 2105.03882}}.

\bibitem{Baggioli:2020ljz}
M.~Baggioli and W.-J. Li, \emph{{Universal Bounds on Transport in Holographic
  Systems with Broken Translations}},
  \href{https://doi.org/10.21468/SciPostPhys.9.1.007}{\emph{SciPost Phys.}
  {\bfseries 9} (2020) 007},
  [\href{https://arxiv.org/abs/2005.06482}{{\ttfamily 2005.06482}}].

\bibitem{Blake:2017ris}
M.~Blake, H.~Lee and H.~Liu, \emph{{A quantum hydrodynamical description for
  scrambling and many-body chaos}},
  \href{https://doi.org/10.1007/JHEP10(2018)127}{\emph{JHEP} {\bfseries 10}
  (2018) 127}, [\href{https://arxiv.org/abs/1801.00010}{{\ttfamily
  1801.00010}}].

\bibitem{Blake:2018leo}
M.~Blake, R.~A. Davison, S.~Grozdanov and H.~Liu, \emph{{Many-body chaos and
  energy dynamics in holography}},
  \href{https://doi.org/10.1007/JHEP10(2018)035}{\emph{JHEP} {\bfseries 10}
  (2018) 035}, [\href{https://arxiv.org/abs/1809.01169}{{\ttfamily
  1809.01169}}].

\bibitem{Grozdanov:2019uhi}
S.~Grozdanov, P.~K. Kovtun, A.~O. Starinets and P.~Tadi\'c, \emph{{The complex
  life of hydrodynamic modes}},
  \href{https://doi.org/10.1007/JHEP11(2019)097}{\emph{JHEP} {\bfseries 11}
  (2019) 097}, [\href{https://arxiv.org/abs/1904.12862}{{\ttfamily
  1904.12862}}].

\bibitem{Roberts:2016wdl}
D.~A. Roberts and B.~Swingle, \emph{{Lieb-Robinson Bound and the Butterfly
  Effect in Quantum Field Theories}},
  \href{https://doi.org/10.1103/PhysRevLett.117.091602}{\emph{Phys. Rev. Lett.}
  {\bfseries 117} (2016) 091602},
  [\href{https://arxiv.org/abs/1603.09298}{{\ttfamily 1603.09298}}].

\bibitem{Gu:2016oyy}
Y.~Gu, X.-L. Qi and D.~Stanford, \emph{{Local criticality, diffusion and chaos
  in generalized Sachdev-Ye-Kitaev models}},
  \href{https://doi.org/10.1007/JHEP05(2017)125}{\emph{JHEP} {\bfseries 05}
  (2017) 125}, [\href{https://arxiv.org/abs/1609.07832}{{\ttfamily
  1609.07832}}].

\bibitem{Jeong:2021zhz}
H.-S. Jeong, K.-Y. Kim and Y.-W. Sun, \emph{{Bound of diffusion constants from
  pole-skipping points: spontaneous symmetry breaking and magnetic field}},
  \href{https://arxiv.org/abs/2104.13084}{{\ttfamily 2104.13084}}.

\bibitem{Kovtun:2004de}
P.~Kovtun, D.~T. Son and A.~O. Starinets, \emph{{Viscosity in strongly
  interacting quantum field theories from black hole physics}},
  \href{https://doi.org/10.1103/PhysRevLett.94.111601}{\emph{Phys. Rev. Lett.}
  {\bfseries 94} (2005) 111601},
  [\href{https://arxiv.org/abs/hep-th/0405231}{{\ttfamily hep-th/0405231}}].

\bibitem{Shenker:2013pqa}
S.~H. Shenker and D.~Stanford, \emph{{Black holes and the butterfly effect}},
  \href{https://doi.org/10.1007/JHEP03(2014)067}{\emph{JHEP} {\bfseries 03}
  (2014) 067}, [\href{https://arxiv.org/abs/1306.0622}{{\ttfamily 1306.0622}}].

\bibitem{Romatschke:2005pm}
P.~Romatschke and R.~Venugopalan, \emph{{Collective non-Abelian instabilities
  in a melting color glass condensate}},
  \href{https://doi.org/10.1103/PhysRevLett.96.062302}{\emph{Phys. Rev. Lett.}
  {\bfseries 96} (2006) 062302},
  [\href{https://arxiv.org/abs/hep-ph/0510121}{{\ttfamily hep-ph/0510121}}].

\bibitem{Mrowczynski:1993qm}
S.~{Mr{\'o}wczy{\'n}ski}, \emph{{Plasma instability at the initial stage of
  ultrarelativistic heavy ion collisions}},
  \href{https://doi.org/10.1016/0370-2693(93)91330-P}{\emph{Phys. Lett. B}
  {\bfseries 314} (1993) 118--121}.

\bibitem{Romatschke:2003ms}
P.~Romatschke and M.~Strickland, \emph{{Collective modes of an anisotropic
  quark gluon plasma}},
  \href{https://doi.org/10.1103/PhysRevD.68.036004}{\emph{Phys. Rev. D}
  {\bfseries 68} (2003) 036004},
  [\href{https://arxiv.org/abs/hep-ph/0304092}{{\ttfamily hep-ph/0304092}}].

\bibitem{Arnold:2003rq}
P.~B. Arnold, J.~Lenaghan and G.~D. Moore, \emph{{QCD plasma instabilities and
  bottom up thermalization}},
  \href{https://doi.org/10.1088/1126-6708/2003/08/002}{\emph{JHEP} {\bfseries
  08} (2003) 002}, [\href{https://arxiv.org/abs/hep-ph/0307325}{{\ttfamily
  hep-ph/0307325}}].

\bibitem{Romatschke:2006wg}
P.~Romatschke and A.~Rebhan, \emph{{Plasma Instabilities in an Anisotropically
  Expanding Geometry}},
  \href{https://doi.org/10.1103/PhysRevLett.97.252301}{\emph{Phys. Rev. Lett.}
  {\bfseries 97} (2006) 252301},
  [\href{https://arxiv.org/abs/hep-ph/0605064}{{\ttfamily hep-ph/0605064}}].

\bibitem{Rebhan:2009ku}
A.~Rebhan and D.~Steineder, \emph{{Collective modes and instabilities in
  anisotropically expanding ultrarelativistic plasmas}},
  \href{https://doi.org/10.1103/PhysRevD.81.085044}{\emph{Phys. Rev. D}
  {\bfseries 81} (2010) 085044},
  [\href{https://arxiv.org/abs/0912.5383}{{\ttfamily 0912.5383}}].

\bibitem{Berges:2013eia}
J.~Berges, K.~Boguslavski, S.~Schlichting and R.~Venugopalan, \emph{{Turbulent
  thermalization process in heavy-ion collisions at ultrarelativistic
  energies}}, \href{https://doi.org/10.1103/PhysRevD.89.074011}{\emph{Phys.
  Rev. D} {\bfseries 89} (2014) 074011},
  [\href{https://arxiv.org/abs/1303.5650}{{\ttfamily 1303.5650}}].

\bibitem{Amoretti:2018tzw}
A.~Amoretti, D.~Are\'an, B.~Gout\'eraux and D.~Musso, \emph{{Universal
  relaxation in a holographic metallic density wave phase}},
  \href{https://doi.org/10.1103/PhysRevLett.123.211602}{\emph{Phys. Rev. Lett.}
  {\bfseries 123} (2019) 211602},
  [\href{https://arxiv.org/abs/1812.08118}{{\ttfamily 1812.08118}}].

\bibitem{Ammon:2019apj}
M.~Ammon, M.~Baggioli, S.~Gray and S.~Grieninger, \emph{{Longitudinal Sound and
  Diffusion in Holographic Massive Gravity}},
  \href{https://doi.org/10.1007/JHEP10(2019)064}{\emph{JHEP} {\bfseries 10}
  (2019) 064}, [\href{https://arxiv.org/abs/1905.09164}{{\ttfamily
  1905.09164}}].

\bibitem{Donos:2019txg}
A.~Donos, D.~Martin, C.~Pantelidou and V.~Ziogas, \emph{{Hydrodynamics of
  broken global symmetries in the bulk}},
  \href{https://doi.org/10.1007/JHEP10(2019)218}{\emph{JHEP} {\bfseries 10}
  (2019) 218}, [\href{https://arxiv.org/abs/1905.00398}{{\ttfamily
  1905.00398}}].

\bibitem{Baggioli:2020nay}
M.~Baggioli, \emph{{Homogeneous holographic viscoelastic models and
  quasicrystals}},
  \href{https://doi.org/10.1103/PhysRevResearch.2.022022}{\emph{Phys. Rev.
  Res.} {\bfseries 2} (2020) 022022},
  [\href{https://arxiv.org/abs/2001.06228}{{\ttfamily 2001.06228}}].

\bibitem{Ghosh:2020lel}
J.~K. Ghosh, R.~Loganayagam, S.~G. Prabhu, M.~Rangamani, A.~Sivakumar and
  V.~Vishal, \emph{{Effective field theory of stochastic diffusion from
  gravity}},  \href{https://arxiv.org/abs/2012.03999}{{\ttfamily 2012.03999}}.

\bibitem{Janik:2016btb}
R.~A. Janik, J.~Jankowski and H.~Soltanpanahi, \emph{{Quasinormal modes and the
  phase structure of strongly coupled matter}},
  \href{https://doi.org/10.1007/JHEP06(2016)047}{\emph{JHEP} {\bfseries 06}
  (2016) 047}, [\href{https://arxiv.org/abs/1603.05950}{{\ttfamily
  1603.05950}}].

\bibitem{Gursoy:2016ggq}
U.~G\"ursoy, A.~Jansen and W.~van~der Schee, \emph{{New dynamical instability
  in asymptotically anti\textendash{}de Sitter spacetime}},
  \href{https://doi.org/10.1103/PhysRevD.94.061901}{\emph{Phys. Rev. D}
  {\bfseries 94} (2016) 061901},
  [\href{https://arxiv.org/abs/1603.07724}{{\ttfamily 1603.07724}}].

\bibitem{Hubeny:2011hd}
V.~E. Hubeny, S.~Minwalla and M.~Rangamani, \emph{{The fluid/gravity
  correspondence}},  in \emph{{Theoretical Advanced Study Institute in
  Elementary Particle Physics}: {String theory and its Applications: From meV
  to the Planck Scale}}, 7, 2011,
  \href{https://arxiv.org/abs/1107.5780}{{\ttfamily 1107.5780}}.

\bibitem{Romatschke:2009kr}
P.~Romatschke, \emph{{Relativistic Viscous Fluid Dynamics and Non-Equilibrium
  Entropy}}, \href{https://doi.org/10.1088/0264-9381/27/2/025006}{\emph{Class.
  Quant. Grav.} {\bfseries 27} (2010) 025006},
  [\href{https://arxiv.org/abs/0906.4787}{{\ttfamily 0906.4787}}].

\bibitem{Banerjee:2012iz}
N.~Banerjee, J.~Bhattacharya, S.~Bhattacharyya, S.~Jain, S.~Minwalla and
  T.~Sharma, \emph{{Constraints on Fluid Dynamics from Equilibrium Partition
  Functions}}, \href{https://doi.org/10.1007/JHEP09(2012)046}{\emph{JHEP}
  {\bfseries 09} (2012) 046},
  [\href{https://arxiv.org/abs/1203.3544}{{\ttfamily 1203.3544}}].

\bibitem{Iyer:2009in}
R.~Iyer and A.~Mukhopadhyay, \emph{{An AdS/CFT Connection between Boltzmann and
  Einstein}}, \href{https://doi.org/10.1103/PhysRevD.81.086005}{\emph{Phys.
  Rev. D} {\bfseries 81} (2010) 086005},
  [\href{https://arxiv.org/abs/0907.1156}{{\ttfamily 0907.1156}}].

\bibitem{Iyer:2011qc}
R.~Iyer and A.~Mukhopadhyay, \emph{{Homogeneous Relaxation at Strong Coupling
  from Gravity}}, \href{https://doi.org/10.1103/PhysRevD.84.126013}{\emph{Phys.
  Rev. D} {\bfseries 84} (2011) 126013},
  [\href{https://arxiv.org/abs/1103.1814}{{\ttfamily 1103.1814}}].

\bibitem{Chesler:2008hg}
P.~M. Chesler and L.~G. Yaffe, \emph{{Horizon formation and
  far-from-equilibrium isotropization in supersymmetric Yang-Mills plasma}},
  \href{https://doi.org/10.1103/PhysRevLett.102.211601}{\emph{Phys. Rev. Lett.}
  {\bfseries 102} (2009) 211601},
  [\href{https://arxiv.org/abs/0812.2053}{{\ttfamily 0812.2053}}].

\bibitem{Chesler:2010bi}
P.~M. Chesler and L.~G. Yaffe, \emph{{Holography and colliding gravitational
  shock waves in asymptotically AdS$_{5}$ spacetime}},
  \href{https://doi.org/10.1103/PhysRevLett.106.021601}{\emph{Phys. Rev. Lett.}
  {\bfseries 106} (2011) 021601},
  [\href{https://arxiv.org/abs/1011.3562}{{\ttfamily 1011.3562}}].

\bibitem{Chesler:2013lia}
P.~M. Chesler and L.~G. Yaffe, \emph{{Numerical solution of gravitational
  dynamics in asymptotically anti-de Sitter spacetimes}},
  \href{https://doi.org/10.1007/JHEP07(2014)086}{\emph{JHEP} {\bfseries 07}
  (2014) 086}, [\href{https://arxiv.org/abs/1309.1439}{{\ttfamily 1309.1439}}].

\end{thebibliography}\endgroup

\end{document}